\PassOptionsToPackage{x11names, table}{xcolor}
\documentclass[11pt]{article}
\pdfoutput=1
\usepackage{dcolumn}
\usepackage{bm}

\usepackage[square,numbers,compress,sort]{natbib}
\usepackage{graphicx}
\usepackage{amssymb,amsmath}
\usepackage{multirow}
\usepackage{slashed,xfrac}
\usepackage[makeroom]{cancel}
\usepackage{rotating}
\usepackage{booktabs}
\usepackage{color,url}
\usepackage[colorlinks=true
,urlcolor=blue
,anchorcolor=blue
,citecolor=blue
,filecolor=blue
,linkcolor=blue
,menucolor=blue
,linktocpage=true
,pdfproducer=medialab
,pdfa=true
]{hyperref}

\usepackage{mciteplus}
\usepackage{epsfig,psfrag,rotating,soul}
\usepackage{rotfloat}
\usepackage[normalem]{ulem}
\usepackage{framed,fancybox,xcolor,xfrac,geometry}
\usepackage[x11names,table]{xcolor}
\usepackage{hhline}
\usepackage[textwidth=1.7cm,color=green!85!black,textsize=tiny]{todonotes}

\renewcommand\eqref[1]{Eq.~(\ref{#1})}
\newcommand\eqrefs[2]{Eqs.~(\ref{#1})-(\ref{#2})}
\newcommand\figref[1]{Fig.~\ref{#1}}

\newcommand\tabref[1]{Table~\ref{#1}}

\evensidemargin \oddsidemargin
\allowdisplaybreaks

\newcommand{\be}{\begin{equation}}
\newcommand{\ee}{\end{equation}}
\newcommand{\bear}{\begin{eqnarray}}
\newcommand{\eear}{\end{eqnarray}}

\newcommand{\mO}{\mathcal{O}}

\newcommand{\ifb}{${\rm fb}^{-1}$}
\newcommand{\bbbbjj}{$pp\to b\bar{b}b\bar{b}jj$ }

\def\thefootnote{\fnsymbol{footnote}}

\let\OLDthebibliography\thebibliography
\renewcommand\thebibliography[1]{
  \OLDthebibliography{#1}
  \setlength{\parskip}{0pt}
  \setlength{\itemsep}{0pt plus 0.3ex}
}

\begin{document}
\thispagestyle{empty}

\begin{flushright}
IFT-UAM/CSIC-18-66  \\
FTUAM-18-17\\
\end{flushright}
~
~
\vspace{0.5cm}

\begin{center}

\begin{Large}
\textbf{\textsc{Probing the Higgs self-coupling through double Higgs production\\
[.25em] in vector boson scattering at the LHC}}
\end{Large}

\vspace{1cm}

{\sc
Ernesto Arganda$^{1,2}$%
\footnote{\tt \href{mailto:ernesto.arganda@fisica.unlp.edu.ar}{ernesto.arganda@fisica.unlp.edu.ar}}%
, Claudia Garcia-Garcia$^2$%
\footnote{\tt \href{mailto:claudia.garcia@uam.es}{claudia.garcia@uam.es}}%
}
and \sc{ Maria Jose Herrero$^2$%
\footnote{\tt \href{mailto:maria.herrero@uam.es}{maria.herrero@uam.es}}%
}

\vspace*{.7cm}

{\sl

$^1$IFLP, CONICET - Dpto. de F\'isica, Universidad Nacional de La Plata,\\
C.C. 67, 1900 La Plata, Argentina

\vspace*{0.1cm}

$^2$Departamento de F\'{\i}sica Te\'orica and Instituto de F\'{\i}sica Te\'orica, IFT-UAM/CSIC,\\
Universidad Aut\'onoma de Madrid, Cantoblanco, 28049 Madrid, Spain
}

\end{center}

\vspace*{0.1cm}

\begin{abstract}
In this work we explore the sensitivity  to the Higgs self-coupling $\lambda$ in the production of two Higgs bosons via vector boson scattering at the LHC. Although these production channels, concretely $W^+W^- \to HH$ and $ ZZ \to HH$,  have lower rates than gluon-gluon fusion, they benefit from being  tree level processes, being independent of top physics and having very distinctive kinematics that allow us to obtain very clean experimental signatures. This makes them competitive channels concerning the sensitivity to the Higgs self-coupling. In order to give predictions for the sensitivity to this coupling, we first study the role of $\lambda$ at the subprocess level, both in and beyond the Standard Model, to move afterwards to the LHC scenario. We characterize the $pp\to HHjj$ case first and then provide quantitative results for the values of $\lambda$ that can be probed at the LHC in vector boson scattering processes after considering the Higgs boson decays. We focus mainly on $pp\to b\bar{b}b\bar{b}jj$, since it has the largest signal rates, and also comment on the potential of other channels, such as $pp\to b\bar{b}\gamma\gamma jj$, as they lead to cleaner, although smaller, signals. Our whole study is performed for a center of mass energy of $\sqrt{s}=14$ TeV and for various future expected LHC luminosities.

\end{abstract}

\def\thefootnote{\arabic{footnote}}
\setcounter{page}{0}
\setcounter{footnote}{0}

\newpage


\section{Introduction}
\label{intro}

The observation of the Higgs boson by the ATLAS and CMS experiments \cite{Aad:2012tfa, Chatrchyan:2012xdj} in 2012 confirmed the prediction of the last particle of the Standard Model (SM) of fundamental interactions. Although this discovery allowed us to answer many important and well established questions about elementary particle physics, it also posed a lot of new mysteries concerning the scalar sector of the SM.

One of these mysteries is that of the true value of the Higgs self-coupling $\lambda$, involved in trilinear and quartic Higgs self-interactions, and its relation to other parameters of the SM.  Particularly, understanding and testing experimentally the relation between $\lambda$ and the Higgs boson mass, $m_H$, will provide an excellent insight into the real nature of the Higgs particle. This relation, given in the SM at the tree level by $m_H^2=2 v^2 \lambda$, with $v=246$ GeV, arises from the Brout-Englert-Higgs (BEH) mechanism \cite{Higgs:1964ia, Englert:1964et, Higgs:1964pj, Higgs:1966ev}, so to really test this theoretical framework one needs to measure $\lambda$ independently of the Higgs mass. Unfortunately, the  value of the Higgs self-coupling has not been established yet with precision at colliders, but there is (and will be in the future) a very intense experimental program focused on the realization of this measurement (for a review, see for instance \cite{Simon:2012ik, Dawson:2013bba, Baer:2013cma, Abramowicz:2016zbo, deFlorian:2016spz}).

The Higgs trilinear coupling can be probed in double Higgs production processes at the LHC, process that have been extensively studied both theoretically 
in \cite{Glover:1987nx, Dicus:1987ic, Plehn:1996wb, Dawson:1998py, Djouadi:1999rca, Baur:2003gp, Grober:2010yv, Dolan:2012rv, Papaefstathiou:2012qe, Baglio:2012np, Yao:2013ika, deFlorian:2013jea, Dolan:2013rja, Barger:2013jfa,  Frederix:2014hta, Liu-Sheng:2014gxa, Goertz:2014qta, Azatov:2015oxa, Dicus:2015yva, Dawson:2015oha, He:2015spf, Dolan:2015zja, Cao:2015oaa, Cao:2015oxx, Huang:2015tdv, Behr:2015oqq, Kling:2016lay, Borowka:2016ypz, Bishara:2016kjn, Cao:2016zob, deFlorian:2016spz, Adhikary:2017jtu, Kim:2018uty, Banerjee:2018yxy, Goncalves:2018qas, Bizon:2018syu, Borowka:2018pxx, Gorbahn:2019lwq}, and experimentally in \cite{Aaboud:2016xco, CMS:2016foy,  CMS:2017ihs, Aaboud:2018knk,  Sirunyan:2018iwt, Aaboud:2018ftw}. At hadron colliders, these processes can take place through a variety of production channels, being gluon-gluon fusion (GGF) and vector boson scattering (VBS), also called vector boson fusion (VBF) in the literature, the main ones regarding the sensitivity to the Higgs self-coupling. Focusing on the LHC case, on which we will base our posterior study, the dominant contribution to double Higgs production comes from GGF, which for $\sqrt{s}=14$ TeV is about a factor 17 larger than from VBS \cite{Frederix:2014hta}. Because of this, most of the works present nowadays in the literature focus on this particular $HH$  production channel, GGF, to study the sensitivity to $\lambda$. In fact, all these works and the best present measurement at the LHC have made possible to constraint this parameter in the range $\lambda\in[-8.2,13.2]\cdot \lambda_{\rm SM}$ at the 95\% CL \cite{Aaboud:2018ftw}.

Although GGF benefits from the highest statistics and rates, it suffers the inconveniences of having large uncertainties, being a one loop process initiated by gluons, and being dependent of the top Yukawa coupling. Double Higgs production via VBS~\cite{Grober:2010yv, Dawson:2013bba, Baglio:2012np, Frederix:2014hta, Liu-Sheng:2014gxa, Dicus:2015yva, He:2015spf, Bishara:2016kjn, deFlorian:2016spz} is, in contrast, a tree level process not initiated by gluons and it is independent of top physics features, leading therefore to smaller uncertainties. Also, at a fundamental level, it is well known that VBS processes involving longitudinally polarized gauge bosons, like the process $V_L V_L\to HH$ that we are interested in, probe genuinely the self interactions of the scalar sector of the SM. This would happen specially at high energies, such as those available at the LHC, since, in this regime, each $V_L$ behaves as its corresponding would-be-Goldstone boson $\phi$. Therefore, testing $V_L V_L\to HH$ is closely related to testing $\phi \phi HH$ interactions. In this way, a new window, qualitatively different than GGF, would be open with VBS to test $\lambda$, meaning that being able to measure these processes for the first time will be a formidable test of the SM itself, and it could even lead to the discovery of physics beyond the Standard Model (BSM). Moreover, the VBS production channel is the second largest contribution to Higgs pair production, and the VBS topologies have very characteristic kinematics, which allow us to select these processes very efficiently as well as to reject undesired backgrounds. In fact, the selection techniques for VBS configurations at the LHC have experienced a great development in the last years and have improved considerably, especialy in the context of electroweak (EW) vector boson scattering $VV \to VV$ \cite{Doroba:2012pd, Szleper:2014xxa, deFlorian:2016spz, Fabbrichesi:2015hsa, Delgado:2017cls}.  Thus, in summary, VBS double Higgs production might be very relevant to study the sensitivity to the Higgs self-coupling, despite the fact that it is considerably smaller in size than GGF, since it could lead to a cleaner experimental signal. Besides, it will be a complementary measurement to that of GGF and will, in any case, help to improve the determination of this $\lambda$ coupling with better precision.  

In this work, motivated by the above commented advantages, we analyze in full detail Higgs pair production via VBS at the LHC to probe the Higgs self-coupling. To this end, we first explore and characterize the subprocesses of our interest, $VV\to HH$ with $V=W,Z$, both for the SM with $\lambda= \lambda_{SM}$ and for BSM scenarios with $\lambda=\kappa\,\lambda_{SM}$, and consider values of $\kappa$ between 10 and -10. For this study, we fix $m_H$ to its experimental value, $m_H=125.18\pm 0.16$ GeV  \cite{PDG2018}, and set the Higgs vacuum expectation value (vev) to $v=246$ GeV. In this way, studying the sensitivity to $\lambda$ in VBS will provide the desired independent test of this coupling. 

Once we have deeply studied double Higgs production at the subprocess level, we then explore in this work the LHC scenario. First we analyze the process $pp\to HHjj$, to fully understand the properties of this scattering, and then we study and give quantitative results for the sensitivity to the Higgs self-coupling after the Higgs decays. The production of $HHjj$ at the LHC, including VBS and GGF, has been studied previously in \cite{Dolan:2013rja, Dolan:2015zja}, where they focus on $b\bar{b}\tau\bar{\tau} jj$ final states. Our main study is performed, in contrast, in the four bottoms and two jets final state, $pp\to b\bar{b}b\bar{b}jj$, since it benefits from the highest rates. We also make predictions for the interesting $pp\to b\bar{b}\gamma\gamma j j$ process which, although with lower rates, leads to cleaner signatures. We would like to point out that all computations and simulations are performed at the parton level with no hadronization or detector response simulation taken into account, since the work is aimed to be a first and simple approximation to the sensitivity to $\lambda$ in VBS processes at the LHC.
 
 The paper is organized as follows: In Section \ref{Subprocess} we study VBS double Higgs production at the subprocess level in and beyond the SM. Afterwards, in Section \ref{LHC}, we move on to the LHC case, exploring first the $pp\to HHjj$ scattering in Subsection \ref{HHjj} and considering later the Higgs decays, both leading $H \to b \bar b$ decay and subleading $H \to \gamma \gamma$ one. Then, we study both signal and background rates for  $pp\to b\bar{b}b\bar{b}jj$ in Subsection \ref{4b2j} and $pp\to b\bar{b}\gamma\gamma j j$ in Subsection \ref{2b2a2j}, providing our results for the sensitivity to $\lambda$ in VBS Higgs pair production at the LHC for a center of mass energy of $\sqrt{s}=14$ TeV and for different and future expected luminosities. Section \ref{Conclusions} summarizes our main conclusions.


\section{Double Higgs production in vector boson scattering}
\label{Subprocess}
As already stated in the paragraphs above, we are interested in exploring the sensitivity to the Higgs self-coupling, $\lambda$, through VBS processes, in particular at the LHC. For that purpose, we have to study and characterize first the subprocess that leads to the specific signal we will be dealing with once we perform the full collider analysis. This subprocess will be, in our case, the production of two Higgs bosons in the final state from the scattering of two EW gauge bosons, $VV\to HH$, with $V=W,Z$. Within this context, in this section we aim to understand the role of the Higgs trilinear coupling in the SM and beyond, as well as the generic characteristics of the scattering processes $W^+W^-\to HH$ and $ZZ\to HH$.

\begin{figure}[t!]
\begin{center}
\includegraphics[width=\textwidth]{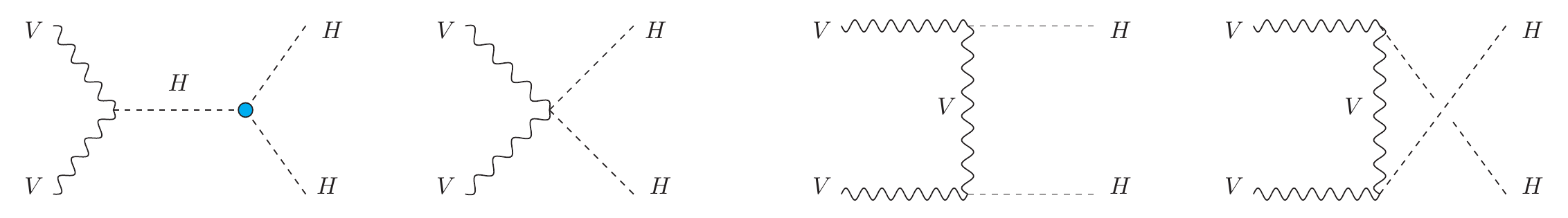}
\caption{Tree level diagrams that contribute to double Higgs production in vector boson scattering in the Unitary gauge. The cyan circle represents the presence of the Higgs self-coupling in the interaction vertex.}
\label{fig:subprocess_diagrams}
\end{center}
\end{figure}

The Higgs self-coupling is only present, at the tree level and in the Unitary gauge, in the $s-$channel diagram of  the studied processes, so the sensitivity to $\lambda$ will only depend on this particular configuration. However, a contact diagram, a $t-$channel diagram and a $u-$channel diagram have to be taken into account too as shown in \figref{fig:subprocess_diagrams}, in which we display all the possible tree level contributions to the mentioned scattering processes in the Unitary gauge. Each of these diagrams has its own energy dependence and its own relative size, so they participate differently in the total amplitude $A\big(V_1(p_1,\varepsilon_1) V_2(p_2,\varepsilon_2)  \to H_1(k_1) H_2(k_2)\big)$. This can be seen in \eqrefs{amplitudeHHs}{amplitudeHHu}, where we show the amplitude of each diagram of the process $W^+W^-\to HH$, $A_d$, with $d=s,c,t,u$ from $s$, contact, $t$ and $u$ channels respectively, computed consistently in the Unitary gauge:
\begin{align}
A_s(W^+W^-\to HH)=&~3 g^2v^2\dfrac{ \lambda}{s-m_H^2}(\varepsilon_1 \cdot \varepsilon_2)\, ,\label{amplitudeHHs} \\
A_c(W^+W^-\to HH)=& ~\dfrac{g^2}{2}(\varepsilon_1 \cdot \varepsilon_2)\, , \label{amplitudeHHc}\\
A_t(W^+W^-\to HH)=&~\dfrac{g^2}{t-m_W^2}(m_W^2(\varepsilon_1 \cdot \varepsilon_2) +(\varepsilon_1 \cdot k_1) (\varepsilon_2 \cdot k_2) )\, , \label{amplitudeHHt}\\
 A_u(W^+W^-\to HH)=&~\dfrac{g^2}{u-m_W^2}(m_W^2(\varepsilon_1 \cdot \varepsilon_2) +(\varepsilon_1 \cdot k_2) (\varepsilon_2 \cdot k_1) )\, .\label{amplitudeHHu}
\end{align}
Here, $g$ is the EW coupling constant, $m_W$ is the mass of the $W$ boson, and $s,t$ and $u$ are the usual Mandelstam variables. The amplitudes for the $ZZ\to HH$ case are identical except for a global factor $1/c_{\rm w}^2$ (with $c_{\rm w}=\cos\theta_{\rm w}$ and with $\theta_{\rm w}$ being the weak angle), that has to be included in each amplitude, and the substitution of $m_W^2$ by $m_Z^2$ in the $t$ and $u$ channel expressions.

\begin{figure}[t!]
\begin{center}
\includegraphics[width=0.49\textwidth]{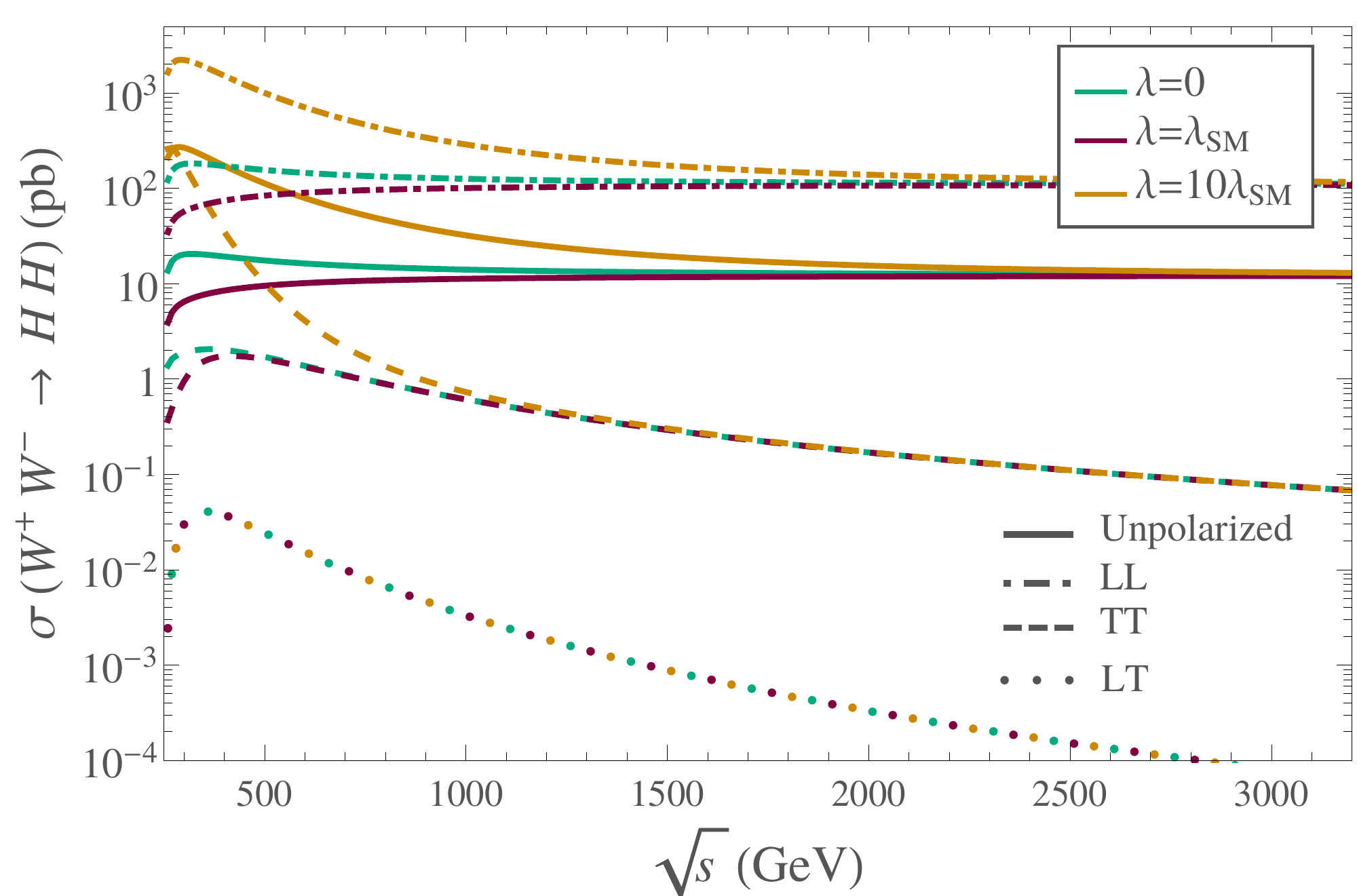}
\includegraphics[width=0.49\textwidth]{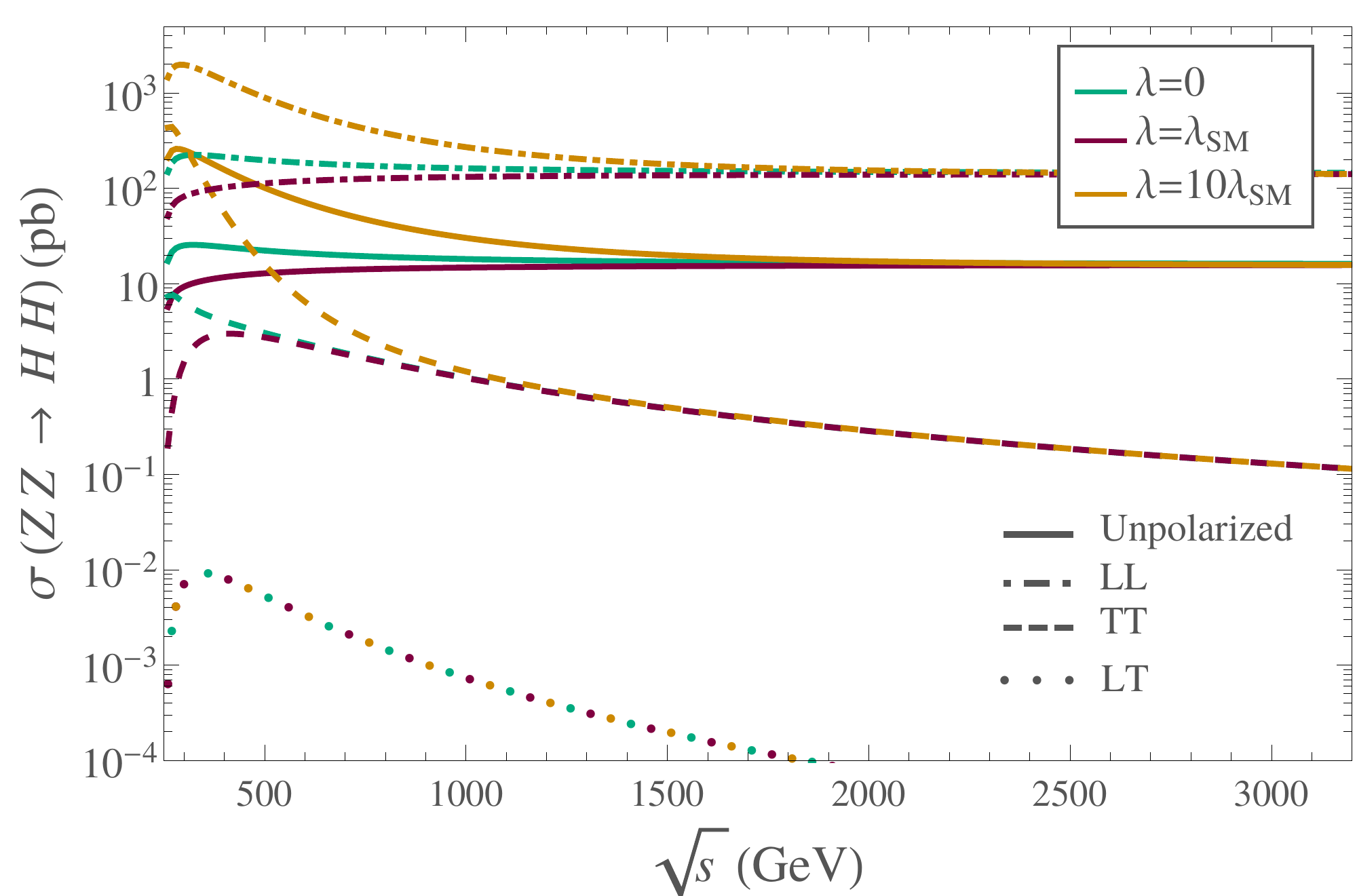}
\caption{Predictions of the cross sections of $W^+W^-\to HH$ (left panel) and $ZZ\to HH$ (right panel) as a function of the center of mass energy $\sqrt{s}$ for three different values of $\lambda$ and for different polarizations of the initial gauge bosons: $V_LV_L$ (upper dot-dashed lines), $V_TV_T$ (middle dashed lines) and $V_LV_T+V_TV_L$ (lower dotted lines). The unpolarized cross section is also included (solid lines). Each polarized cross section contributes with a factor 1/9 to the unpolarized (averaged) cross section.}
\label{fig:subprocess_polarizations}
\end{center}
\end{figure}

On the other hand, the contribution of each polarization state of the initial EW gauge bosons behaves differently, not only energetically, but also in what concerns to the sensitivity to $\lambda$. There are only two polarization channels that do depend on $\lambda$: the purely longitudinal, $V_LV_L$, and the purely transverse in which both vector bosons have the same polarization, $V_{T^+}V_{T^+}$ and $V_{T^-}V_{T^-}$. All the other channels have vanishing $s$-channel contributions and will not actively participate, therefore, in the study of the Higgs trilinear coupling, although all polarization states contribute to the total cross section. Moreover, this total cross section is dominated, specially at high energies, by the purely longitudinal $V_LV_L$ configuration, and so is each diagram contribution. All these features can be seen in \figref{fig:subprocess_polarizations}, where we display the predictions for the cross sections of $W^+W^-\to HH$ and $ZZ\to HH$ as a function of the center of mass energy for three different values of $\lambda$ separated by polarizations of the gauge bosons, including, also, the unpolarized cross section. In this figure two things are manifest: the first one is that the $V_LV_T$ configuration is indeed independent of $\lambda$. The second one is that the total cross section is clearly strongly dominated by the purely longitudinal contribution at all energies. This is a very interesting result, since it means that, if this process was measured, we would be being sensitive to the purely longitudinal configurations of the gauge bosons, and therefore to the heart of the self-interactions of the SM scalar sector.

\begin{figure}[t!]
\begin{center}
\includegraphics[width=0.49\textwidth]{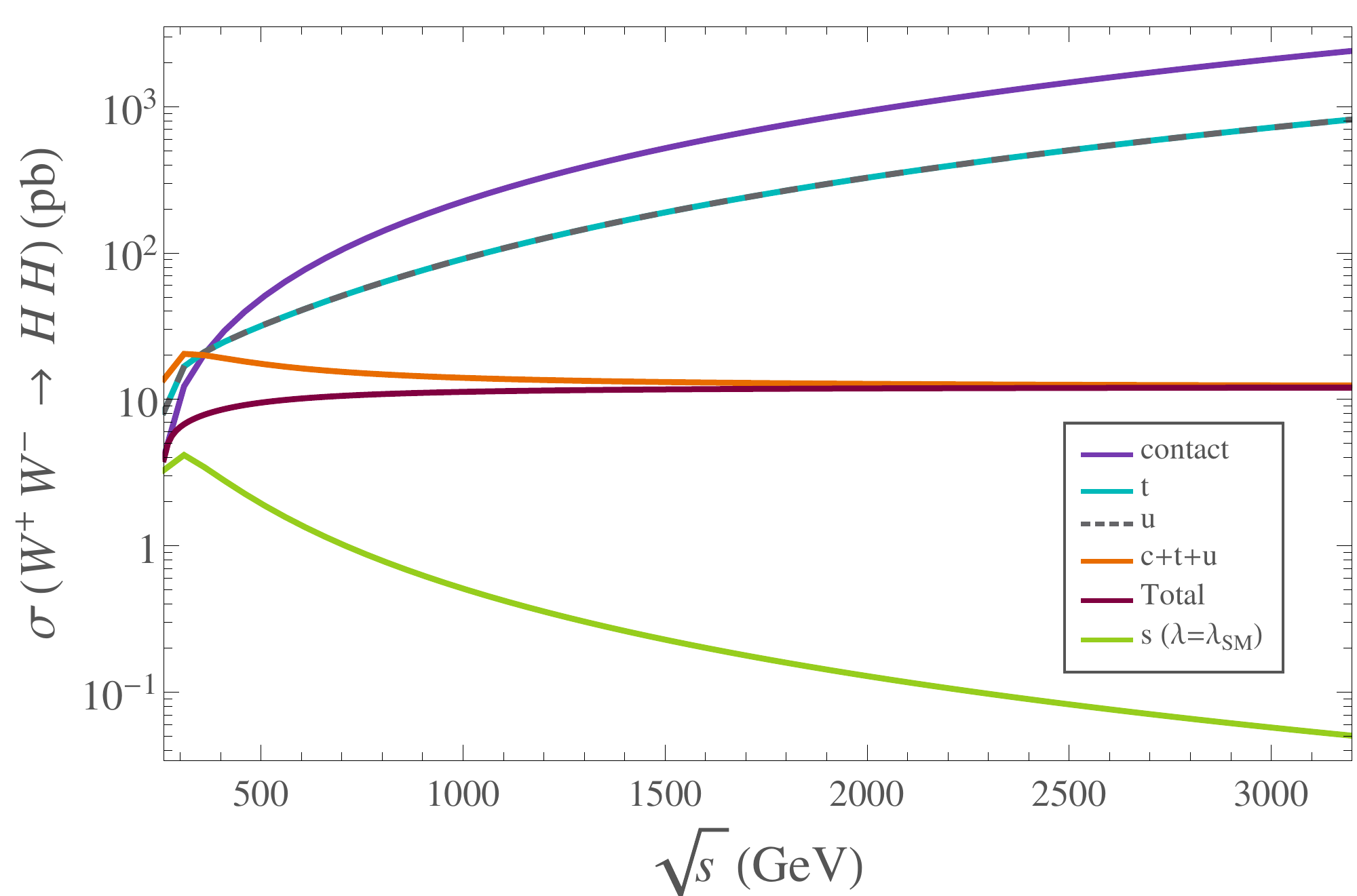}
\includegraphics[width=0.49\textwidth]{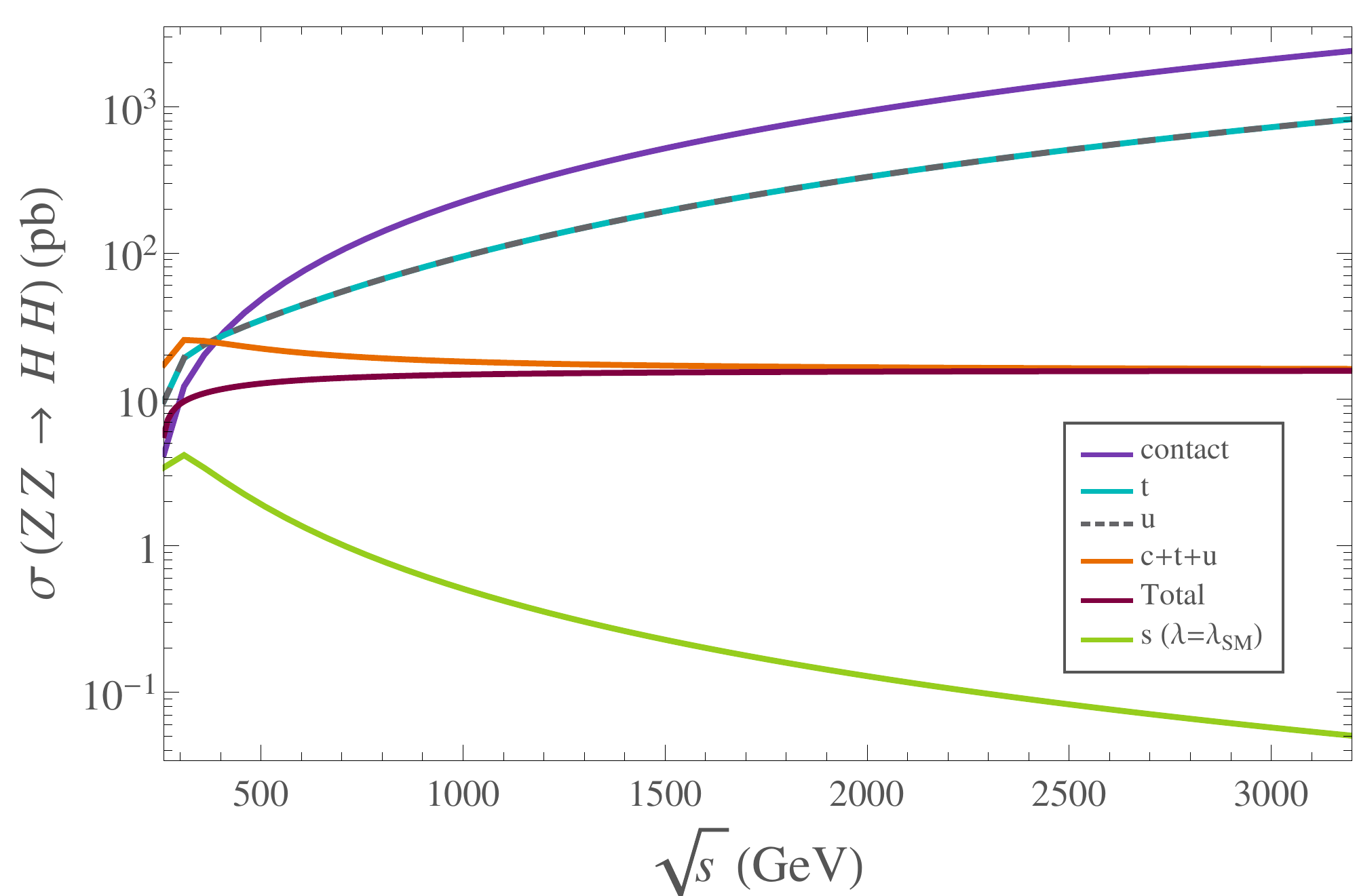}
\caption{Contribution to the total cross section of $W^+W^-\to HH$ (left panel), and of $ZZ\to HH$ (right panel) in the SM, i.e., $\lambda=\lambda_{SM}$, of each diagram displayed in \figref{fig:subprocess_diagrams} as a function of the center of mass energy $\sqrt{s}$. The sum of the contributions of the contact, $t$-channel and $u$-channel diagrams as well as the sum of all diagrams that contribute are also presented.}
\label{fig:subprocess_diagrams_effect}
\end{center}
\end{figure}

The $V_LV_L$ dominance can be understood through the inspection of the energy dependence of the longitudinal polarization vectors, $\varepsilon_V$, at high energies. They are all proportional, for $\sqrt{s}\gg m_V$, to a power of the energy over the mass, $E_V/m_V$. This leads to a behavior of the amplitudes, presented in Eqs.(\ref{amplitudeHHc})-(\ref{amplitudeHHu}), for the contact, $t$ and $u$ channels respectively, proportional to $s$,  and to a constant behavior with energy of the $s$-channel amplitude given in Eq.(\ref{amplitudeHHs}). Including the extra $1/s$ suppression factor to compute the cross section from the squared amplitude one obtains the energy dependence seen in \figref{fig:subprocess_diagrams_effect}, where we present the contribution of each diagram to the total cross sections of $W^+W^-\to HH$ and $ZZ\to HH$ in the SM, as well as the sum of the contact, $t$-channel and $u$-channel diagrams, $(c+t+u)$, and the total cross section taking all diagrams into account. In this figure, we see clearly that the sum of the contact, $t$ and $u$ channels tends at high energy to a constant value. This happens because in the SM there is a cancellation of the linear terms in $s$ among these three channels. In contrast, the $s$-channel contribution decreases as $1/s$ and is subleading numerically in the SM with respect to the other $(c+t+u)$ contributions. It is only, at lower energies near the production threshold of two Higgs bosons, where the s-channel contribution is numerically comparable to the other channels and, in fact, a mild cancellation occurs between this $s$-channel and the rest $(c+u+t)$. Therefore, the s-channel and in consequence $\lambda$, do not effectively participate in  the constant behavior at high energies of the total cross section in the SM. At this point, it is worth recalling that these constant behaviors of the cross sections with energy are characteristic of VBS processes at high energies, precisely because of the above commented dominance of the longitudinal configurations.

When going beyond the SM by taking $\lambda \neq \lambda_{SM}$, the previously described dependence with energy and the delicate cancellations commented above among the various contributing diagrams may change drastically. In fact, varying the size of the Higgs trilinear coupling could modify the relative importance of the contributing diagrams and, in particular, it could allow for the s-channel contribution to be very relevant or even dominate the scattering. This could happen not only at low energies close to the threshold of $HH$ production, but also at larger energies, where the pattern of cancellations among diagrams could be strongly modified. This may lead to a different high energy behavior, and, hence, to a different experimental signature. The crucial point is that such a large  deviation in $\lambda$ with respect to the SM value is still experimentally possible, as the present bounds on the trilinear coupling are not yet very tight. The best bounds at present set $\kappa = \lambda/\lambda_{SM}\in[-8.2,13.2]$ \cite{Aaboud:2018ftw}, so values of order 10 times the SM coupling are still allowed by LHC data. Then, if in the future the LHC could improve this sensitivity to lower values of $\lambda$ it would be a formidable test of the presence of new physics beyond the SM. We will show next that this sensitivity can be indeed reached in the future by means of VBS.

\begin{figure}[t!]
\begin{center}
\includegraphics[width=0.49\textwidth]{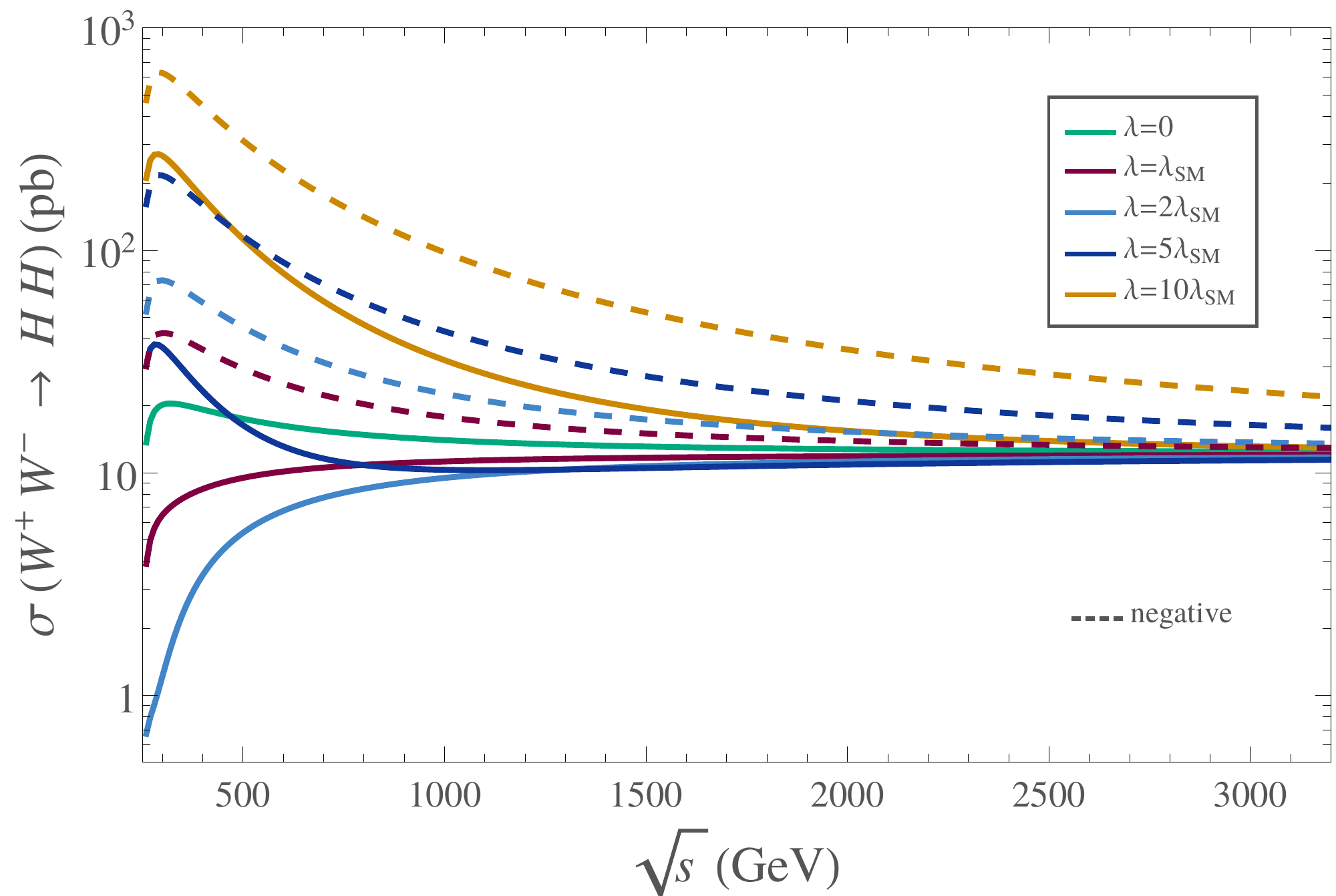}
\includegraphics[width=0.49\textwidth]{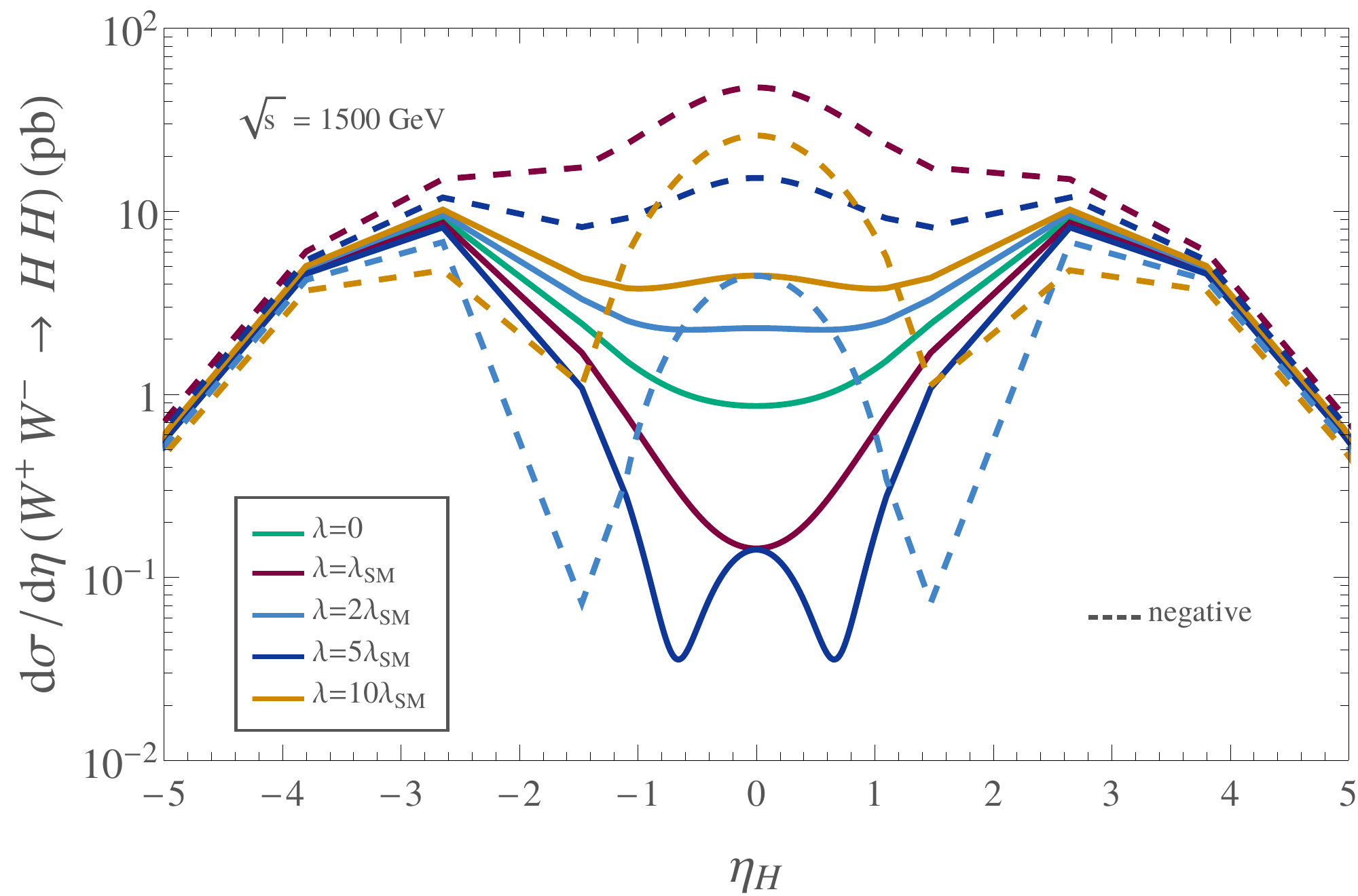}
\caption{Predictions for the total cross section of the process $W^+W^-\to HH$,   as a function of the center of mass energy $\sqrt{s}$ (left panel) and as a function of the pesudorapidity of one of the final $H$ at a fixed center of mass energy of $\sqrt{s}=1500$ GeV (right panel) for different values of the Higgs self-coupling $\lambda$. Solid (dashed) lines correspond to positive (negative) values of $\lambda$.}
\label{fig:subprocess_varylambda}
\end{center}
\end{figure}

It is important to understand in more detail at this point the implications of setting $\lambda$ to a different value than $\lambda_{SM}$ in the kinematical properties of the VBS processes we are studying here. For this purpose, we present in \figref{fig:subprocess_varylambda} the total cross section of the process $W^+W^-\to HH$ as a function of the center of mass energy $\sqrt{s}$  and the differential cross section with respect to the pseudorapidity $\eta_H$ of one of the final Higgs bosons (notice that the distribution with respect to the pseudorapidity of the other Higgs particle is the same) for different values of positive, vanishing and negative $\lambda$\footnote{We assume here a phenomenological approach when setting $\lambda\neq\lambda_{SM}$ , meaning that it is not our aim to understand the theoretical implications of such a result like potential instabilities for negative values of $\lambda$, etc. We understand that the deviations in this coupling would come together with other BSM Lagrangian terms that would make the whole framework consistent.}. The results for $ZZ\to HH$ (not shown) are very similar to those of $W^+W^-\to HH$. From this figure, it can be seen that, first and most evidently, the total cross section changes in magnitude and in energy dependence with respect to the SM one, as already announced. This happens especially near the $HH$ production threshold, confirming that the sensitivity to deviations in $\lambda$ with respect to the SM value is larger in this region. For the case of positive $\lambda$ the total BMS cross section can be larger or lower than that in the SM, depending on the size of the deviations in $\lambda$ with respect to $\lambda_{SM}$, since in this case there is a destructive interference between the $s$ channel contribution and the rest $(c+t+u)$. In contrast, for the case of negative $\lambda$ values, the sum of diagrams is always constructive and one obtains bigger cross sections than the SM one independently of the absolute value of the coupling. The details of these features will be extended when commenting the next figure.      
Regarding the angular dependence of the differential cross section, or correspondingly the distribution respect to $\eta _H$ also shown in \figref{fig:subprocess_varylambda}, we see clearly that it also changes in the BSM scenarios respect to the SM one. We particularly learn from this figure that for central values of the Higgs pseudorapidity, concretely for $|\eta_H|<2.5$, it is much easier to distinguish between different values of $\lambda$. Therefore, this suggests the kind of optimal cuts in this variable $\eta _H$, or the equivalent one in terms of the final particles from the Higgs decays, we should be giving to enhance the sensitivity to the signal when moving to the realistic case of the $pp$ collisions at the LHC.

\begin{figure}[b!]
\begin{center}
\includegraphics[width=0.49\textwidth]{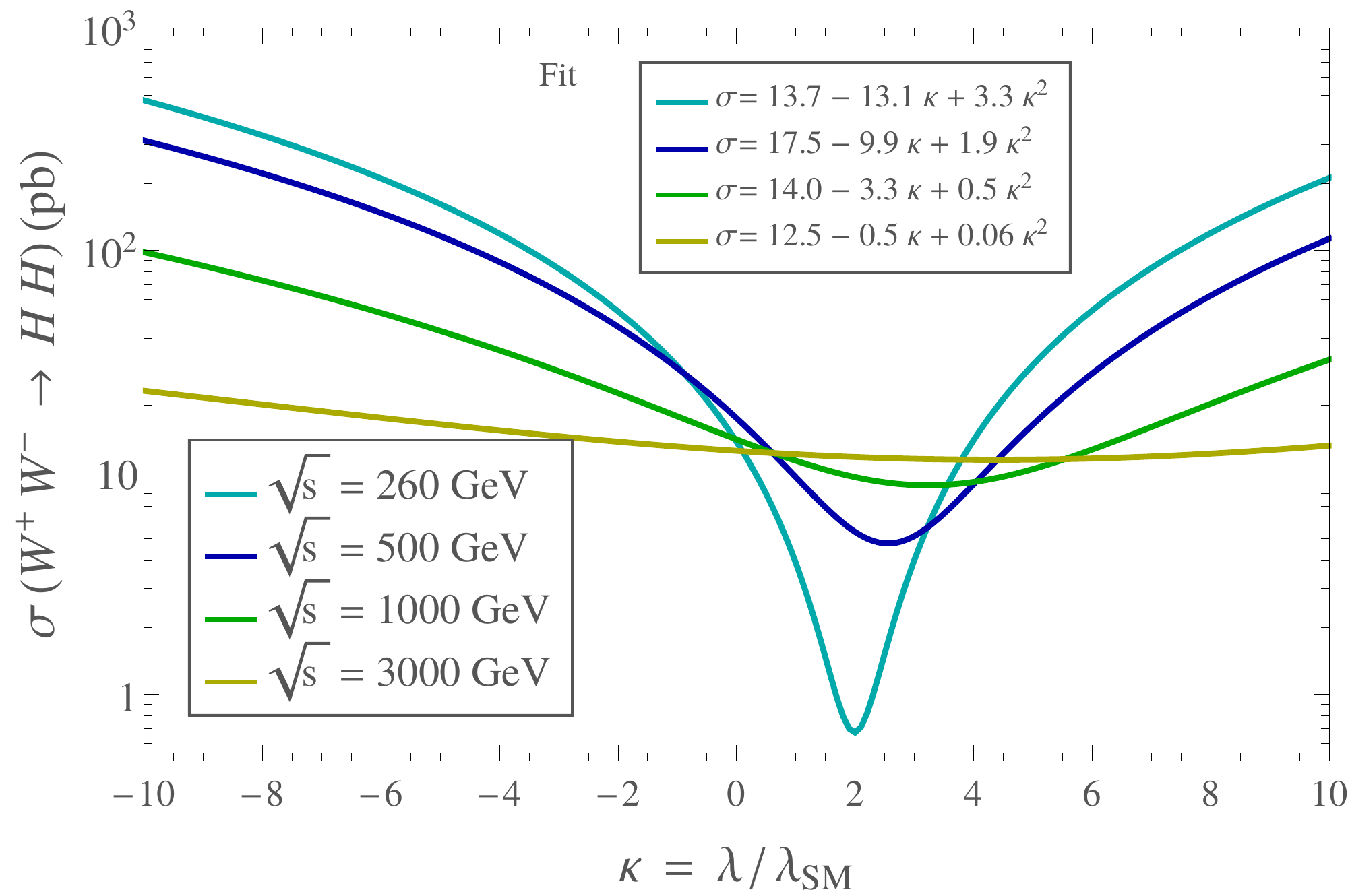}
\includegraphics[width=0.49\textwidth]{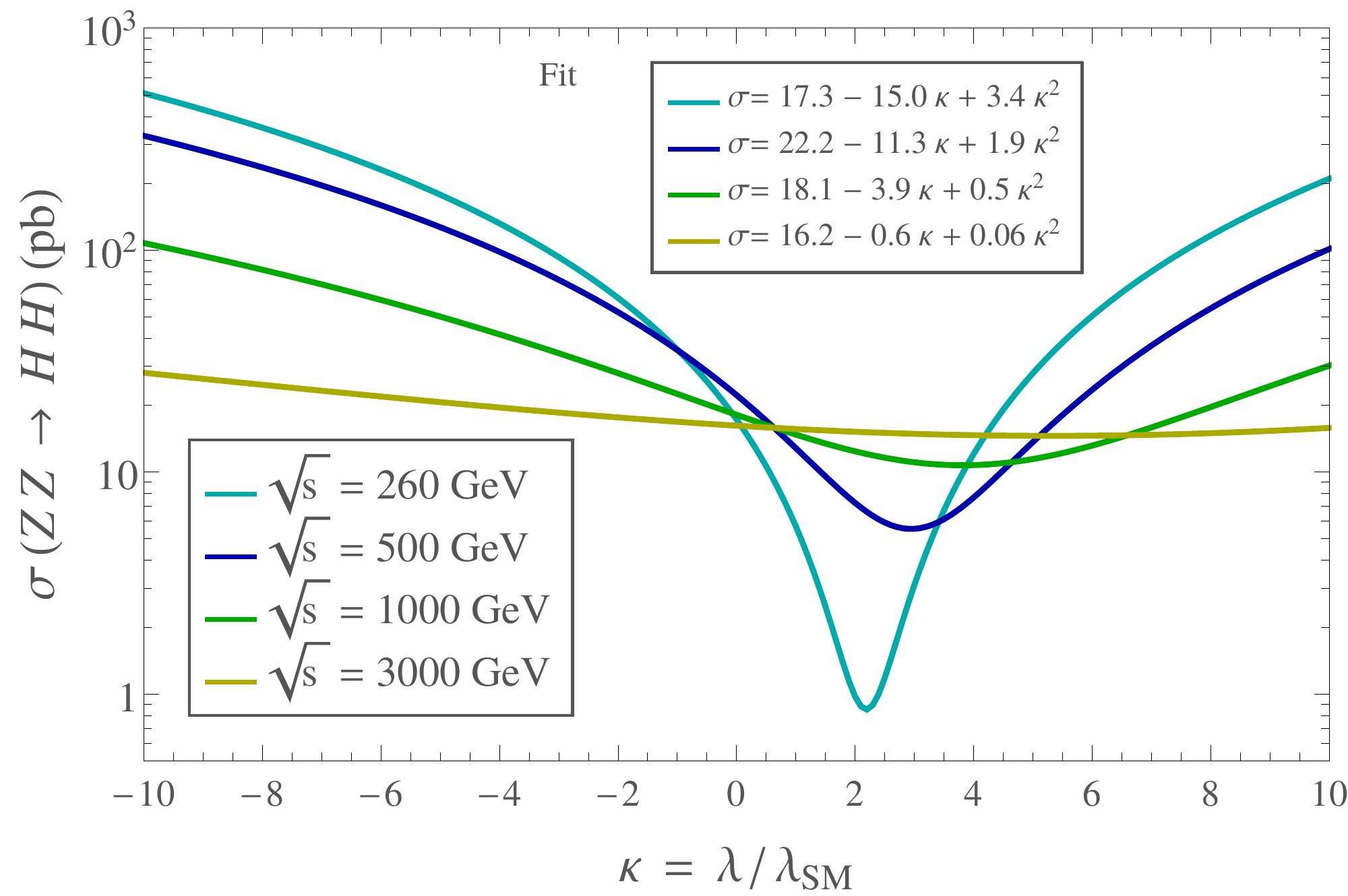}
\caption{Prediction for the total cross section of the VBS process $W^+W^- \to HH$ (left panel) and of $ZZ\to HH$ (right panel) as a function of the ratio of a generic $\lambda$ value over the SM value for four different center of mass energies: $\sqrt{s}=$ 260, 500, 1000 and 3000 GeV.}
\label{fig:subprocess_cancellation_energy}
\end{center}
\end{figure}

In \figref{fig:subprocess_cancellation_energy} we display our predictions for the total cross section of the two relevant VBS processes as a function of $\kappa$ for four different values of fixed center of mass energy $\sqrt{s}=260,\,500,\,1000,\,3000$ GeV. We also display the parabolic fits that allow us to describe each of the curves to have a more analytical insight into the details of how the above commented cancellations among diagrams do actually occur. The formulas of the fits in this figure manifest that, in general, the cross section has a quadratic, a constant and a linear term in $\kappa$, coming, respectively, from the $s$-channel contribution, from the $(c+t+u)$ contribution and from the interference of these two. The sign of the interference is negative for positive values of $\kappa$ and positive for  negative values of $\kappa$. This destructive interference for $\lambda>0$ produces that the minima of these lines are placed at $\lambda > \lambda_{SM}$. Besides, depending on the energy and on the size of $\kappa$, the behavior of the cross section will be dominantly constant, linear or quadratic in $\lambda$, and therefore the sensitivity to $\lambda$ will vary accordingly. Near the production threshold, i.e., at energies around 250 GeV, two issues can be seen. The first one is that, as we already saw in \figref{fig:subprocess_varylambda}, the differences in the cross section when we vary $\lambda$ are maximal, and so will be the sensitivity to differences in this coupling. The second one is that, at these low energies, the SM, corresponding to $\kappa=1$, suffers, as already said, a mild cancellation between the linear and the constant terms, and therefore the sensitivity to $\lambda$ will be mainly quadratic. We can also see that the minima of the parabolas soften, in the sense that the variations in the cross section when we vary $\lambda$ become smaller, and that their position moves from $\lambda/\lambda_{SM}$ close to 2 to larger values as the energy is increased. Because of this, the bigger the energy, the bigger the value of $\lambda$ that maximizes the cancellations. Thus, as a first conclusion at this point, we will have to keep in mind, once we perform the full collider analysis, that the sensitivity to different values of the trilinear coupling and the issue of delicate cancellations among diagrams in VBS are clearly correlated and this will affect  the final results at the LHC.

A final comment has to be made in this section, and it is that of a potential unitarity violation problem for large $|\lambda|$ values in the processes of our interest here, $VV \to HH$. To check this unitarity issue, we have evaluated the partial waves $a_J$ of the dominant polarization channels for this VBS, which, as we have said, are the longitudinal ones, i.e., $V_LV_L \to HH$. These $a_J$ of fixed angular momentum $J$ are evaluated as usual, by computing:
\begin{align}
a_J&=\dfrac{1}{64\pi}\int_{-1}^1 d\cos\theta~ A(V_LV_L\to HH)\, P_J(\cos\theta) \,, \label{unitarity}
\end{align}
where $P_J(\cos\theta)$ are the Legendre polynomials. For a given energy, $\sqrt{s}$, we then define the unitarity violation limit as the value of $\lambda$ for which $|a_J(s)|=1$. By doing this exercise, we find that all the partial waves $|a_J|$ that we have computed are below 0.1 for values of $\lambda$ between -10 and 10 times the SM value at all energies. So, for the present study, we are safe from unitarity violation problems. For completeness, we have also made a fast estimate of the value of $\lambda$ that would be required to violate unitarity in this process. For large values of $|\lambda|$, the dominant contribution to the total amplitude comes from the $s$-channel. This contribution, as we mentioned before, behaves, at high energies and for the purely longitudinal case, as a constant. In particular, one obtains that $A_s(V_LV_L\to HH)\sim 6\,\lambda$ for $\sqrt{s}\gg m_H$. With this amplitude, one can compute the value of $\lambda$ for which the biggest partial wave (in this case we have checked that it is the one corresponding to $J=0$) becomes one. We obtain $\lambda_{\rm unit}\sim 17$. Notice that this upper limit of $\lambda$ is above the perturbativity limit given naively by $\lambda_{\rm pert}\sim\sqrt{16\pi}\sim 7$. 
  
With all these features in mind, we can move on from the subprocess level to the full process at the LHC to study the sensitivity of this collider to the Higgs self coupling in VBS processes.

 
 \section{Sensitivity to the Higgs self-coupling at the LHC}
 \label{LHC}
 
Once we have characterized completely the scattering $VV \to HH$, it is time to explore the full process at the LHC to quantify how sensitive this machine could be to the Higgs trilinear coupling in VBS processes. 
 At this point, we would like to stress again the fact that this double Higgs production channel, via the scattering of two EW gauge bosons, has been poorly studied previously in the literature, due to the fact that it provides less statistics than the GGF one. Nevertheless, now that the LHC is close to reach its nominal energy, $\sqrt{s}=14$ TeV, and that it is already achieving high integrated luminosities, close to $L = 40$ \ifb, the possibility of measuring VBS processes, that were inaccessible before, opens up. In fact, several VBS measurements have been already performed at this collider by ATLAS \cite{Aad:2014zda, Aad:2016ett, Aaboud:2016uuk, Aaboud:2016ffv, Aaboud:2017pds, ATLAS:2018ogo, ATLAS:2018ucv} and CMS \cite{Khachatryan:2014sta, Khachatryan:2016vif, Khachatryan:2017jub, Sirunyan:2017fvv, Sirunyan:2017ret, Sirunyan:2017jej, CMS:2018hlo, CMS:2018ysc}.  Taking this into account, and the fact that the kinematics of the VBS processes are incredibly characteristic and allow for a very efficient signal selection and background rejection, a dedicated study of the sensitivity to $\lambda$ via VBS processes is on demand.

This is precisely the aim of this section, in which we first promote the analysis of Section \ref{Subprocess} to that of its LHC signal, $pp\to HHjj$, so that we can fully understand its behavior and properties, and then we give more quantitative and realistic results for the sensitivity to $\lambda$ once the Higgs bosons have decayed. Specifically, we will focus first on the dominant Higgs decays to bottoms, leading to the process \bbbbjj. This process benefits from having more statistics due to the large branching ratios involved, and, because of this, it is presumably the one that will lead to the best sensitivities. We will also present results on other channels, concretely for $p p \to b \bar{b} \gamma\gamma jj$, where one of the two Higgs bosons has decayed to photons, that, despite their smaller number of events, might also provide interesting results since they suffer from less severe backgrounds.

For all computations and results of the signal events we use MadGraph5$@$NLO \cite{Alwall:2014hca}, setting the factorization scale to $Q^2=m_Z^2$ and using the set of PDF's NNPDF2.3~\cite{Ball:2013hta}. We have found that changing the chosen value of $Q^2$ does not lead to relevant changes in the signal rates. Concerning the backgrounds, all of them are simulated with the same settings and PDF's as the signal, using  MadGraph5$@$NLO as well. 
For the case of the multijet QCD background in the $pp \to b \bar{b} b \bar{b} jj$ channel, due to its complexity, we have simulated events using both MadGraph5 with the previous mentioned settings and PDF's, and AlpGen \cite{Mangano:2002ea}, this time choosing $Q^2=(p_{T_b}^2+p_{T_{\bar{b}}}^2+\sum p_{T_j}^2)/6$
 and selecting the set of PDF's CTEQ5L \cite{Lai:1999wy}. We have found agreement between the results of these two Monte Carlos, within the provided errors, 
in the total normalization of the cross section with the basic cuts, and in the shape of the relevant distributions. All results are presented for a center of mass energy of $\sqrt{s}= 14$ TeV.

Our study is aimed to be a first and simple approach to the sensitivity to $\lambda$ in VBS processes at the LHC. This means that, in order to simplify the procedure, the analysis is done at the parton level, and no hadronization or detector response simulation are performed, leaving always room for more expert improvement towards a full and dedicated experimental study.

\begin{figure}[t!]
\begin{center}
\includegraphics[width=0.3\textwidth]{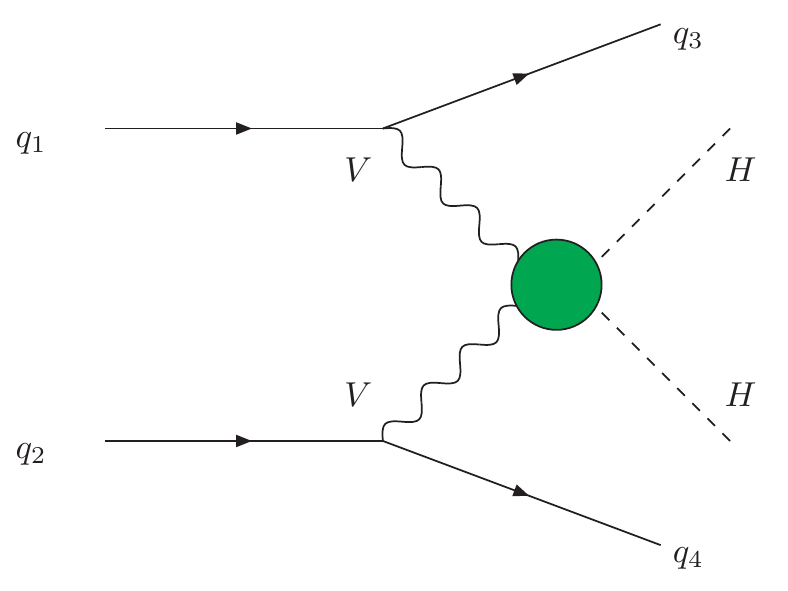}
\caption{Schematic representation of  partonic double Higgs production though VBS at the LHC. The green blob represents the presence of the Higgs self-coupling $\lambda$ in the process, although all diagrams in Fig.(\ref{fig:subprocess_diagrams}) are considered.}
\label{fig:LHC_diagrams}
\end{center}
\end{figure}

 
 \subsection{Study and characterization of  $pp\to HHjj$ signal events}
 \label{HHjj}
 
In order to be able to estimate the sensitivity to the Higgs self-coupling in VBS at the LHC, we need to understand how the results of the previous section translate into the full process when we start with protons as initial particles. This full process, $pp\to HHjj$,  can be produced via many different channels, and not only in VBS configurations. In fact, it is well known that this VBS subset of diagrams contributing to $q_1 q_2 \to q_3 q_4 H H$ is not gauge invariant by itself and all kinds of contributing diagrams have to be included to get gauge invariant result. This is indeed what we are doing here, since when we use MadGraph to compute the signal all kind of diagrams are included. 

The crucial point regarding the phenomenological interest of VBS, that indeed motivates this work, is that the specific VBS configuration can be very efficiently selected by choosing the appropriate kinematic regions of the two extra jets variables, as it is well known \cite{Goncalves:2018qas, Doroba:2012pd, Szleper:2014xxa, Delgado:2017cls}. In particular, at the LHC, the VBS topologies are characterized by large separations in pseudorapidity of the jets, $|\Delta\eta_{jj}|=|\eta_{j_1}-\eta_{j_2}|$, and by large invariant masses of the dijet system, $M_{jj}$. Imposing proper cuts over these two variables makes possible to obtain events that come dominantly from VBS processes and, as we will see later on, also to reject many background events.

\begin{figure}[t!]
\begin{center}
\includegraphics[width=0.49\textwidth]{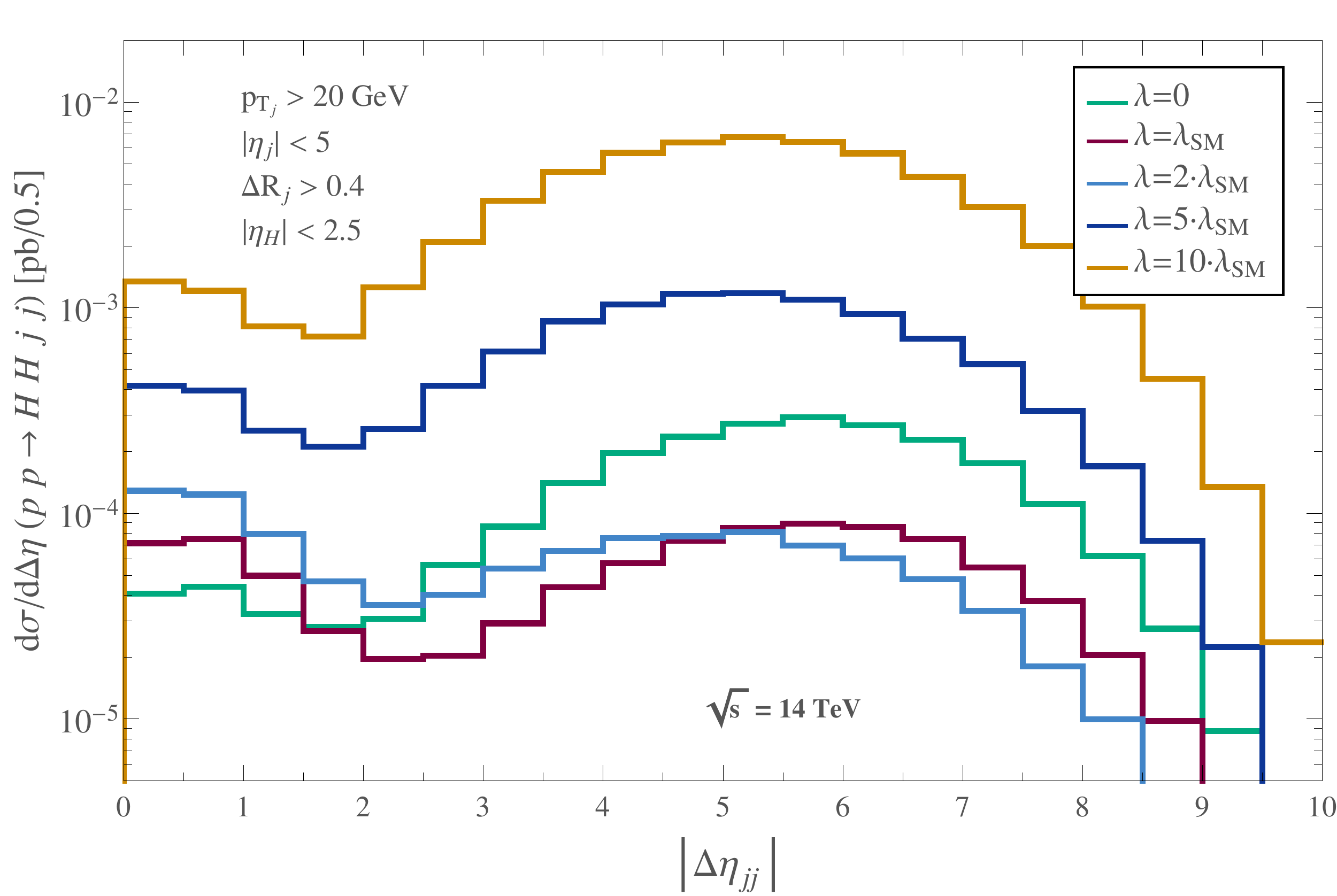}
\includegraphics[width=0.49\textwidth]{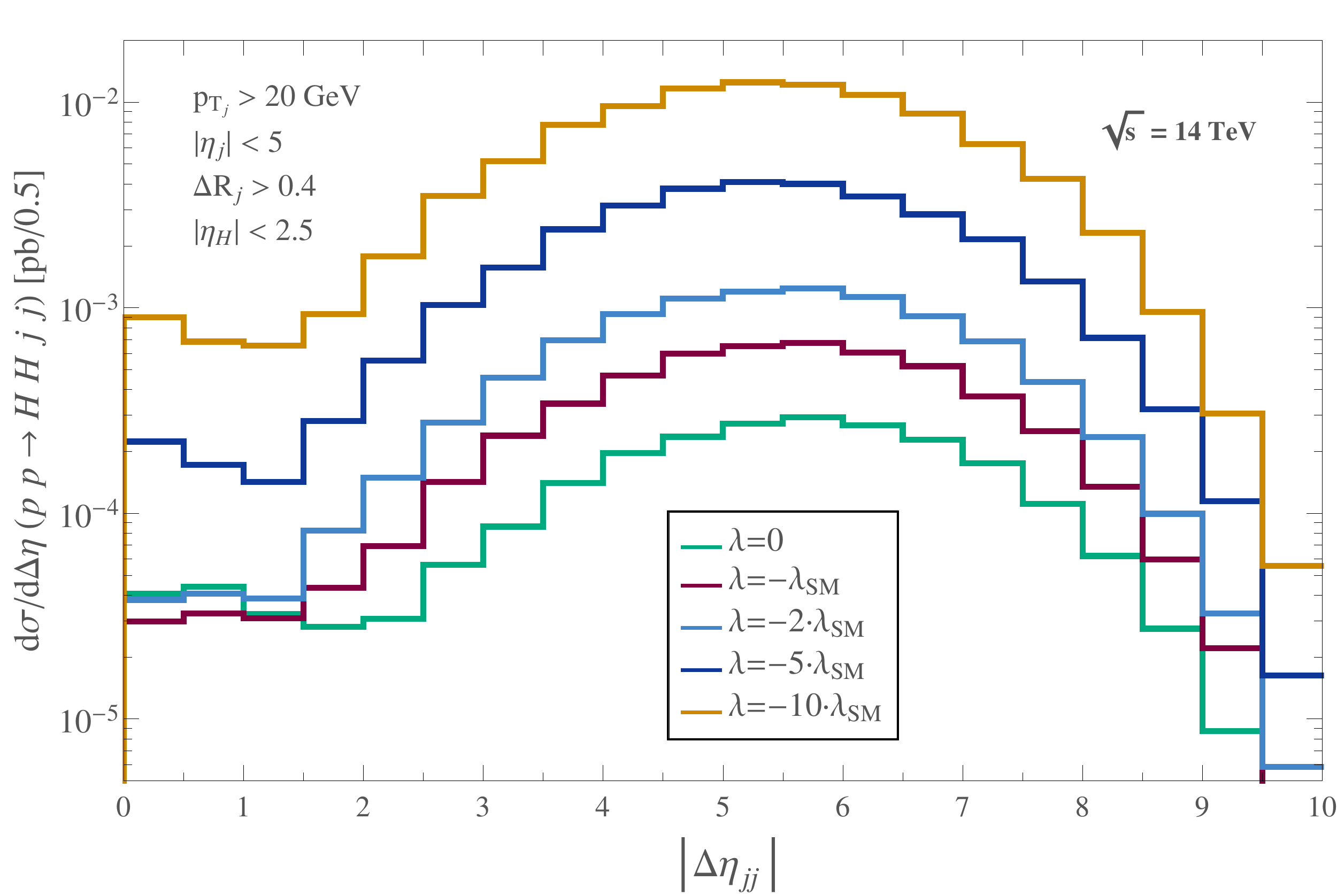}
\includegraphics[width=0.49\textwidth]{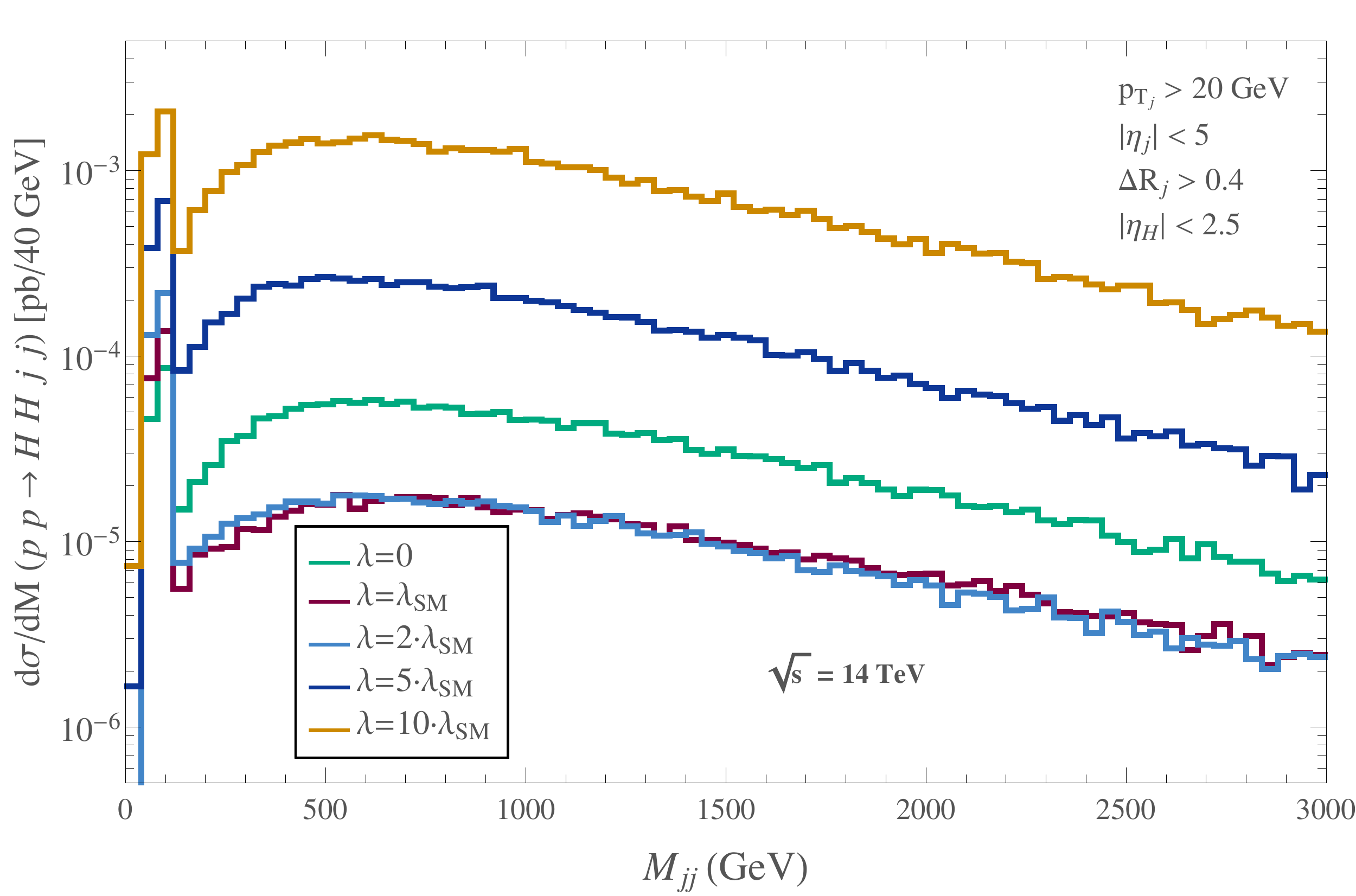}
\includegraphics[width=0.49\textwidth]{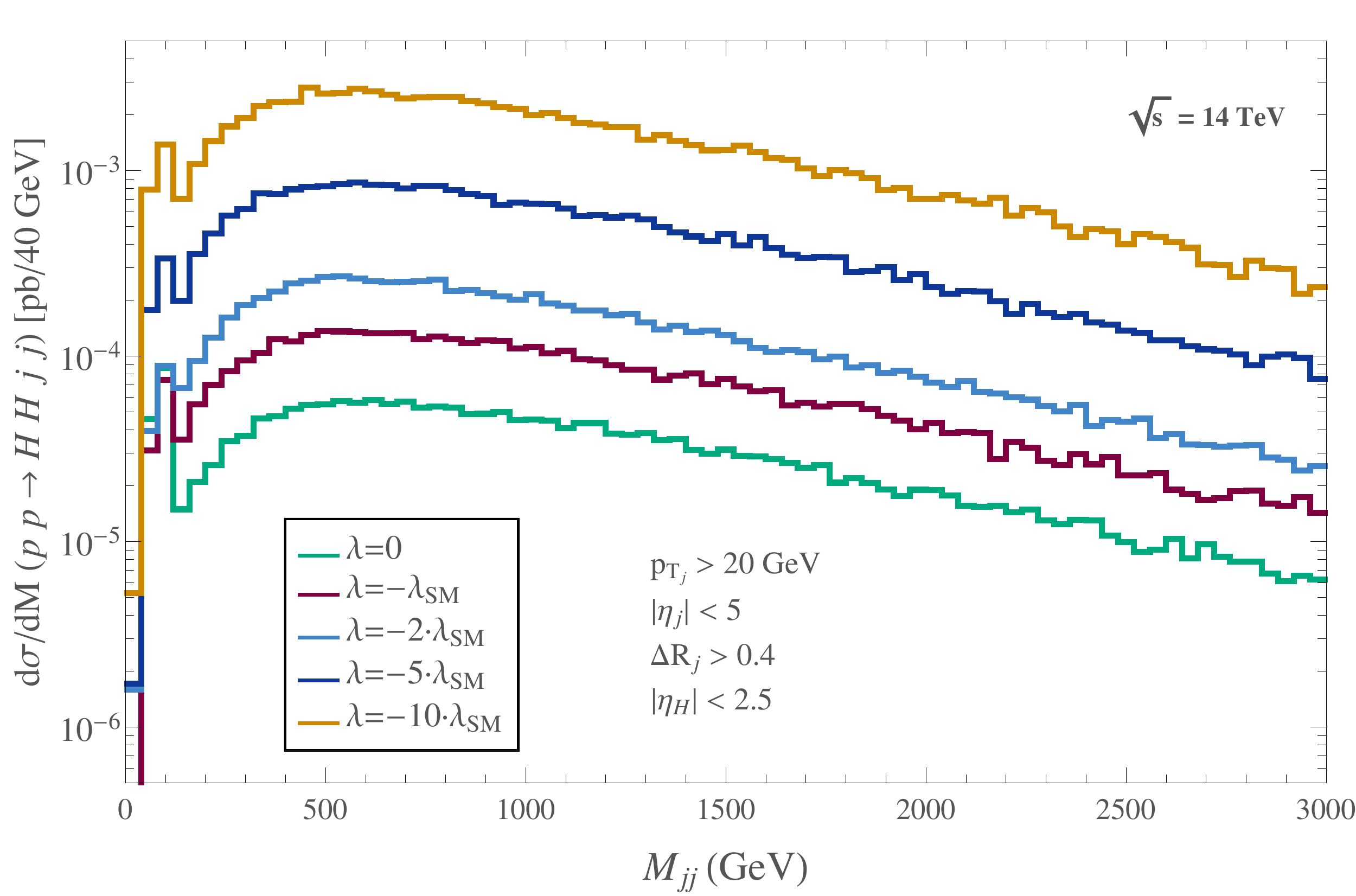}
\caption{Predictions for the total cross section of the process $pp\to HHjj$ as a function of the absolute value of the difference between pseudorapidities of the two jets $|\Delta\eta_{jj}|$ (upper panels) and as a function of the invariant mass of the two jets $M_{jj}$ (lower panels) for different values of the Higgs self-coupling $\lambda$. We display positive (left panels) and negative (right panels) values of $\lambda$ for comparison. We also include the case $\lambda=0$. Cuts in Eq.(\ref{basiccutsHHjj}) have been applied and the center of mass energy has been set to $\sqrt{s}=14$ TeV.}
\label{fig:HHjj_VBSdistributions}
\end{center}
\end{figure}

 The VBS processes involved in $pp\to HHjj$ can be seen schematicaly in \figref{fig:LHC_diagrams}, where the green blob represents all diagrams in \figref{fig:subprocess_diagrams}, including the presence of the $s$-channel with the generic Higgs trilinear coupling $\lambda$. This kind of processes will inherit the properties of the sub-scatterings we have studied, but will also have differences with respect to them due to the fact that we now have protons in the initial state. Then, it is important to know at this stage how close to the ``pure'' VBS configuration our $pp\to HHjj$ signal is. To this end, we have generated with MadGraph5 $pp\to HHjj$ signal events for this process for different values of $\lambda$ with a set of basic cuts that allow for the detection of the final particles, given by: 
\begin{align}
p_{T_j} & > 20~ {\rm GeV}\,,~~~ |\eta_j|<5\,,~~~ \Delta R_{jj} > 0.4\,,~~~ |\eta_H|<2.5\,,\label{basiccutsHHjj} 
\end{align}
where $p_{T_j}$ is the transverse momentum of the jets, $\eta_{j,H}$ is the pseudorapidity of the jets or of the Higgs bosons, and $\Delta R_{jj}$ is the angular separation between two jets defined as $\Delta R_{jj}=\sqrt{\Delta\eta_{jj}^2+\Delta\phi_{jj}^2}$, with $\Delta\eta_{jj}$ and $\Delta\phi_{jj}$ being the angular separation in the longitudinal and transverse planes, respectively.

With these generated events, we have studied some relevant distributions for the signal cross section that we have found give the most efficient access to the VBS configuration in $pp \to HH jj$ events: distributions with $M_{jj}$, $\Delta\eta_{jj}$ and $M_{HH}$.

 In \figref{fig:HHjj_VBSdistributions} we present the predictions for the cross section of the process $pp\to HHjj$ for different values of $\lambda$ as a function of the separation in pseudorapidity of the final jets $|\Delta\eta_{jj}|$ and as a function of the invariant mass of these two jets $M_{jj}$. In these plots we can see that our signal is indeed dominated by the VBS configuration, since a very large fraction of the events populate the kinematic regions that correspond to VBS topologies. To have a quantitative estimation, we can take, for instance, the VBS selection cuts proposed in \cite{Delgado:2017cls} and impose them to the events shown in \figref{fig:HHjj_VBSdistributions}. Thus, by imposing these cuts:
\begin{align}
{\rm VBS\,\, CUTS :\,}\,\,\,\,\,|\Delta\eta_{jj}|&>4\,,~~~ M_{jj}>500~ {\rm GeV}\,,\label{VBSselectioncuts}
\end{align}
 we obtain that between 50\% and 75\% (depending on the value of $\lambda$, with closer values to 75\% for the larger values of $|\lambda|$) of the events are accepted within them, which means that the VBS topologies amount\footnote{In the sense of the fraction of events that pass the VBS cuts with respect to the total number of events.} , at least, to half of the total cross section of $pp\to HHjj$. This is indeed a very interesting result, since, as we will see in the forthcoming section, the VBS cuts allow us to reduce some backgrounds even in two orders of magnitude. The fact that the signal is practically left unaffected by these cuts is an excellent outcome as the signal to background ratio will favor a better sensitivity to $\lambda$.  
 
 \begin{figure}[t!]
\begin{center}
\includegraphics[width=0.49\textwidth]{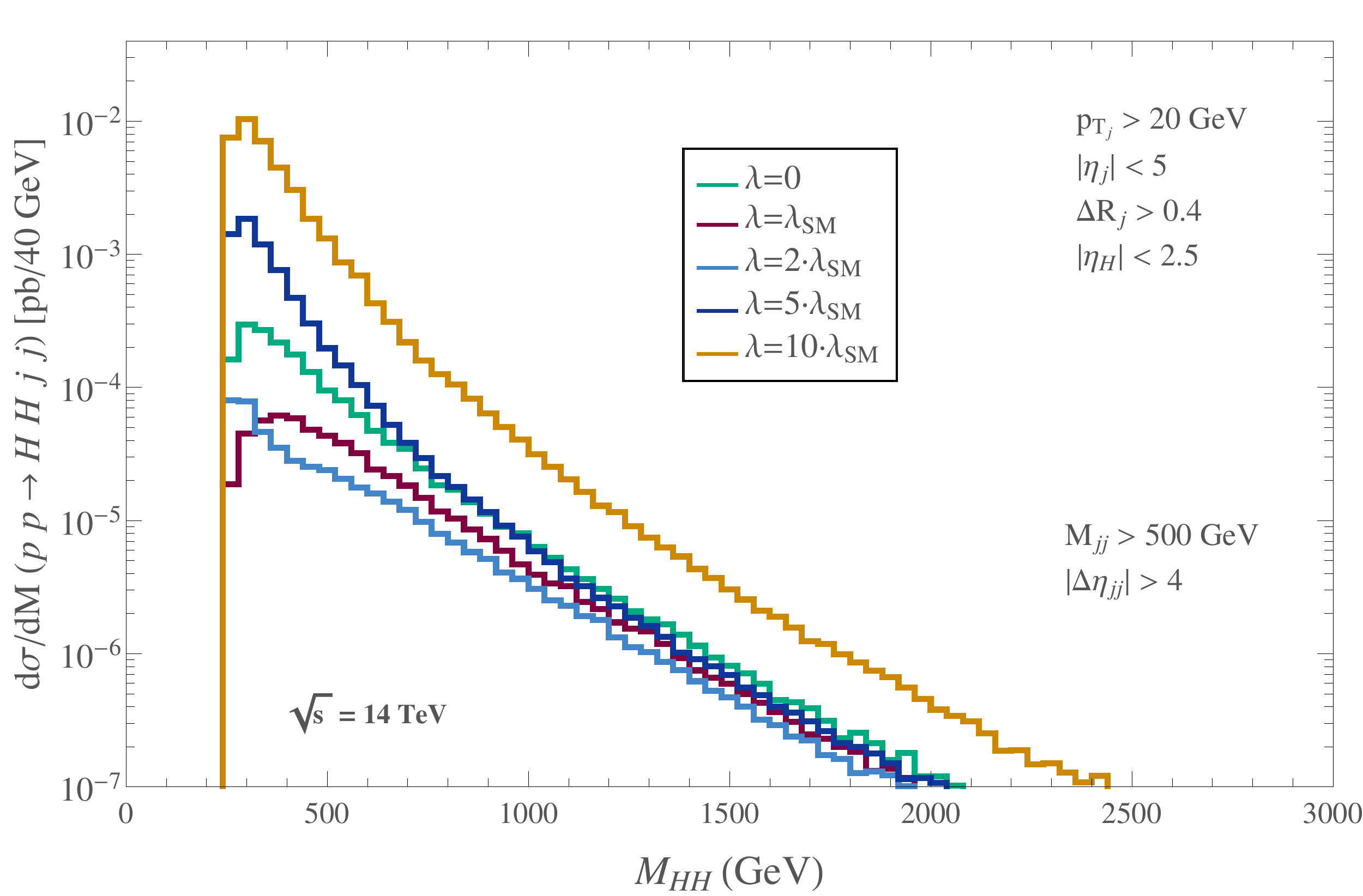}
\includegraphics[width=0.49\textwidth]{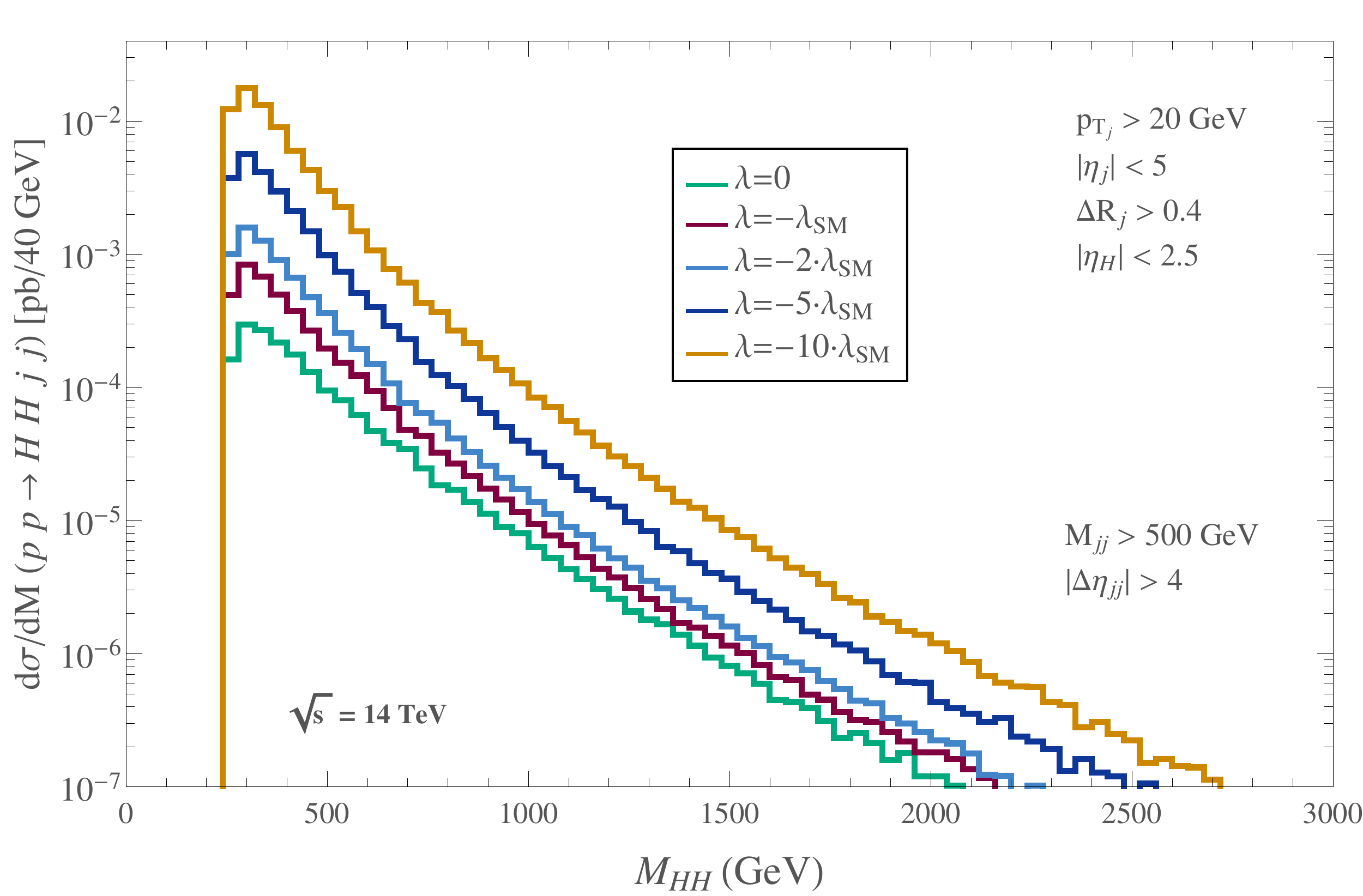}
\caption{Predictions for the total cross section of the process $pp\to HHjj$ as a function of the invariant mass of the di-Higgs system $M_{HH}$ for different values of the Higgs self-coupling $\lambda$. We display positive (left panel) and negative (right panel) values of $\lambda$ for comparison. We also include the case $\lambda=0$. Cuts in Eq.(\ref{basiccutsHHjj}) and VBS selection cuts presented in Eq.(\ref{VBSselectioncuts}) have been applied. The center of mass energy has been set to $\sqrt{s}=14$ TeV.}
\label{fig:MHH_distributions}
\end{center}
\end{figure}

Furthermore, knowing that the process of our interest at the LHC has a dominant VBS configuration, we would expect  the translation from the subprocess results to the complete ones at this level to be straightforward. This appears to be the case, as shown in \figref{fig:MHH_distributions}, where we display the predictions for the total cross section of the process $pp\to HHjj$ as a function of the invariant mass of the diHiggs system, $M_{HH}$, for different values of the Higgs self-coupling after imposing the cuts given in Eqs. (\ref{basiccutsHHjj}) and (\ref{VBSselectioncuts}). In these plots, it is manifest that the curves follow the same tendency as the subprocess when we vary $\lambda$. Near the $HH$ production threshold the difference in the cross sections for different values of the coupling is more pronounced, and one can see again that the cancellations play a role in the same way we learnt at the subprocess level. The SM cross section ($\kappa=1$, in red) lies between the $\kappa=0$ (in green) one, which is bigger, and the $\kappa=2$ (in light blue) one, which is smaller. Again, for negative values of $\kappa$ the cross section is always larger than the SM one, so we will have, for the same absolute value of the coupling, better sensitivities for negative $\lambda$ values.

The issue of the cancellations that take place between the diagram that depends on $\lambda$ and the rest is shown in more detail in \figref{fig:cancelations}. In this figure, we present the predictions for the total cross section for $pp\to HHjj$, and for the ratio of the total cross section over its SM value as a function of the Higgs self-coupling. We also compare the results with and without imposing the VBS cuts given in Eq.(\ref{VBSselectioncuts}) to explore how the cancellation happens at the LHC, and how it depends on the selection of the VBS topologies. We learn again, that, for the same absolute value of $\lambda$, negative values give rise to larger cross sections, and therefore to better sensitivities. The smallest cross section corresponds roughly to $\kappa\sim1.6$, which is the value that will be harder to reach at the LHC.
One may notice that this value does not coincide exactly with that in \figref{fig:subprocess_cancellation_energy}, even for the dominant contribution close to the threshold. This slight displacement of the minimum is due to the fact that many different topologies in addition to those of VBS contribute to this final state, in contrast with the results in  \figref{fig:subprocess_cancellation_energy} that took into account only VBS configurations. In fact, once we apply the VBS cuts the minimum gets closer to that of \figref{fig:subprocess_cancellation_energy}.
 Besides, and interestingly, the effect of imposing the VBS selection cuts can ameliorate the sensitivity to $\lambda$ . Although the cross sections reduce in value after applying the cuts, the ratio of the total cross section for a given trilinear coupling over the SM cross section increases when we are away from the region in which the cancellations are relevant, i.e., for $\kappa>3$ and $\kappa<1$.
 
 \begin{figure}[t!]
\begin{center}
\includegraphics[width=0.49\textwidth]{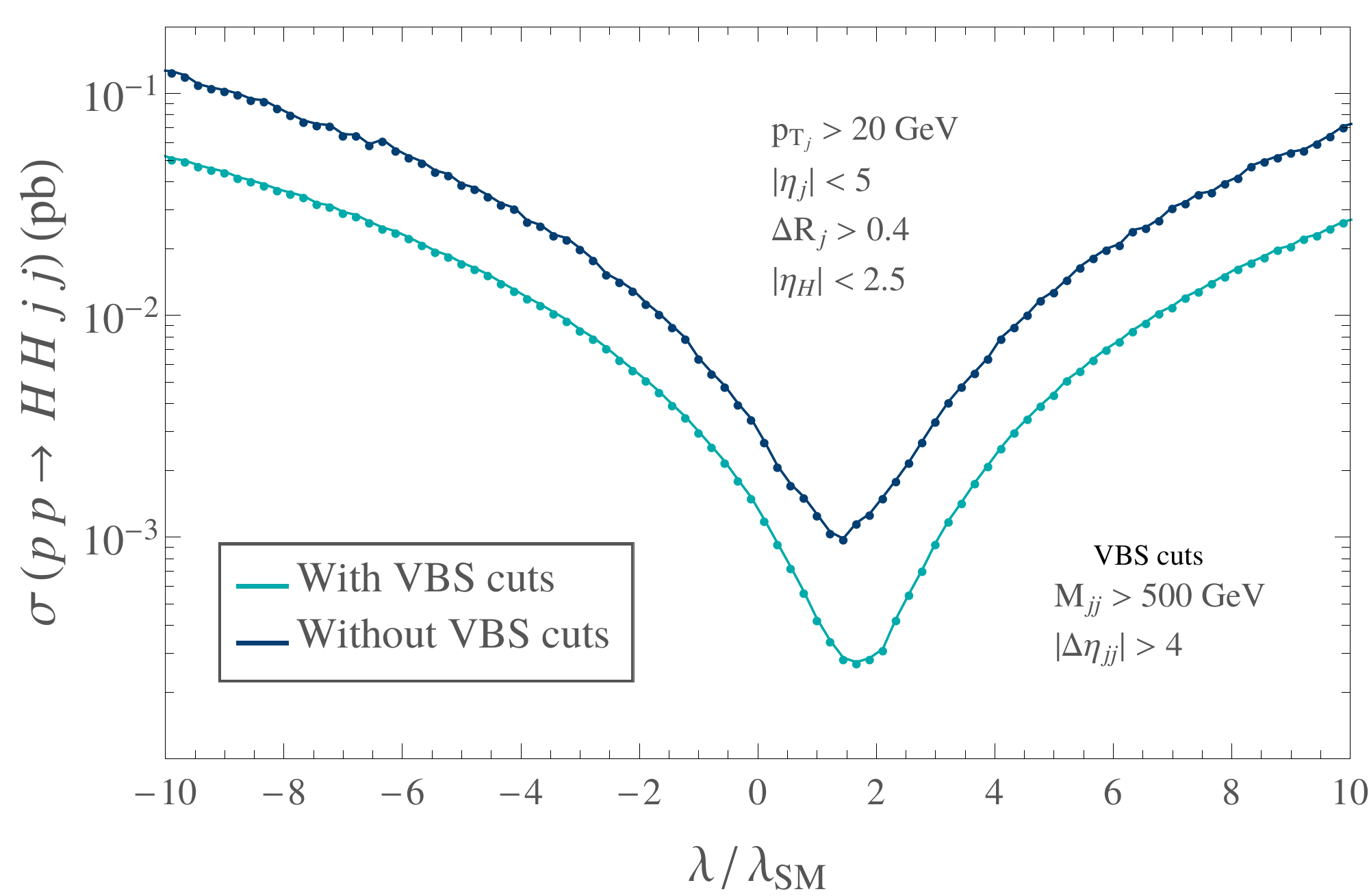}
\includegraphics[width=0.49\textwidth]{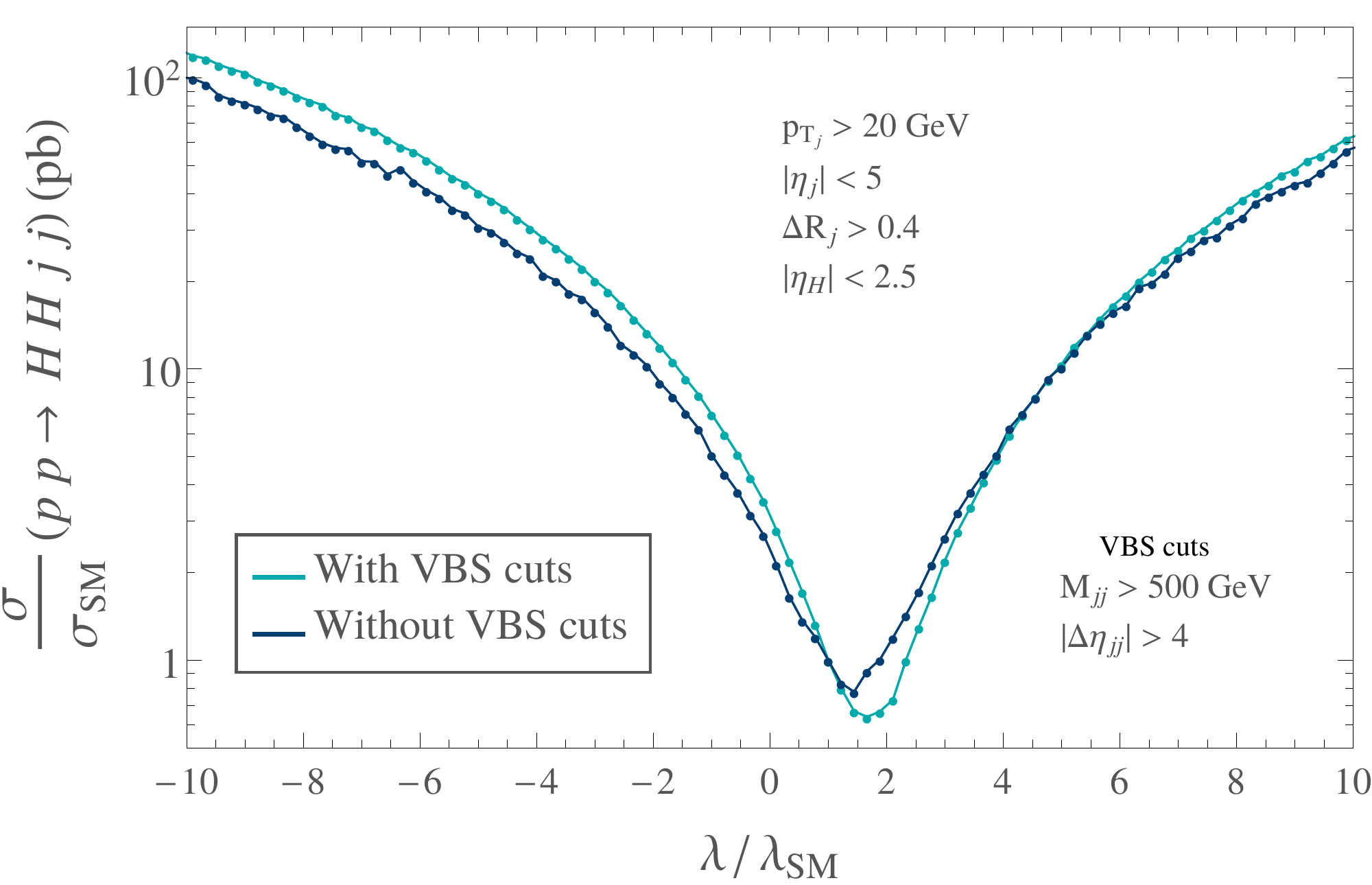}
\caption{Predictions for the total cross section (left panel) and for the ratio of the total cross section over its SM value (right panel) as a function of the Higgs self-coupling $\lambda$ with and without imposing the VBS selection cuts given in Eq.(\ref{VBSselectioncuts}). Cuts in Eq.(\ref{basiccutsHHjj}) have been applied and the center of mass energy has been set to $\sqrt{s}=14$ TeV. }
\label{fig:cancelations}
\end{center}
\end{figure}

The last issue we would like to point out in this section refers to the kinematical behavior of the VBS subsystem, that is then translated to the kinematics of the final Higgs bosons. Usually, in vector boson scattering processes at the LHC, most of the energy of the initial $pp$ state is transmitted to the radiated EW gauge bosons. This leads, as a consequence, to a very boosted system of final $HH$ pairs, which can be profitable to select these kind of events against backgrounds. If the final Higgs particles are very boosted, their decay products, will have, in general, small angular separations. This, together with the fact that the invariant mass of the two particles that come from the Higgs decay has to lie near the Higgs mass, will allow us to characterize very efficiently the Higgs boson candidates as we will see in the next section. With this and the VBS topologies under control, we can study the full processes in which the Higgs bosons have decayed, and compute the sensitivities to $\lambda$ in these realistic BSM scenarios.


 \subsection{Analysis after Higgs boson decays: sensitivity to $\lambda$ in $pp\to b\bar{b}b\bar{b}jj$ events}
 \label{4b2j}

 As previously mentioned, once we have fully characterized our most basic process, $pp\to HHjj$, we need to take into account the Higgs decays to perform a realistic analysis at the LHC. The channel we are going to focus on is $p p\to b\bar{b}b\bar{b}jj$, since the decay of the Higgs boson to a bottom-antibottom pair benefits from the biggest branching ratio, BR($H\to b\bar{b}$) $\sim$ 60 \%. Because of this, we will obtain the largest possible rates for our signal, which will allow us to probe the broadest interval of deviations in the Higgs self-coupling. 
 
 Although this process is really interesting because of its large statistics, it is important to mention that it also suffers from having a severe background: the one coming from pure multijet QCD events. This QCD background, of  $\mathcal{O}(\alpha_S^3)$ at the amplitude level, leads to the same final state as our signal, $p p\to b\bar{b}b\bar{b}jj$, and, although in general they have very different kinematics, their rates are so high that some of the events can mimic the signal coming from the decay of two Higgs particles. For this reason, we need to be very efficient when applying selection cuts and criteria to be able to reject this particular background.
 
 We learnt in the previous sections that our signal is very dominated by the VBS configuration. Oppositely, the multijet QCD background is composed primarily by topologies that do not share kinematical properties with VBS processes. This is the reason why we will first select those QCD events that can be misidentified as those signal events coming from VBS, and take them as a starting point to perform our more refined study of the signal and background.

To have a first insight on how efficient the VBS selection criteria are, we have generated with MadGraph5 ten thousand events for our signal, $p p\to HHjj\to b\bar{b}b\bar{b}jj$ in the SM, i.e., $\kappa=1$, and for the multijet QCD background with a set of basic cuts that ensure the detection of the final state particles:
 \begin{align}
p_{T_{j,b}}>20~ {\rm GeV}\,;~|\eta_j|<5\,;~|\eta_b|<2.5\,;~\Delta R_{jj,jb}>0.4\,;~\Delta R_{bb}>0.2\,.\label{basiccuts4b2j}
 \end{align}
where $p_{T_j,b}$ is the transverse momentum of the jets and bottoms, $\eta_{j,b}$ are the pseudorapidities of the jets or of the bottom particles, and $\Delta R_{ij}$ is the angular separation between the $i$ and $j$ particles.

\begin{figure}[t!]
\begin{center}
\includegraphics[width=0.49\textwidth]{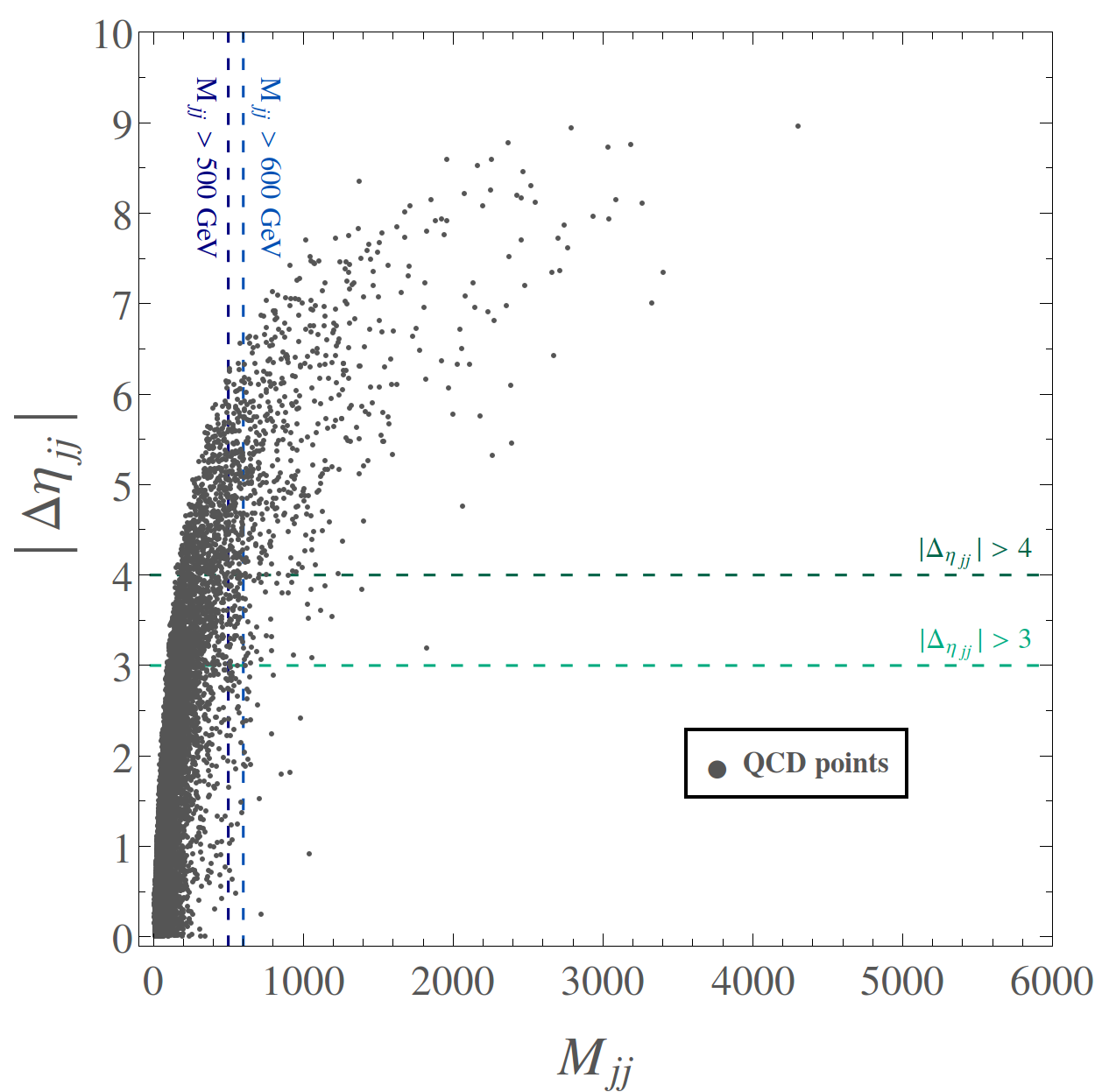}
\includegraphics[width=0.49\textwidth]{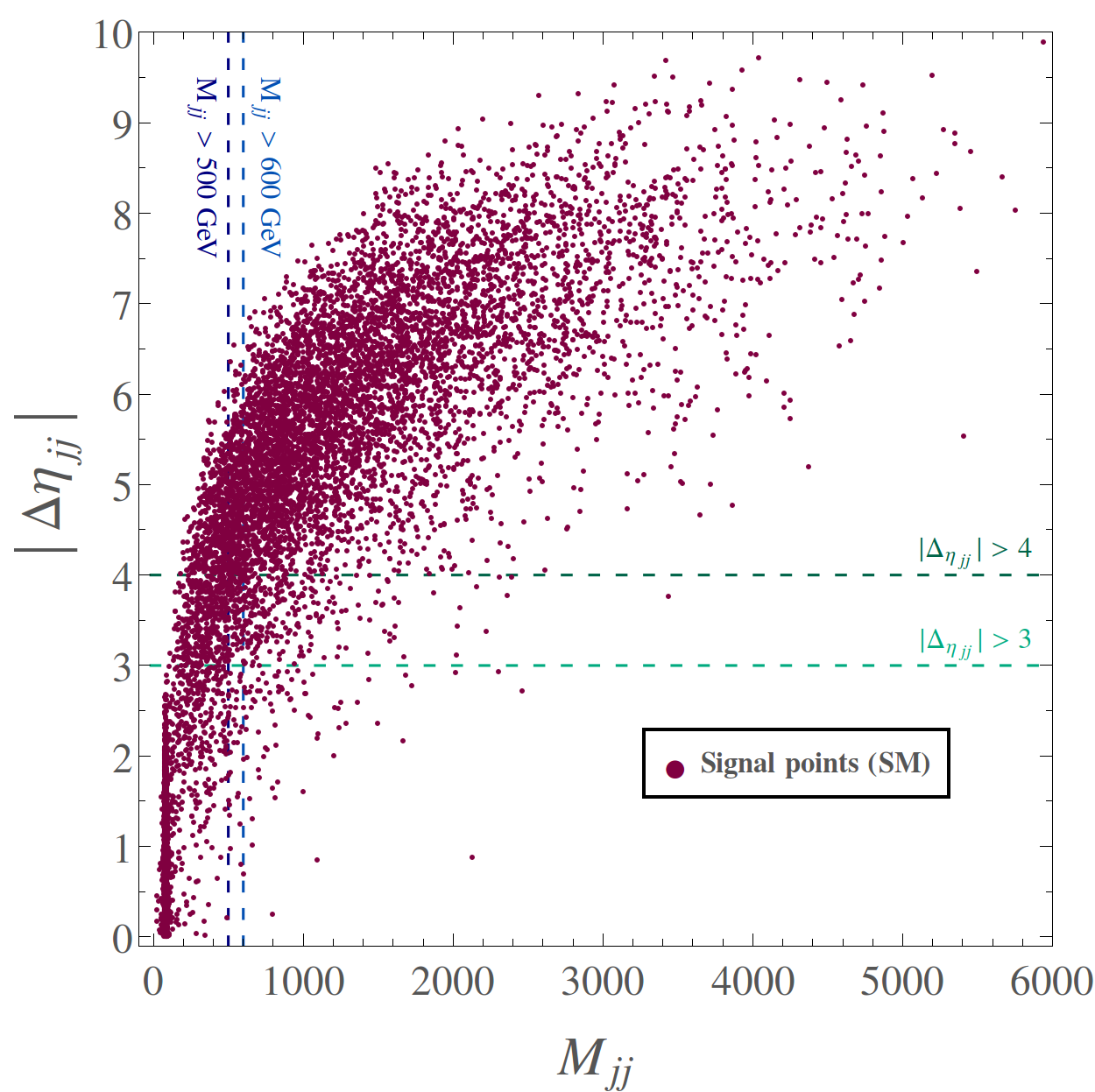}
\caption{Distribution of 10000 Monte Carlo events of multijet QCD background $pp\to b\bar{b}b\bar{b}jj$ (left panel) and signal $pp\to HHjj\to b\bar{b}b\bar{b}jj$ (right panel) in the plane of the absolute value of the difference between pseudorapidities of the two jets $|\Delta\eta_{jj}|$ versus the invariant mass of the two jets $M_{jj}$. Cuts in \eqref{basiccuts4b2j} have been implemented. The center of mass energy has been set to $\sqrt{s}=$14 TeV.}
\label{fig:4b2j_VBSvariables2D}
\end{center}
\end{figure}

In \figref{fig:4b2j_VBSvariables2D} we display the localization of these events in the $|\Delta\eta_{jj}|-M_{jj}$ plane, the two variables that better characterize the VBS processes. One can see, indeed, that the QCD events populate mostly the region of small invariant masses of the dijet system and of small differences in pseudorapidity of the jets, as opposed, precisely, to the signal events. Thus, imposing the proper VBS cuts, like those in \eqref{VBSselectioncuts}, should relevantly reduce the QCD background leaving the signal nearly unaffected.

In \figref{fig:4b2j_2D_Mbb} we aim precisely to see this effect, since we present the same set of events as in \figref{fig:4b2j_VBSvariables2D} for the QCD background and for the signal highlighting in orange those events that fulfill the VBS selection criteria given in \eqref{VBSselectioncuts} as an example. This time we show the results in the $M_{bb_1}-M_{bb_2}$ plane, where $M_{bb_{1,2}}$ are the corresponding invariant masses of the two bottom pairs that are the best candidates to come from the decay of a Higgs boson, as we will see later.

The first thing one can observe in both plots of \figref{fig:4b2j_2D_Mbb} is that very few QCD events survive the imposition of the VBS cuts, whereas practically all events of the signal do. The concrete fraction of the events ($\mathcal{A}$) that survive in both cases is also presented in the plots. We call $\mathcal{A_{\rm VBS}}$ the acceptance of the VBS cuts, defined as
\begin{equation}
\mathcal{A}_{\rm VBS}\equiv\dfrac{\sigma(pp\to b\bar{b}b\bar{b}jj)|_{\rm VBS}}{\sigma(pp\to b\bar{b}b\bar{b}jj)}\,,
\end{equation}
i.e., the ratio between the cross section of the process after applying the VBS cuts like those in \eqref{VBSselectioncuts} over the cross section of the process without having applied them. The basic cuts are imposed in all cases. Taking a look at these numbers, we see that 60\% of the signal events pass these cuts while only 9\% of the QCD events do. At this point, one might wonder wether these results are very dependent on the specific VBS cuts we impose or not. In \tabref{table:VBS} we show the predictions for the acceptances, $\mathcal{A}_{\rm VBS}$, of different sets of VBS selection cuts, i.e., different cuts in $|\Delta\eta_{jj}|$ and in $M_{jj}$, for both the multijet QCD background and the signal with $\kappa=1$. From those predictions we can see that all the sets of cuts considered lead to very similar results: around 60\% of the signal fulfills the VBS selection criteria whereas a 5-10\% of the multijet QCD background does. We have checked that for other values of $\kappa$ the acceptance for the signal varies between a 55\% and a 75\%. From now on we will apply the VBS selection cuts given in \eqref{VBSselectioncuts}, since this set is well explored in the literature and qualitatively provides the same results as the other sets of cuts that we have analyzed.

\begin{figure}[t!]
\begin{center}
\includegraphics[width=0.49\textwidth]{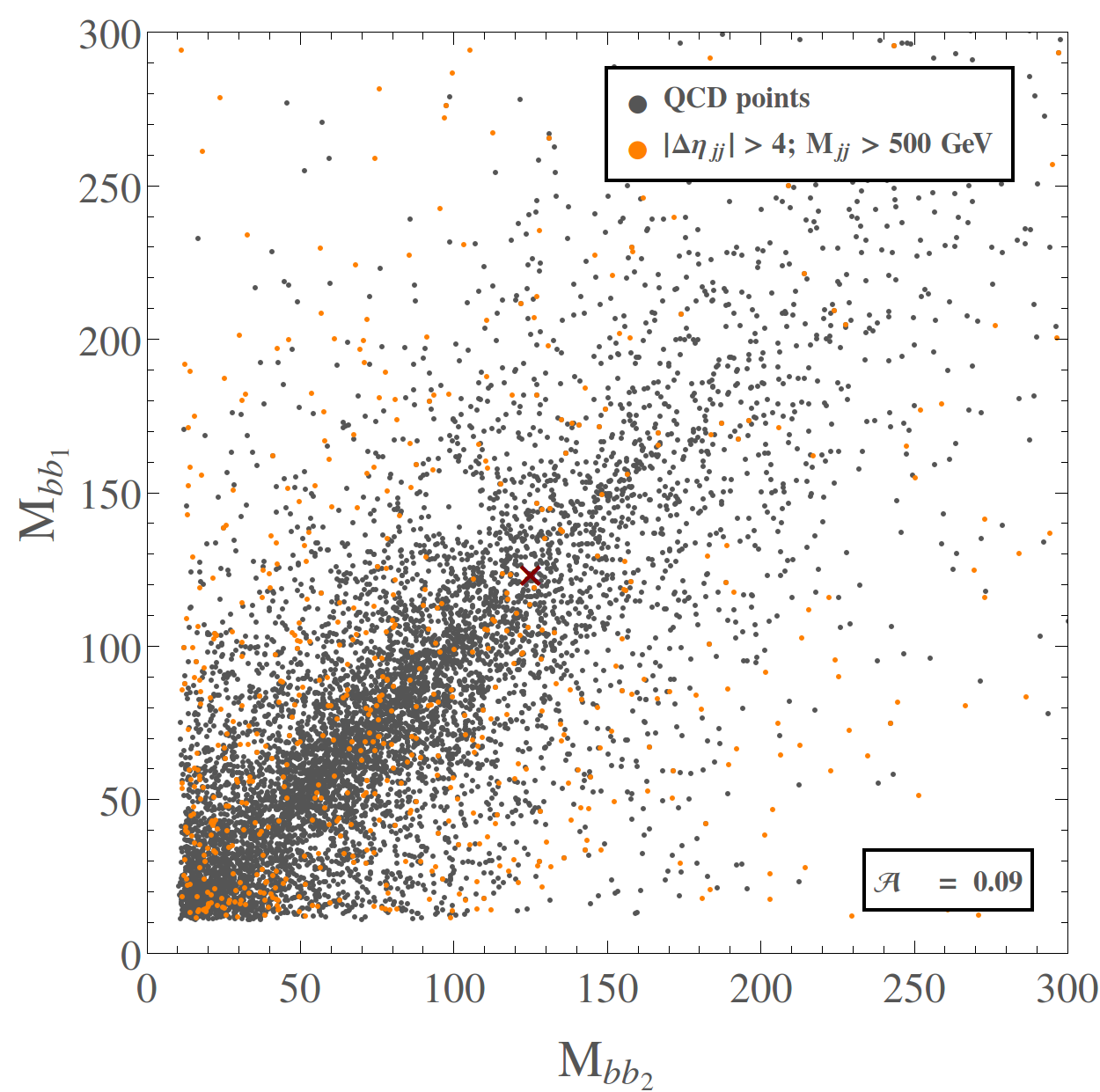}
\includegraphics[width=0.49\textwidth]{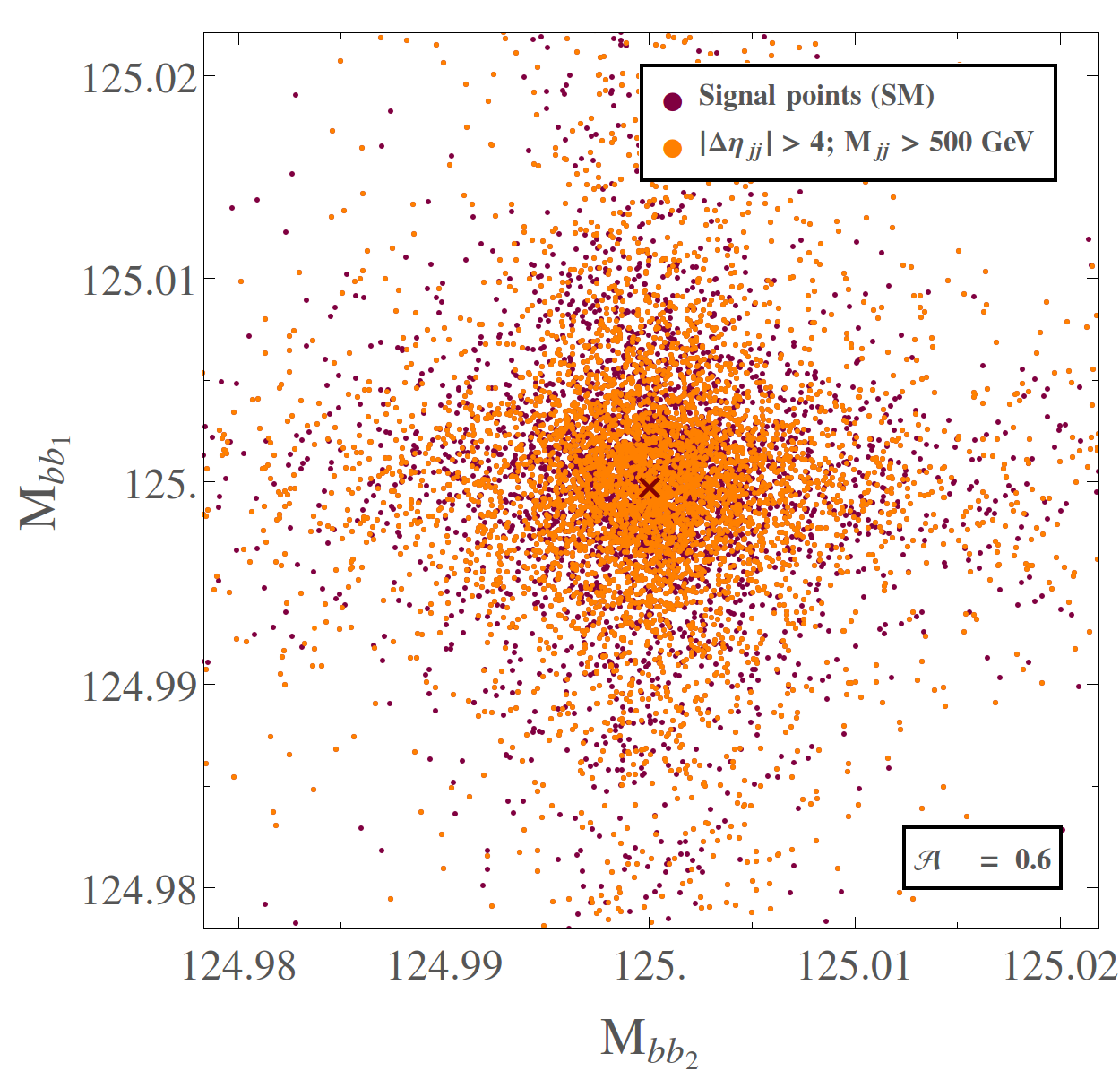}
\caption{Distribution of 10000 Monte Carlo events of multijet QCD background $pp\to b\bar{b}b\bar{b}jj$ (left panel) and signal $pp\to HHjj\to b\bar{b}b\bar{b}jj$ (right panel) in the plane of the invariant mass of one bottom pair identified as a Higgs candidate following the criteria presented in the text $M_{bb_1}$ versus the invariant mass of the other bottom pair identified as the other Higgs candidate $M_{bb_2}$. Orange dots correspond to those events that pass the implemented VBS selection cuts given in Eq.(\ref{VBSselectioncuts}). Cuts in \eqref{basiccuts4b2j} have been implemented. The value of the acceptance $\mathcal{A}$ is also included. The red cross represents the value of the Higgs mass The center of mass energy has been set to $\sqrt{s}=$14 TeV.}
\label{fig:4b2j_2D_Mbb}
\end{center}
\end{figure}

The second issue that we can notice about \figref{fig:4b2j_2D_Mbb} is that, again, the QCD events populate a very different region of this plane than those of the signal. QCD events tend to lie on low values of ${M_{bb_i}}$, somehow away from the  region close to the $[M_{bb_1}=m_H,M_{bb_2}=m_H]$ point in the $M_{bb_1}-M_{bb_2}$ plane, in which most of our signal settles. Evidently, two particles coming from the decay of a Higgs boson should have a total invariant mass value near the Higgs boson mass, $m_H$, as our signal does. This motivates the next selection criteria we are going to apply, following the search strategies of ATLAS \cite{Aaboud:2018knk} and CMS \cite{CMS:2016foy} for double Higgs production, that are aimed to efficiently identify the $HH$ candidates.

The $HH$ candidate identification criteria are also based on what we have learned in the previous sections. Logically, each $H$ candidate corresponds to a $b$-quark pair, and therefore we first need to define how we are going to pair the final $b$-quarks. From now on, it is worth mentioning that we will not distinguish between bottom and anti-bottom, similarly to what is done in experimental analyses. Therefore, with four bottom-like particles in the final state we have three possible double pairings. Among these three possibilities, we select the one in which the values of the invariant masses of the pairs are closer, i.e., the one that minimizes $|M_{bb_1}-M_{bb_2}|$, where $M_{bb_1}$ is the invariant mass of one of the $bb$ pairs and $M_{bb_2}$ is the invariant mass of the other pair. Once we have defined the $b$-quark pairing, we can profit from the fact that, as mentioned before, if two $b$-quarks come from the decay of a boosted Higgs boson, as it happens in VBS processes, the angular separation between them should be small. Thus, we should look for pairs of $b$-quarks with small (and yet  enough to resolve the particles) $\Delta R_{bb}$. Furthermore, we have already discussed that our signal is characterized by the fact that the invariant mass of each $b$-quark pair should be around the Higgs mass, $m_H$. Therefore, imposing this criterion will ensure that we are maximizing the selection of events that come from the decays of two Higgs bosons.

\begin{table}[t!]
\begin{center}
\vspace{.2cm}
\begin{tabular}{ c @{\extracolsep{1cm}} c  @{\extracolsep{1cm}} c }
\toprule
\toprule
Set of VBS cuts & $\mathcal{A}_{\rm VBS}^{\rm QCD}$  & $\mathcal{A}_{\rm VBS}^{{\rm Signal};\kappa=1}$\\
\midrule
$|\Delta\eta_{jj}|>4,~M_{jj} > 500$ GeV &  0.086 & 0.631\\
\rule{0pt}{1ex}
$|\Delta\eta_{jj}|>4,~M_{jj} > 600$ GeV   &  0.066  &   0.597\\
\rule{0pt}{1ex}
$|\Delta\eta_{jj}|>4,~M_{jj} > 700$ GeV & 0.054   &     0.558\\
\rule{0pt}{1ex}
 $|\Delta\eta_{jj}|>3,~M_{jj} > 500$ GeV&  0.098  &     0.669\\
\rule{0pt}{1ex}
$|\Delta\eta_{jj}|>3,~M_{jj} > 600$ GeV &  0.071  & 0.626\\
\rule{0pt}{1ex}
$|\Delta\eta_{jj}|>3,~M_{jj} > 700$ GeV&   0.057 &     0.580\\
\bottomrule
\bottomrule
\end{tabular}
\caption{\small Predictions for the acceptance of different sets of VBS cuts, including those in \eqref{VBSselectioncuts}, for the multijet QCD background and for the signal with $\kappa=1$. Signal acceptances for the other values of $\kappa$ considered in the present work, $\kappa\in[-10,10]$, vary between 0.5 and 0.75.}
\label{table:VBS}
\end{center}
\end{table}


With all these features in mind, and guided by the ATLAS search strategies \cite{Aaboud:2018knk}, we define the following set of cuts as the requirements to efficiently select the candidates to Higgs boson pairs: 
\begin{align}
&HH \,\,\, {\rm CANDIDATE\,\,\, CUTS \,\,\,:} \nonumber \\[0.3em]
&p_{T_b}> 35~ {\rm GeV}\,,\label{cutsHH1}\\
&\hat{\Delta}R_{bb}\equiv\left\{ \begin{array}{l} 0.2 < \Delta R_{bb^l} < \frac{653}{M_{4b}}+0.475\,;~0.2 < \Delta R_{bb^s} < \frac{875}{M_{4b}}+0.35\,,~M_{4b}<1250~{\rm GeV}\,, \\0.2 < \Delta R_{bb^l} < 1\,;~0.2 < \Delta R_{bb^s} < 1\,,~M_{4b}>1250~{\rm GeV}\,, \end{array} \right.\label{cutsHH2}\\
&\hat{p}_{T_{bb}}\equiv p_{T_{bb^l}}>M_{4b}/2-103 {\rm GeV}\,;~ p_{T_{bb^s}}>M_{4b}/3-73{\rm GeV}\,,\label{cutsHH3}\\
&\chi_{HH}\equiv\sqrt{\left(\dfrac{M_{bb^l}-m_H}{0.05\,m_H}\right)^2+\left(\dfrac{M_{bb^s}-m_H}{0.05\,m_H}\right)^2}<1\,,\label{cutsHH}
\end{align}
where the super-indices $l$ and $s$ denote, respectively, leading and subleading, defining the leading $b$-quark pair as the one with largest scalar sum of $p_T$. One might notice that the requirement of small angular separation between the two $b$-quarks of a pair, and the fact that the invariant mass of each $b$-quark pair has to lie near the mass of the Higgs, are encoded in the $\hat{\Delta}R_{bb}$ and in the $\chi_{HH}$ cuts, respectively. The latter is equivalent to impose that the events in the $M_{bb_1}-M_{bb_2}$ plane have to lie inside a circle of radius $0.05\,m_H=6.25$ GeV centered in the point $[M_{bb_1}=m_H,M_{bb_2}=m_H]$.

Nevertheless, although multijet QCD events represent the most severe background, there are other processes that can fake our signal. One of them is the $t\bar{t}$ background, with the subsequent decays of the top quarks and $W$ bosons, $t\bar{t}\to b W^+ \bar{b} W^-\to   b\bar{b}b\bar{b}jj$. This is, however, a very controlled background, since it is well suppressed by non-diagonal CKM matrix elements and its kinematics are radically different than those of VBS. Starting from a cross section of $5.4\cdot 10^{-5}$ pb with all the basic cuts in \eqref{basiccuts4b2j} applied, one ends up in $1.7\cdot 10^{-7}$ pb after applying the $HH$ candidate cuts, and in $2.0 \cdot 10^{-10}$ pb after applying the VBS cuts afterwards. Therefore, since this background is five orders of magnitude smaller than the smallest of our signals, we will neglect it from now onwards.
Finally, we still have to deal with other potentially important backgrounds corresponding to $pp\to HZjj \to b\bar{b}b\bar{b}jj$ and $pp\to ZZjj \to b\bar{b}b\bar{b}jj$. These two $HZ$ and $ZZ$ production processes, receiving contributions of order ($\alpha\cdot\alpha_S$) and  ($\alpha^2$) at the amplitude level, also drive to the same final state as our signal and may give rise to similar kinematics, since they can also take place through VBS configurations. In fact, their rates are very close to those of our signal after applying the VBS selection cuts, that reduce these backgrounds less efficiently than the multijet QCD one. However, we can again take advantage of the fact that the $b$-quark pairs have to come from a Higgs boson with a well defined mass. Therefore the $HH$ candidate cuts should allow us to reject these backgrounds.

\begin{table}[b!]
\begin{center}
\vspace{.2cm}
\begin{tabular}{ c @{\extracolsep{0.8cm}} c @{\extracolsep{0.8cm}} c @{\extracolsep{0.8cm}} c }
\toprule
\toprule
Cut & $\sigma_{\rm QCD}$  [pb]& $\sigma_{ ZHjj,ZZjj}$ [pb] & $\sigma_{{\rm Signal};\kappa=1}$ [pb]\\
\midrule
Basic detection cuts in \eqref{basiccuts4b2j} & 602.72  & 0.028 &5.1$\cdot 10^{-4}$\\
$p_{T_b}>$ 35 GeV, \eqref{cutsHH1}  &  98.31  &  0.01 &3.0$\cdot 10^{-4}$\\
$\hat{\Delta}R_{bb}$, \eqref{cutsHH2}  & 33.80   &  $6.3\cdot 10^{-3}$   &1.1$\cdot 10^{-4}$\\
$\hat{p}_{T_{bb}}$, \eqref{cutsHH3}&  29.77  &  5.8$\cdot 10^{-3}$   &9.0$\cdot 10^{-5}$\\
$\chi_{HH}<1$, \eqref{cutsHH} &  $7.9\cdot10^{-2}$  &  8.6$\cdot 10^{-6}$   &9.0$\cdot 10^{-5}$\\
VBS cuts in \eqref{VBSselectioncuts} &  $6.8\cdot10^{-3}$  &   5.5$\cdot 10^{-6}$  &4.1$\cdot 10^{-5}$\\
\bottomrule
\bottomrule
\end{tabular}
\caption{\small Predictions for the total cross section of the multijet QCD background, of the combined $pp\to HZjj \to b\bar{b}b\bar{b}jj$ and $pp\to ZZjj \to b\bar{b}b\bar{b}jj$ background and of the signal with $\kappa=1$ after imposing each of the cuts given in \eqref{basiccuts4b2j} and in Eqs. (\ref{cutsHH1})-(\ref{cutsHH}) subsequently. We show as well the total cross section after applying, afterwards, the VBS selection cuts in \eqref{VBSselectioncuts}. }
\label{table:cutflow}
\end{center}
\end{table}


In \tabref{table:cutflow} we present the cross sections of the multijet QCD background, of the combined $pp\to HZjj \to b\bar{b}b\bar{b}jj$ and $pp\to ZZjj \to b\bar{b}b\bar{b}jj$ background and of the signal with $\kappa=1$, with the basic cuts already set, after applying each of the cuts in Eqs.(\ref{cutsHH1})-(\ref{cutsHH}) subsequently. This way, we see the reduction factor after each cut, and the total cross section of both signal and background once we have performed our complete $HH$ candidate selection. We show as well the effect of applying the VBS cuts given in \eqref{VBSselectioncuts} afterwards, since we have checked that both sets of cuts ($HH$ candidate cuts and VBS cuts) are practically independent. Thus, we have the total cross sections of the two main backgrounds and of our SM signal after applying all the selection criteria. In \tabref{Tab:allSig_4b2j} we provide the total cross sections of the signal for all the values of $\lambda$ considered in this work,  again after applying all the selection criteria, for comparison.

\begin{table}[t!]
\begin{center}
\begin{tabular}{c| @{\extracolsep{0.8cm}} c @{\extracolsep{0.8cm}} c @{\extracolsep{0.8cm}} c @{\extracolsep{0.8cm}} c @{\extracolsep{0.8cm}} c @{\extracolsep{0.8cm}} c @{\extracolsep{0.8cm}} c @{\extracolsep{0.8cm}} c @{\extracolsep{0.8cm}} c}
\toprule
\toprule
$\kappa$&  $0$& $1$ & $-1$ & $2$& $-2$& $5$& $-5$& $10$& $-10$\\
\midrule
 $\sigma_{\rm Signal}\cdot 10^4$ [pb] &$1.9$ & $0.4$ & $5.0$ & $0.4$ & $9.7$&$10.1$ & $33.2$ & $56.4$ & $102.6$ \\
\bottomrule
\bottomrule
\end{tabular}
\caption{Predictions for the total cross section of the signal $pp\to b\bar{b}b\bar{b} j j$ after imposing all the selection criteria, VBS cuts given in in \eqref{VBSselectioncuts} and $HH$ candidate cuts given in Eqs. (\ref{cutsHH1})-(\ref{cutsHH}) for all the values of $\kappa$ considered in this work: $\kappa=0,\pm 1, \pm 2, \pm 5, \pm 10$. Basic cuts in \eqref{basiccuts4b2j} are also applied.}
\label{Tab:allSig_4b2j}
\end{center}
\end{table}

From the results in \tabref{table:cutflow} we can learn that the sum of the two backgrounds, $ZHjj$+$ZZjj$,  is under control after applying the $HH$ candidate cuts, since its cross section lies an order of magnitud below the SM signal. On the other hand, the multijet QCD background remains being very relevant even after imposing all the selection criteria. However, as we will see later, the total reduction that it suffers still allows to be sensitive to interesting values of $\kappa$ even for low luminosities. This reduction, along with that suffered by the $ZHjj$+$ZZjj$ backgrounds and with that suffered by the SM signal, is presented in \tabref{table:acceptances}. There we show the acceptances of the VBS cuts and the $HH$ candidate cuts separately and together for the multijet QCD background, for the combined $pp\to HZjj \to b\bar{b}b\bar{b}jj$ and $pp\to ZZjj \to b\bar{b}b\bar{b}jj$ background and for the SM signal, for comparison.

\begin{table}[b!]
\begin{center}
\vspace{.2cm}
\begin{tabular}{ c @{\extracolsep{0.8cm}} c @{\extracolsep{0.8cm}} c @{\extracolsep{0.8cm}} c }
\toprule
\toprule
Cut & $\mathcal{A}^{\rm QCD}$& $\mathcal{A}^{ ZHjj,ZZjj}$  & $\mathcal{A}^{{\rm Signal};\kappa=1}$
\\
\midrule
VBS cuts in \eqref{VBSselectioncuts} & 0.086  & 0.630 & 0.631
\\
$HH$ candidate cuts in Eqs. (\ref{cutsHH1})-(\ref{cutsHH})   &  1.3$\cdot 10^{-4}$  &  3.1$\cdot 10^{-4}$ & 0.17
\\
VBS cuts $+$ HH candidate cuts   &  1.1$\cdot 10^{-5}$  &  2.0$\cdot 10^{-4}$ & 0.081
\\
\bottomrule
\bottomrule
\end{tabular}
\caption{\small Predictions for acceptances of the VBS cuts given in \eqref{VBSselectioncuts}, of the $HH$ candidate cuts given in Eqs. (\ref{cutsHH1})-(\ref{cutsHH}), and of both sets of cuts combined  for the multijet QCD background, for the combined $pp\to HZjj \to b\bar{b}b\bar{b}jj$ and $pp\to ZZjj \to b\bar{b}b\bar{b}jj$ background and for the signal with $\kappa=1$. All the results are computed with the basic cuts in \eqref{basiccuts4b2j} already applied.} 
\label{table:acceptances}
\end{center}
\end{table}


 It must be noticed that other backgrounds apart from those having the same final particle content as our signal can contribute relevantly. This would  be the case if some final state particles were misidentified, leading to a ``fake'' $ b\bar{b}b\bar{b}jj$ state.  The most important of these backgrounds is the production of a $t\bar{t}$ pair decaying into two $b$ quarks and four light jets,  $t\bar{t}\to b W^+ \bar{b} W^-\to   b\bar{b}jjjj$ with two of these final light jets being misidentified as two $b$ jets.  In order to estimate the contribution of this background, we have generated with MadGraph5 $t\bar{t}\to b W^+ \bar{b} W^-\to   b\bar{b}jjjj$ events applying first the minimal cuts $|p_{T_{j,b}}|>20$ GeV, $|\eta_{j,b}|<5$ and $\Delta R_{jj,bj,bb}>0.2$, with a total cross section of 246 pb. Applying a mistagging rate of 1\% per each light jet misidentified as a $b$ jet, we obtain $246\cdot(0.01)^2=2.5\cdot 10^{-2}$ pb as starting point to compare to our main multijet QCD $ b\bar{b}b\bar{b}jj$ background. Now we need to apply our selection cuts described in \eqrefs{VBSselectioncuts}{cutsHH} to see their impact on this particular background. We apply first the VBS selection cuts asking that there is at least one pair of light jets fulfilling the criteria in \eqref{VBSselectioncuts}. These cuts reduces the cross section to $1.3\cdot 10^{-4}$ pb. Now analyzing the events that pass the VBS cuts, if there is only one pair of ``VBS-like'' light jets, the other two light jets are identified as $b$ quarks. If there is more than one, we select as $b$ quarks those that minimize $|M_{pp_1}-M_{pp_2}|$, with $p=b,j$ among all possibilities. Once we have characterized our two light jet candidates and our four $b$-quark candidates, we proceed with the $HH$ candidate selection cuts. This way, applying subsequently the criteria explained in \eqrefs{cutsHH1}{cutsHH}, we obtain the following cross sections: $1.2\cdot 10^{-5}$ pb ($p_{T_b}$), $2.5\cdot 10^{-6}$ pb ($\hat{\Delta}R_{bb}$), $4.4\cdot 10^{-7}$ pb ($\hat{p}_{T_{bb}}$) and finally $2.1\cdot 10^{-8}$ pb ($\chi_{HH}$). Therefore, since this $t\bar{t}$ background is five orders of magnitude below our main considered background, whose final cross section given in \tabref{table:cutflow} is $6.8\cdot10^{-3}$ pb, we conclude that it can be safely neglected.

We have also considered the possible backgrounds coming from multijet QCD processes leading to different final states than that of $b\bar{b}b\bar{b}jj$, such as $6j$ and $b\bar{b}jjjj$, in which some of the final state light jets are again misidentified as $b$ jets. To estimate their contribution to the background we have used the total cross sections of these processes given in \cite{Mangano:2002ea}. These are, for a center of mass of 14 TeV, $1.3\cdot10^5$ pb and $7.5\cdot 10^3$ pb, respectively. If we apply now the corresponding misidentification rates we end up with  $1.3\cdot10^5 \cdot (0.01)^4=1.3\cdot10^{-3}$ pb for the case in which we have six light jets, and $7.5\cdot 10^3 \cdot (0.01)^2=7.5\cdot10^{-1}$ pb for the case in which we have two $b$ jets and four light jets. We now assume that the selection cuts we specify in Eqs. (\ref{VBSselectioncuts}) -(\ref{cutsHH}) will have a similar impact on these backgrounds as they do on the multijet QCD production of four $b$ jets and two light jets, since they all take place through similar QCD configurations. Thus, applying the corresponding acceptance factor of these cuts we obtain the following total cross sections: $1.3\cdot10^{-3}\cdot 1.1 \cdot 10^{-5}=1.4 \cdot 10^{-8}$ pb for the six light jets case and  $7.5\cdot10^{-1}\cdot 1.1 \cdot 10^{-5}=8.2 \cdot 10^{-6}$ pb for the two $b$ jets and four light jets case. Both of these cross sections are more than three orders of magnitude below that  of the  $b\bar{b}b\bar{b}jj$ background, so we conclude that they can also be safely neglected without introducing big uncertainties.

Once we have the possible backgrounds under control, we can move on to fully explore the sensitivity to the Higgs self-coupling $\lambda$ in $p p\to b\bar{b}b\bar{b}jj$ events. In \figref{fig:M4b_distributions} we display the predictions for the total cross section of the total SM background (the sum of multijet QCD background and  the combined $ZHjj+ZZjj$ backgrounds) and of the signal for different values of $\lambda$ as a function of the invariant mass of the four-bottom system $M_{b\bar{b}b\bar{b}}$. These distributions should be the analogous to those in \figref{fig:MHH_distributions} after the Higgs boson decays, as it is manifest since the signal curves follow the same tendency and are very similar except for the global factor of the Higgs-to-bottoms branching ratio. In this figure we can also see that the total SM background is of the same order of magnitude than the $\kappa=10$ and $\kappa=-5$ signals, and it is even below the $\kappa=-10$ signal prediction. This is a very interesting result, since it means that if, for example, the true value of $\lambda$ was minus five times that of the SM, the LHC should be able to measure twice as many events as those expected from the SM background only in this VBS configuration. Similar conclusions can be extracted for other values of $\kappa$.

  \begin{figure}[t!]
\begin{center}
\includegraphics[width=0.49\textwidth]{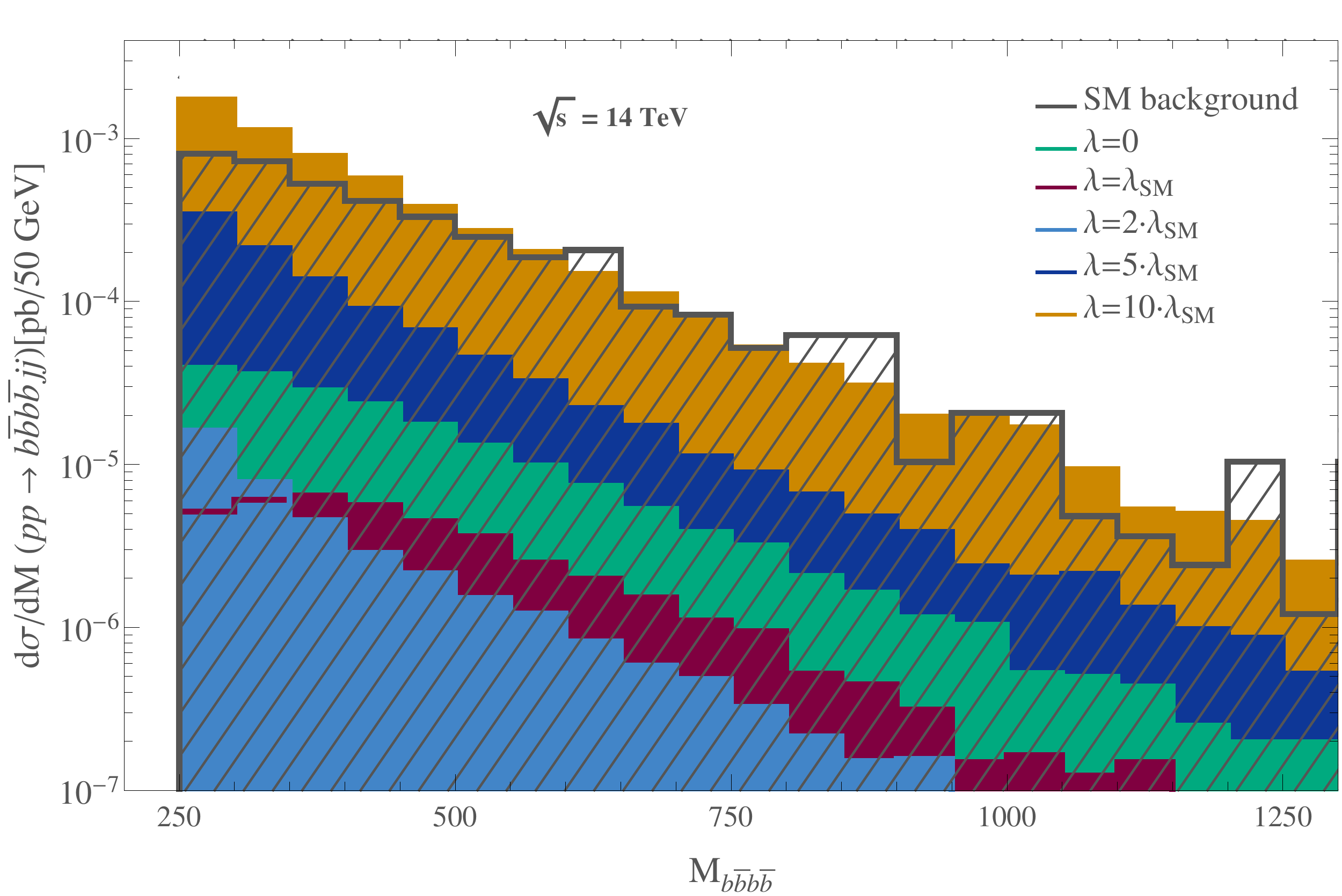}
\includegraphics[width=0.49\textwidth]{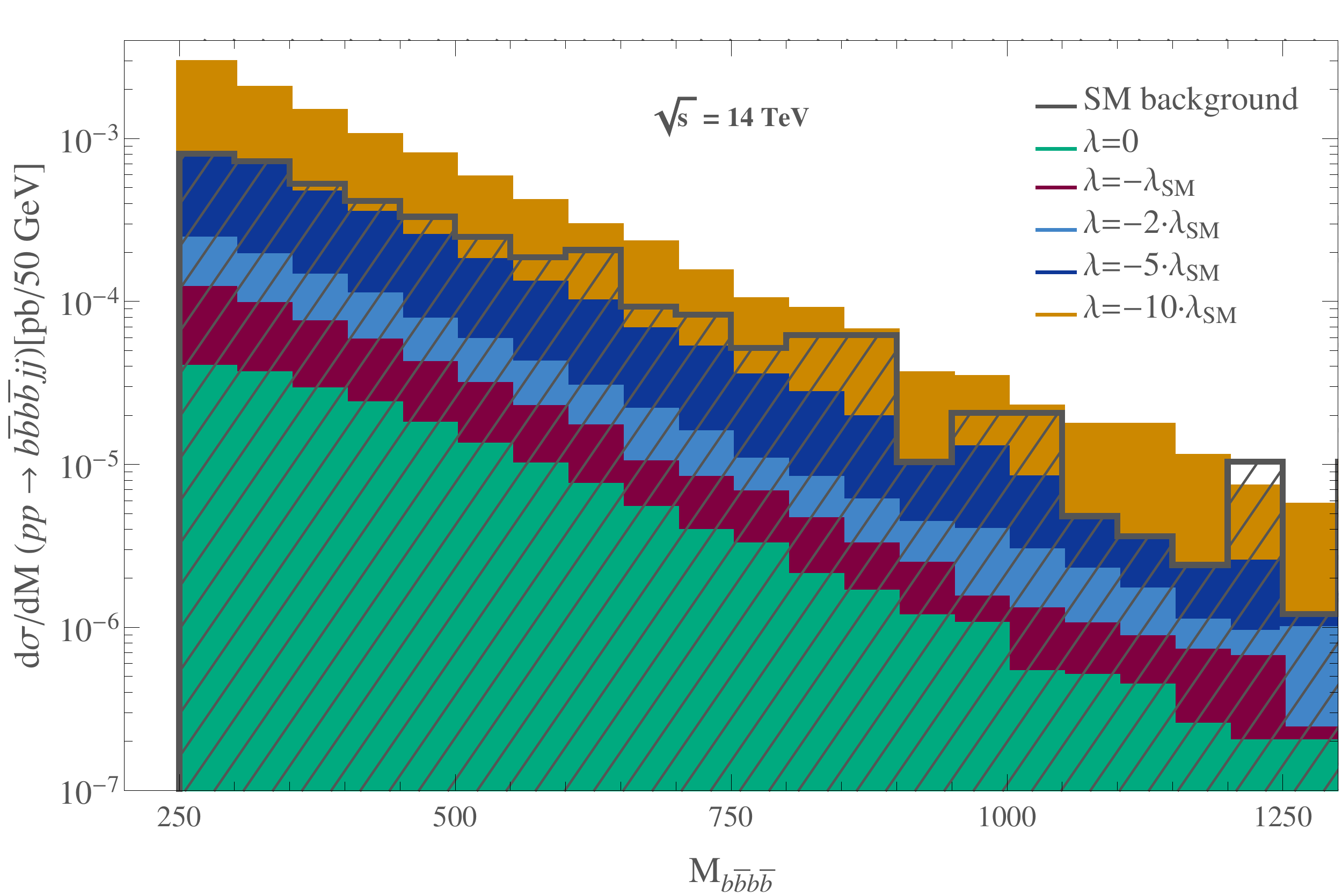}
\caption{Predictions for the total cross section of the process $pp\to b\bar{b}b\bar{b}jj$ as a function of the invariant mass of the four-bottom system $M_{b\bar{b}b\bar{b}}$ for different values of the Higgs self-coupling $\lambda$. We display the predictions for the signal with positive (left panel) and negative (right panel) values of $\lambda$ for comparison, as well as the total SM background given by the sum of $ZHjj$, $ZZjj$ and the multijet QCD background. Cuts in Eq.(\ref{basiccutsHHjj}) and VBS selection cuts presented in Eq.(\ref{VBSselectioncuts}) have been applied. The center of mass energy has been set to $\sqrt{s}=14$ TeV.}
\label{fig:M4b_distributions}
\end{center}
\end{figure}

Given the encouraging previous results, our last step is to give quantitative predictions for the sensitivity to $\lambda$ in $p p\to b\bar{b}b\bar{b}jj$ processes at the LHC. To this end, we compute the statistical significance $\mathcal{S}_{\rm stat}$, as defined in \cite{Cowan:2010js} by: 
\begin{equation}
\mathcal{S}_{\rm stat}=\sqrt{-2\left((N_S+N_B)\log\left(\dfrac{N_B}{N_S+N_B}\right)+N_S\right)}\,,\label{signieq}
\end{equation}
where $N_S$ and $N_B$ are the number of events of signal and background, respectively. Notice that for  $N_S/N_B \ll 1$, this definition of $\mathcal{S}_{\rm stat}$ tends to the usual $N_S/\sqrt{N_B}$ expression. This computation is going to be performed for four different values of the luminosity: $L=50,300,1000,3000$ fb${}^{-1}$, that correspond to a near-future LHC value for the current run (50 fb${}^{-1}$), and to planned luminosities for the third run (300 fb${}^{-1}$) and the High-Luminosity LHC (HL-LHC) (1000 and 3000 fb${}^{-1}$) \cite{Barachetti:2016chu}.

In \figref{fig:significances_4b2j} we present the results of the statistical significance of our signal, $\mathcal{S}_{\rm stat}$, in $pp\to b\bar{b}b\bar{b}jj$ events as a function of the value of $\kappa$, for the four luminosities considered. We  display as well a closer look for the values of $\kappa$ ranging between 0.5 and 2.5, interesting for an elevated number of well motivated BSM models. In the lower part of the left panel we also present the corresponding predictions for the total number of signal events, $N_S$, as a function of $\kappa$. The marked points correspond to our evaluated predictions. We show as well, in the right panel of this figure, our predictions for the value of the total integrated luminosity, $L$, as a function of the value of $\kappa$ as well, that will be required to obtain a sensitivity to a 
given $\kappa$ in $pp\to b\bar{b}b\bar{b} jj$ events at the 3$\sigma$ and  5$\sigma$ level. In this plot,  we have also marked the areas in luminosity where the number of predicted signal events $N_S$ is below 1, 10 and 100, respectively, to get a reference of the statistics obtained. 

 \begin{figure}[t!]
\begin{center}
\includegraphics[width=0.49\textwidth]{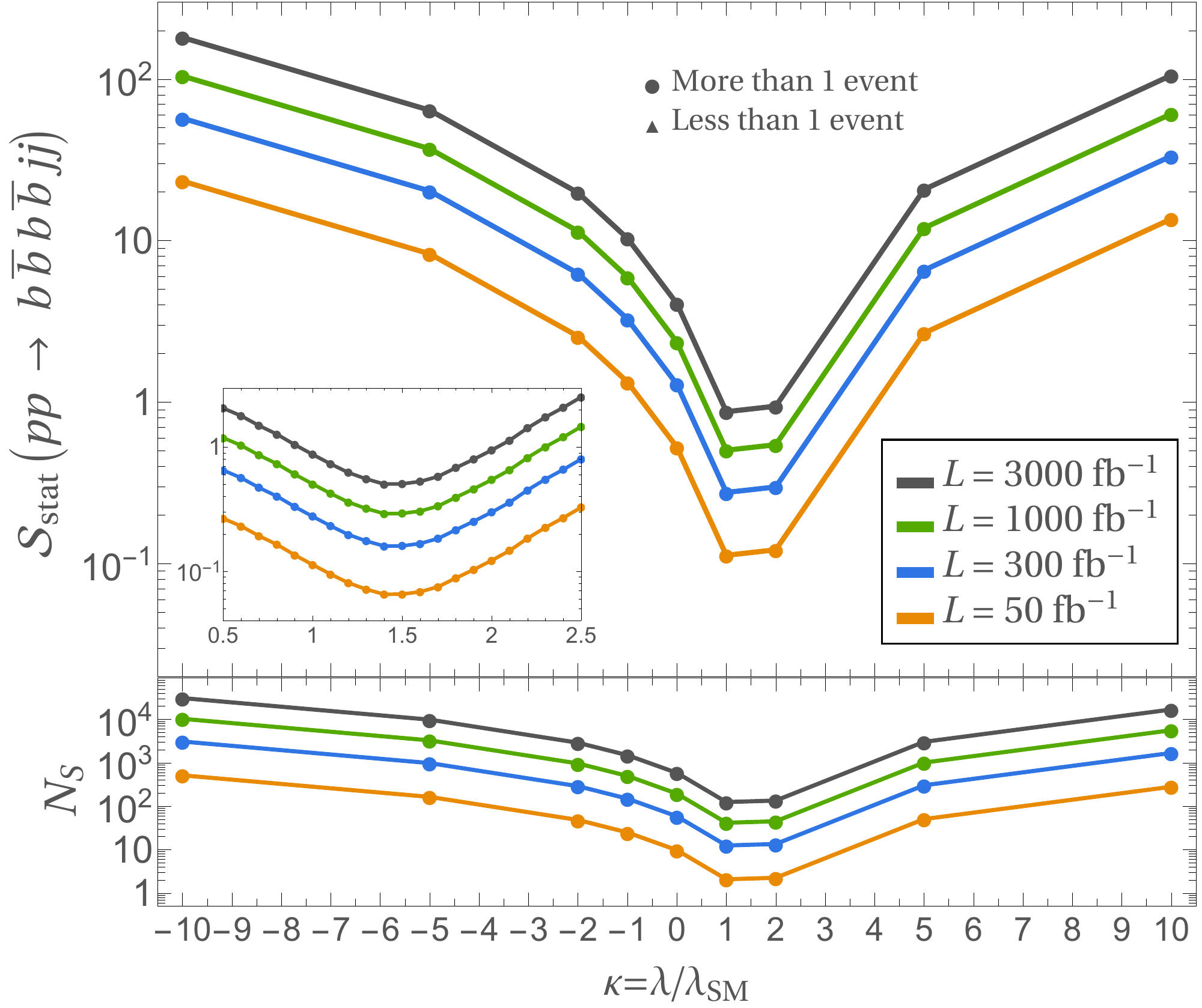}
\includegraphics[width=0.49\textwidth]{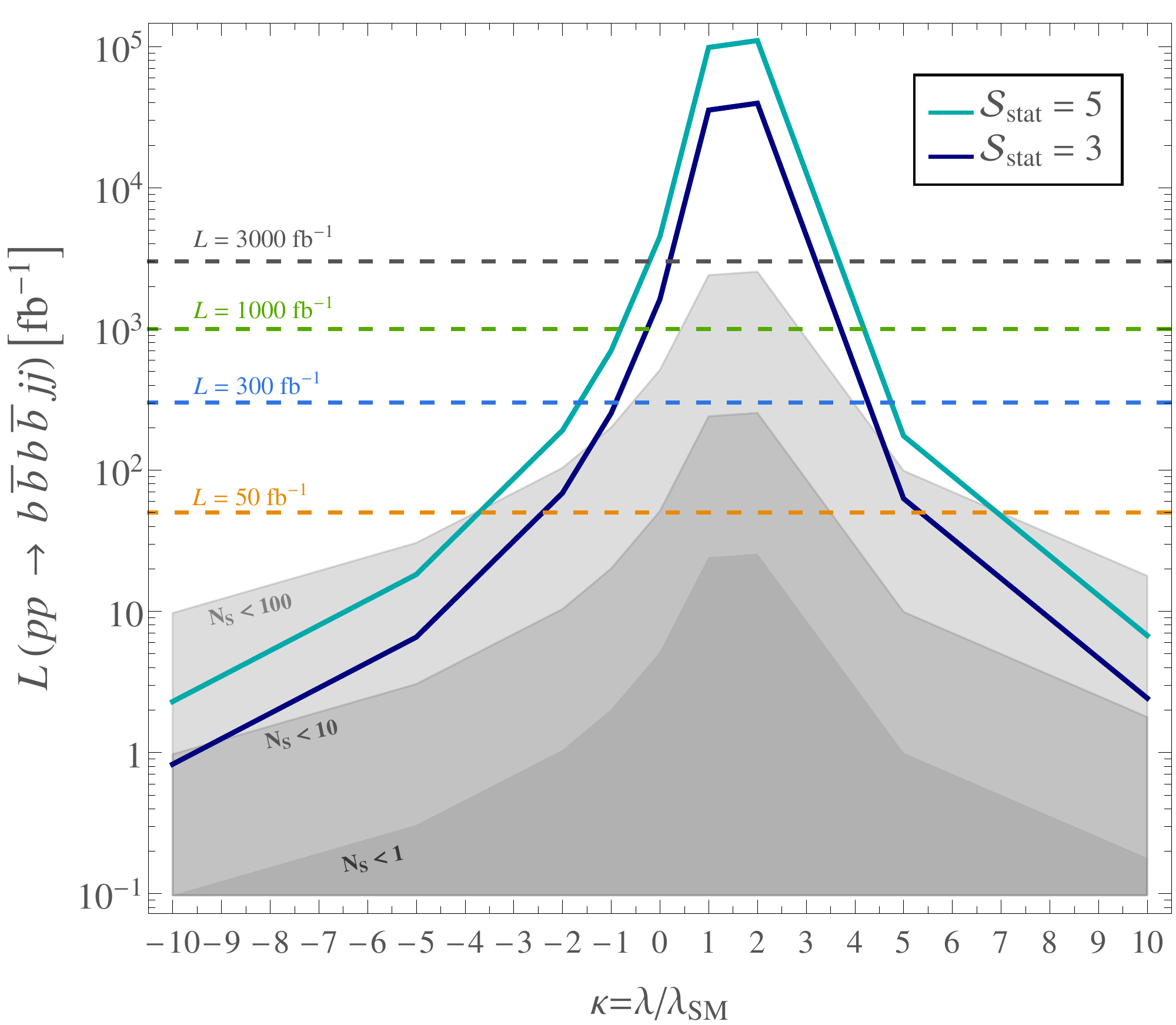}
\caption{Prediction of the statistical significance, $\mathcal{S}_{\rm stat}$, of the process $p p\to b\bar{b}b\bar{b} jj$ for the four luminosities considered $L=50,300,1000,3000$ fb${}^{-1}$ (left panel) and  of the value of the luminosity that will be required to probe a given $\kappa$  at the LHC at  3$\sigma$ and at 5$\sigma$ (right panel),  as a function of the value of $\kappa$. The marked points represent our evaluations.  In the left panel, a zoom is performed on the interesting values of $\kappa$ ranging between 0.5 and 2.5. The shadowed areas in the right panel correspond to the regions where the number of predicted signal events $N_S$ is below 1, 10 and 100. The center of mass energy has been set to $\sqrt{s}=14$ TeV. }
\label{fig:significances_4b2j}
\end{center}
\end{figure}

\begin{table}[t!]
\begin{center}
\begin{tabular}{c| @{\extracolsep{0.8cm}} c @{\extracolsep{0.8cm}} c @{\extracolsep{0.8cm}} c @{\extracolsep{0.8cm}} c}
\toprule
\toprule
$L$ \big[fb${}^{-1}$\big]& 50  & 300  & 1000  & 3000\\
\midrule
$\kappa > 0$ & $\kappa > 5.4 ~(7.0)$ &$\kappa >  4.3  ~(4.8)$ & $\kappa > 3.7 ~ (4.2)$ & $\kappa> 3.2 ~ (3.7)$
\\
$\kappa < 0$ & $\kappa < -2.4 ~ (-3.8)$ &$\kappa <  -1.0  ~(-1.7)$ & $\kappa < -0.3  ~(-0.8)$ & $\kappa < 0 ~ (-0.2)$
\\
\bottomrule
\bottomrule
\end{tabular}
\caption{Predictions for the values of $\kappa\equiv\lambda/\lambda_{SM}$ that the LHC would be able to probe in  $pp\to b\bar{b}b\bar{b}jj$ events, with a sensitivity equal or better than  $3\sigma$ ($5\sigma$) for the four luminosities considered: $L=50,300,1000,3000$ fb${}^{-1}$.}
\label{Tab:range_4b2j}
\end{center}
\end{table}

From these plots, we can extract directly the conclusions on the sensitivity to $\lambda$ in VBS processes at the LHC in $pp\to b\bar{b}b\bar{b}jj$ events. The first thing one might observe is the high statistics and significances of the signal  for most of the studied cases, except for the region close to the SM value, say for $\kappa$ between 1 and 2. Studying carefully this particular region of the parameter space, we conclude that it is the most challenging one to access at the LHC, since all the predicted statistical significances given for $\kappa\in[0.5,2]$ are below 2$\sigma$ even for the highest luminosity considered. The second one is that, for the same absolute value of the coupling, the sensitivities to negative values of $\kappa$ are higher than to positive values of $\kappa$.  The third conclusion is that the LHC should be sensitive to very broad intervals of $\kappa$, even for the lowest luminosity considered, $L=50$ fb${}^{-1}$, with high statistical significance. These means that VBS processes could allow us to probe the value of $\lambda$ with very good accuracy in the near future. More specifically, in \tabref{Tab:range_4b2j} we show the summary of the predictions for the values of $\kappa\equiv\lambda/\lambda_{SM}$ that the LHC would be able to probe in  $pp\to b\bar{b}b\bar{b}jj$ events, with a sensitivity equal or better than  $3\sigma$ ($5\sigma$) for the four luminosities considered: $L=50,300,1000,3000$ fb${}^{-1}$.

These results are indeed very interesting, since the sensitivities to $\lambda$ that one can obtain from studying VBS double Higgs production are very promising even for the lowest luminosity considered $50$ fb${}^{-1}$. The ranges of $\lambda$ that the LHC could be able to probe in this kind of processes indicate that it is worth to study VBS as a viable and useful production mechanism to measure the Higgs trilinear coupling. On the other hand, it can be seen that the HL-LHC should be able to test
very small deviations in the value of the Higgs self-coupling and that it should be sensitive to all the explored negative values for $\kappa$. Although the present work is a naive study, since it is performed at the parton level and does not take into account hadronization and detector response simulation, the results in  \tabref{Tab:range_4b2j} show that the VBS production channel could be very promising to measure the true value of $\lambda$, and, therefore, to understand the nature of the Higgs mechanism.


 \subsection{Analysis after Higgs boson decays: sensitivity to $\lambda$ in $pp\to b\bar{b}\gamma\gamma jj$ events}
 \label{2b2a2j}
 
The $p p\to b\bar{b}b\bar{b} jj$ process is, as we have seen, a very promising channel to study the Higgs self-coupling at the LHC due to its large event rates. However, it is clear that it suffers from quite severe backgrounds, coming specially from multijet QCD events, so one could think of studying complementary channels with smaller rates but with a cleaner experimental signature. This is the reason why we would like to explore the case in which one of the Higgs bosons decays to a $b$-quark pair, as before, while the other one decays to two photons through gauge bosons and fermion loops. This implies a large reduction factor in statistics due to the comparative low branching ratio BR$(H \to \gamma \gamma)\sim 0.2\%$, a factor 0.003 smaller than that of $H \to b\bar{b}$.  

The analysis of the process $pp\to b\bar{b}\gamma\gamma jj$ implies to go through its main backgrounds as well. We will consider in this section the same background $ZH$ of the previous case, since the $ZH$ final state can also lead to processes with two photons and two bottoms, $pp\to HZjj \to b\bar{b}\gamma\gamma jj$, coming from the decays of the $H$ and the $Z$. In addition, we also consider the mixed QCD-EW $pp\to b\bar{b}\gamma\gamma jj$ background, of $\mO(\alpha\cdot\alpha_S^2)$ at the amplitude level, that should be the most severe one.

As we did before, to study signal and background, we first need to establish a set of cuts that ensure particle detection, so we apply the following basic cuts:
\begin{align}
 p_{T_{j,b}}>20~ {\rm GeV}\,;~p_{T_{\gamma}}>18 ~{\rm GeV}\,;~|\eta_j|<5\,;~|\eta_{b,\gamma}|<2.5\,;~\Delta R_{jj,jb,\gamma\gamma,\gamma b,\gamma j}>0.4\,;~\Delta R_{bb}>0.2\,,\label{basiccuts2b2a2j}
 \end{align}
and afterwards, to reject the QCD-EW and the $ZHjj$ backgrounds we will apply first the VBS cuts given in \eqref{VBSselectioncuts} and subsequently the following kinematical cuts given by CMS in \cite{Sirunyan:2018iwt}:
\begin{equation}
p_{T_{\gamma^l}}/M_{\gamma\gamma}>1/3\,;~~~p_{T_{\gamma^s}}/M_{\gamma\gamma}>1/4\,,\label{bkgcuts2b2a2j}
 \end{equation}
where $l$ and $s$ stand for leading (highest $p_T$ value) and subleading photons, and where $M_{\gamma\gamma}$ is the invariant mass of the photon pair. The final ingredient is to apply the $\chi_{HH}$ cut, taking now into account that we have a $b$-quark pair and a photon pair in the final state:
\begin{equation}
\chi_{HH}=\sqrt{\left(\dfrac{M_{bb}-m_H}{0.05\,m_H}\right)^2+\left(\dfrac{M_{\gamma\gamma}-m_H}{0.05\,m_H}\right)^2}<1\label{chiHHphotons}\,.
\end{equation}
 This ensures that the two $b$-quarks and the two photons come from the decay of a Higgs particle.

  Once again, there might be important background contributions from multijet QCD processes leading to different final states than that of $b\bar{b}\gamma\gamma jj$, such as $6j$ and $b\bar{b}jjjj$, in which some of the final state light jets are again misidentified as $b$ jets and some are misidentified as photons. Taking again as the presumably leading QCD background processes the production of six light jets and of two $b$ jets and four light jets, applying a misidentification rate of 0.1\% per each jet misidentified as a photon, and considering a similar reduction factor after our selection cuts as before, since the selection cuts are very similar, we obtain:   $1.3\cdot10^5 \cdot (0.01)^2\cdot (0.001)^2\cdot 1.1 \cdot 10^{-5}= 1.4 \cdot 10^{-10}$ pb for the six light jets case and $7.5\cdot 10^3 \cdot (0.001)^2\cdot 1.1 \cdot 10^{-5}= 8.2 \cdot 10^{-8}$ pb for the $2b4j$ case. Again in both cases the final cross sections are more than one order of magnitude smaller than the main background we have considered, being of order $10^{-6}$ pb, concluding again that they can be neglected as well.
\begin{table}[t!]
\begin{center}
\begin{tabular}{c| @{\extracolsep{0.8cm}} c @{\extracolsep{0.8cm}} c @{\extracolsep{0.8cm}} c @{\extracolsep{0.8cm}} c @{\extracolsep{0.8cm}} c @{\extracolsep{0.8cm}} c @{\extracolsep{0.8cm}} c @{\extracolsep{0.8cm}} c @{\extracolsep{0.8cm}} c}
\toprule
\toprule
$\kappa $&  0& 1 & -1 & 2& -2& 5& -5& 10& -10\\
\midrule
 $\sigma_{\rm Signal}\cdot 10^6$ [pb] & 2.0 & 0.7 & 4.5 & 0.5 & 8.0 & 6.4 & 25.2 & 38.4 & 76.0  \\
\bottomrule
\bottomrule
\end{tabular}
\caption{Predictions for the total cross section of the signal $pp\to b\bar{b}\gamma\gamma j j$ after imposing all the selection criteria, VBS cuts given in \eqref{VBSselectioncuts} and cuts given in Eqs. (\ref{bkgcuts2b2a2j}) and (\ref{chiHHphotons}) for all the values of $\kappa$ considered in this work: $\kappa=0,\pm 1, \pm 2, \pm 5, \pm 10$. The cross section of the SM background for this same cuts amounts to $\sigma_{\rm Background}=1.4\cdot 10^{-6}$ pb. Basic cuts in \eqref{basiccuts2b2a2j} are also applied.}
\label{Tab:allSig_2b2a2j}
\end{center}
\end{table}
  \begin{figure}[b!]
\begin{center}
\includegraphics[width=0.49\textwidth]{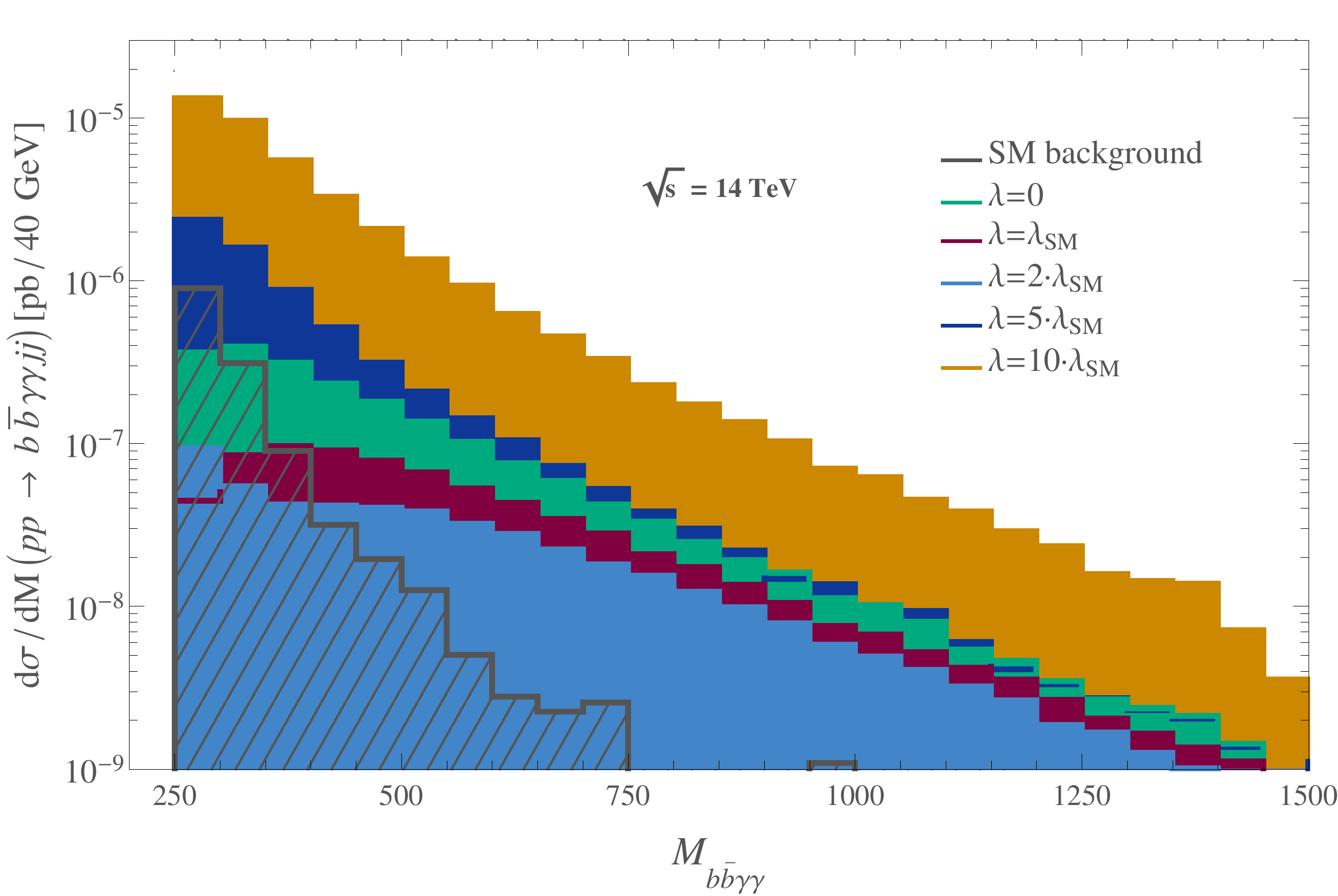}
\includegraphics[width=0.49\textwidth]{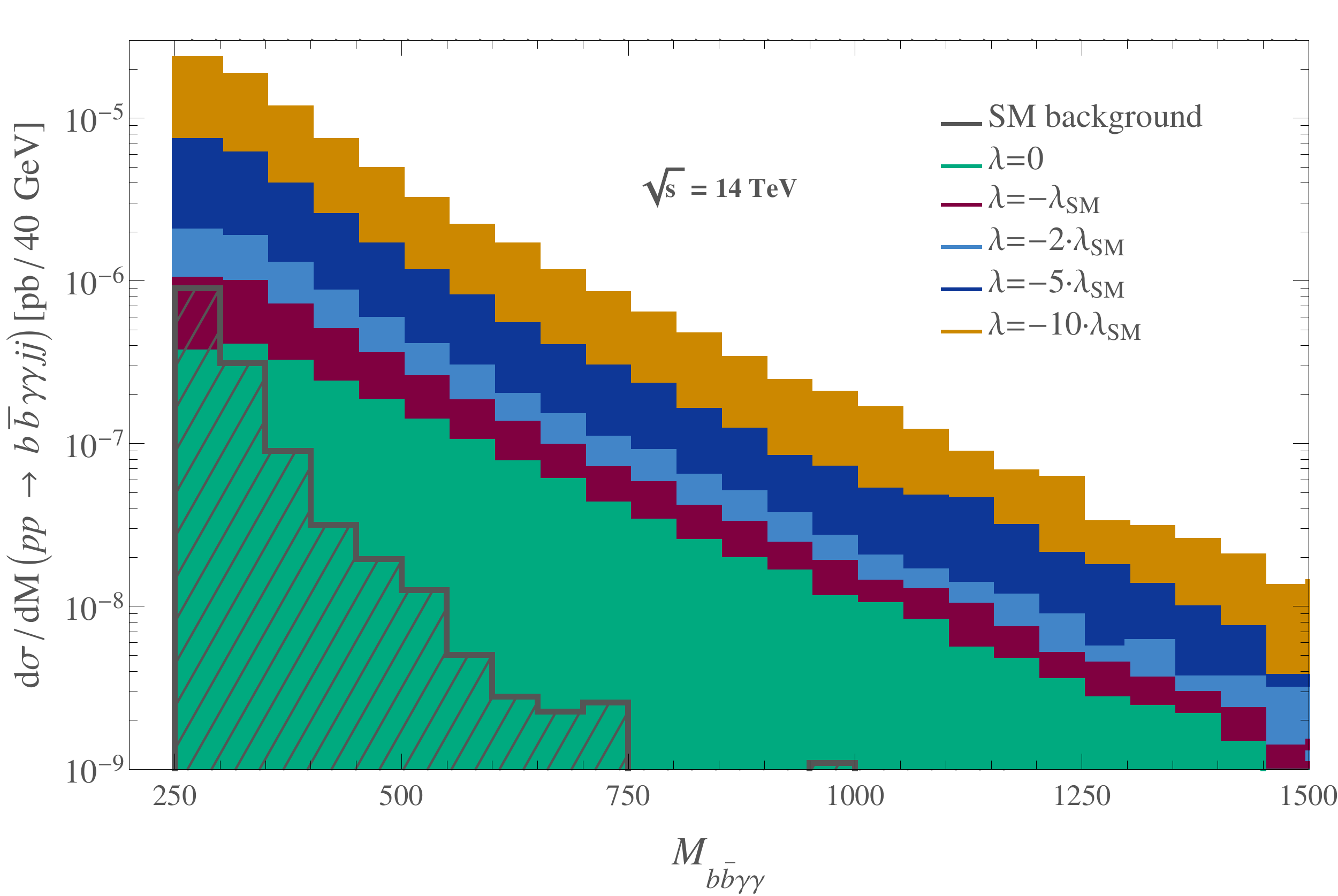}
\caption{Predictions for the total cross section of the process $pp\to b\bar{b}\gamma\gamma jj$ as a function of the invariant mass of the $b\bar{b}\gamma\gamma$ system $M_{b\bar{b}\gamma\gamma}$ for different values of the Higgs self-coupling $\lambda$. We display the predictions for the signal with positive (left panel) and negative (right panel) values of $\lambda$ for comparison, as well as the total SM background. Cuts in Eqs.(\ref{basiccuts2b2a2j})-(\ref{chiHHphotons}) and VBS selection cuts presented in Eq.(\ref{VBSselectioncuts}) have been applied. The center of mass energy is set to $\sqrt{s}=14$ TeV.}
\label{fig:M2b2a_distributions}
\end{center}
\end{figure}

Having all this in mind, we present in \figref{fig:M2b2a_distributions} the predictions for the total cross section of the process $pp\to b\bar{b}\gamma\gamma jj$ as a function of the invariant mass of the $b\bar{b}\gamma\gamma$ system $M_{b\bar{b}\gamma\gamma}$, for different values of the Higgs self-coupling $\lambda$. We also display the prediction for the total SM background (sum of the QCD-EW and the $ZHjj$ background) for comparison. Once again, one can see that the signal distributions for different values of $\kappa$ are very similar to those shown in  \figref{fig:MHH_distributions}, and that the main difference is due to the reduction factor of the branching ratios into photons and into $b$-quarks. They are very similar, too, to the results of the  $b\bar{b}b\bar{b}jj$ final state, in \figref{fig:M4b_distributions}, although two-three orders of magnitude smaller.  The background is, however, very different with respect to the one for  $b\bar{b}b\bar{b}jj$ events. It is smaller in comparison with the signal, specially at high  $M_{b\bar{b}\gamma\gamma}$, since it decreases much more steeply. Therefore, we would expect to have good sensitivities to the Higgs self-coupling despite the lower rates of the process involving photons. For completeness, we display in \tabref{Tab:allSig_2b2a2j} the predictions for the total cross section of the signal, for the set of $\kappa$ values considered, and  after applying all cuts given in Eq. (\ref{VBSselectioncuts}) and in Eqs. (\ref{basiccuts2b2a2j})-(\ref{chiHHphotons}). The prediction for the cross section of the total SM background for this same cuts amounts to $\sigma_{\rm Background}=1.4\cdot 10^{-6}$ pb.

\begin{table}[t!]
\begin{center}
\begin{tabular}{c|@{\extracolsep{0.8cm}}c@{\extracolsep{0.8cm}}c@{\extracolsep{0.8cm}}c@{\extracolsep{0.8cm}}c}
\toprule
\toprule
$L$ \big[fb${}^{-1}$\big]&  50 & 300 & 1000 & 3000\\
\midrule
$\kappa> 0$ & $\kappa >  9.9~(14.2)$ &$\kappa >  6.4~(8.4)$ & $\kappa > 4.6 ~(6.0)$ & $\kappa >  3.8~(4.7)$
\\
$\kappa <0$ & $\kappa <  -6.7~(-10.0)$ & $\kappa <  -2.7~(-4.6)$ & $\kappa <  -1.1~(-2.3)$ & $\kappa <  -0.2~(-1.0)$ 
\\
\bottomrule
\bottomrule
\end{tabular}
\caption{Predictions for the values of $\kappa\equiv\lambda/\lambda_{SM}$ that the LHC would be able to probe in  $pp\to b\bar{b}\gamma\gamma jj$ events, with a sensitivity equal or better than $3\sigma$ ($5\sigma$) for the four luminosities considered: $L=50,300,1000,3000$ fb${}^{-1}$. }
\label{Tab:range_2b2a2j}
\end{center}
\end{table}
  \begin{figure}[t!]
\begin{center}
\includegraphics[width=0.49\textwidth]{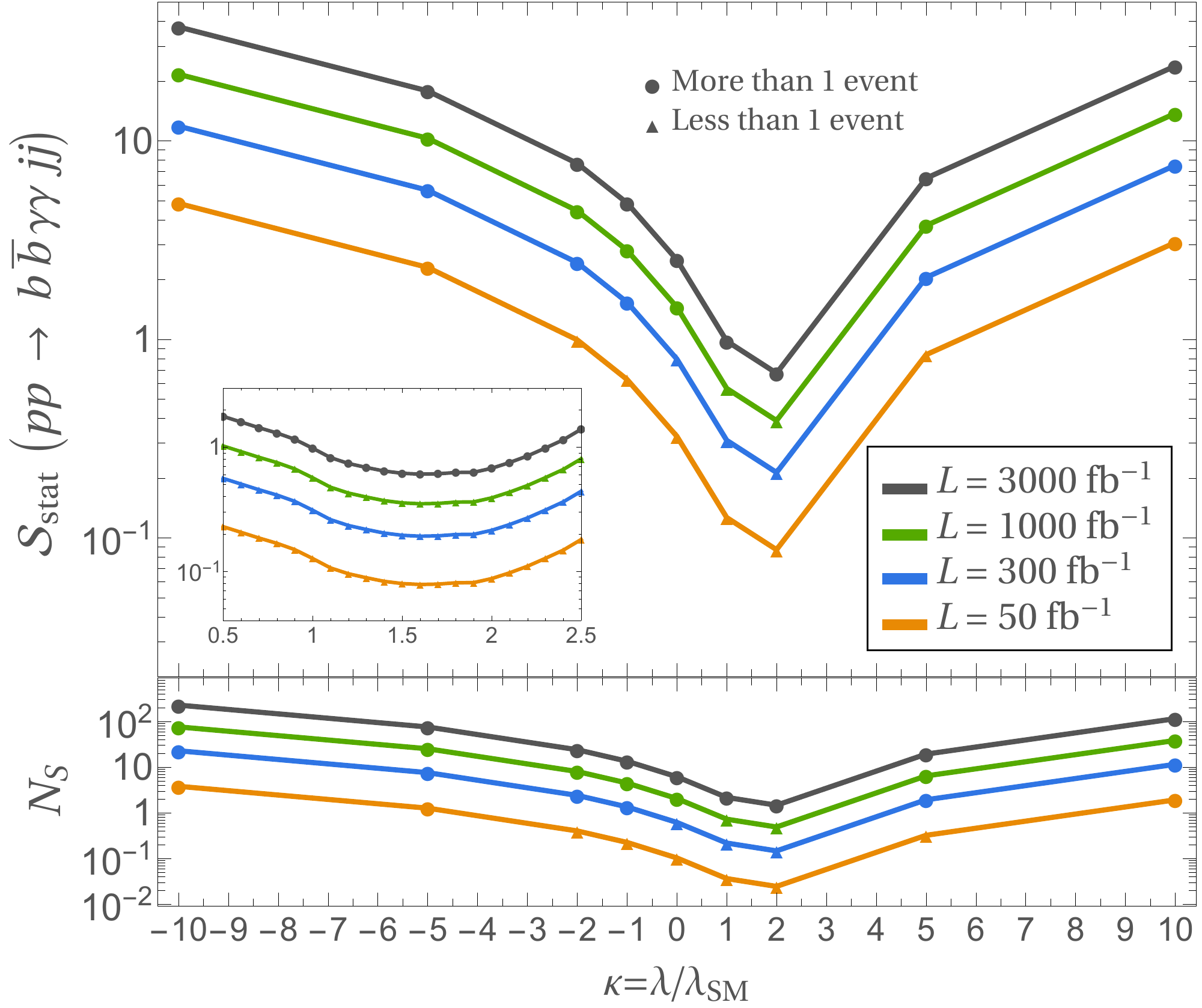}
\includegraphics[width=0.49\textwidth]{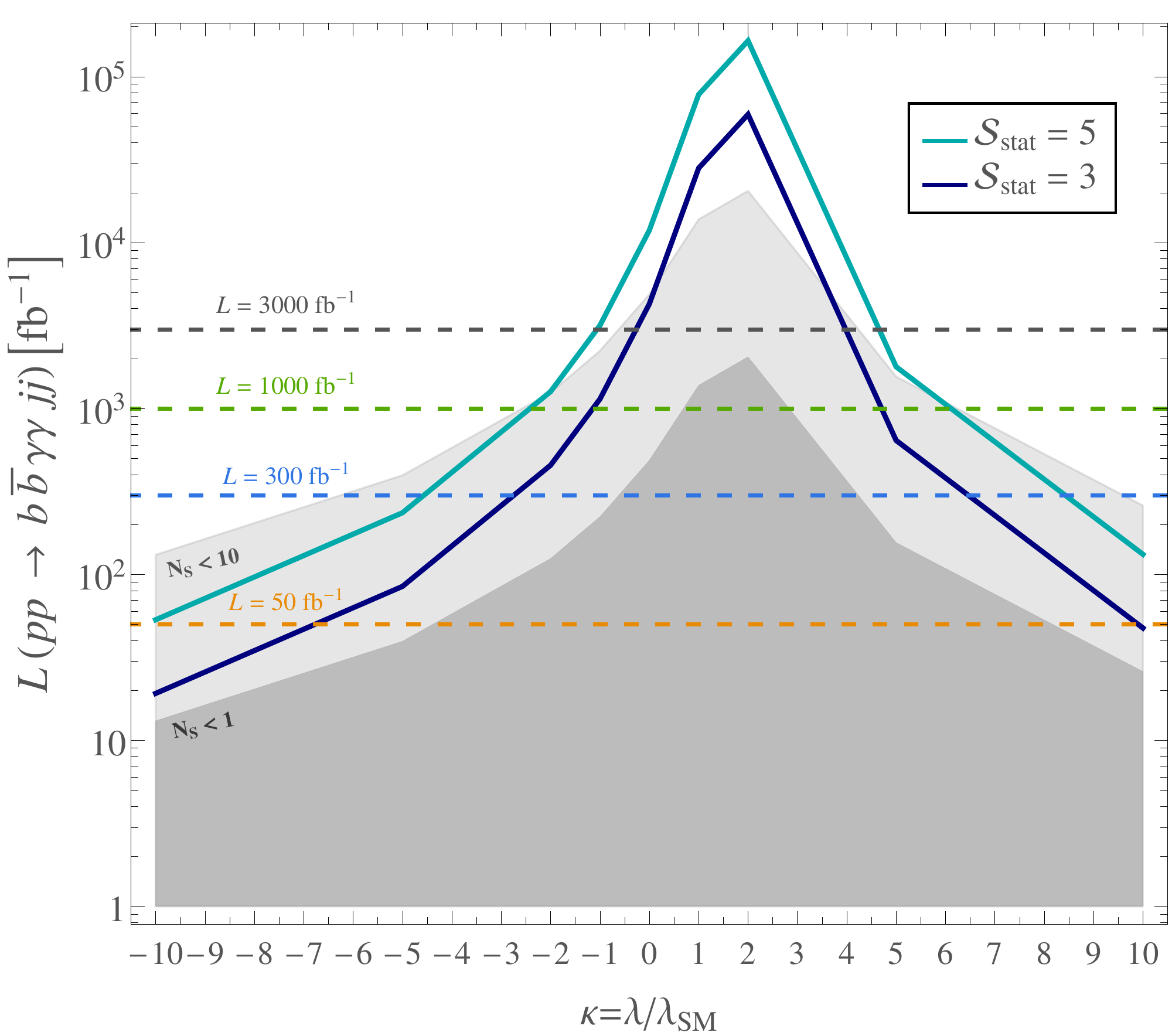}
\caption{Prediction of the statistical significance, $\mathcal{S}_{\rm stat}$, of the process $p p\to b\bar{b}\gamma\gamma jj$ for the four luminosities considered $L=50,300,1000,3000$ fb${}^{-1}$ (left panel) and  of the value of the luminosity that will be required to probe a given $\kappa$  at the LHC at  3$\sigma$ and at 5$\sigma$ (right panel),  as a function of the value of $\kappa$. The marked points represent our evaluations.  In the left panel, a zoom is performed on the interesting values of $\kappa$ ranging between 0.5 and 2.5. The shadowed areas in the right panel correspond to the regions where the number of predicted signal events $N_S$ is below 1, and 10.
The center of mass energy has been set to $\sqrt{s}=14$ TeV.}
\label{fig:significances_2b2a2j}
\end{center}
\end{figure}

In \figref{fig:significances_2b2a2j} we show the predictions for the statistical significance $\mathcal{S}_{\rm stat}$, computed in the same way as in the previous section, making use of \eqref{signieq}, for the four luminosities considered previously, $L=50,300,1000,3000$ fb${}^{-1}$  and taking again a closer look for the values of $\kappa$ ranging between 0.5 and 2.5. We also show the predictions of the final number of signal events, $N_S$ as a function of $\kappa$, for these same luminosities.  On the right panel of this figure we present the prediction for  the value of the luminosity that will be required to probe a given $\kappa$ value with sensitivities at 3$\sigma$ and 5$\sigma$, as a function of the value of $\kappa$. In these plots, due to the lower statistics of this process, some of the computed significances correspond to scenarios in which there is not even one signal event. The concrete predictions for these signal event rates can be read from the lower plot of the left panel.

Taking a look at these figures, we can again extract the conclusions on the sensitivity to the Higgs self-coupling at the LHC in VBS processes, this time in $pp\to b\bar{b}\gamma\gamma j j$ events. One might notice that, although the results are less encouraging than those of  $pp\to b\bar{b}b\bar{b} j j$ events, this channel could also be very useful to measure the value of $\lambda$. Analogously to the previous section, in \tabref{Tab:range_2b2a2j} we present the values of $\kappa\equiv\lambda/\lambda_{SM}$ that would  be accessible at the LHC in these type of events, $pp\to b\bar{b}\gamma\gamma jj$, with a statistical significance equal or better than $3\sigma$($5\sigma$), for the four luminosities considered.

These results show again that the values of $\kappa$ that can be probed in the future at LHC  through the study of VBS processes leading to the final state $ b\bar{b}\gamma\gamma jj$ could be very competitive as well.  Except for the lowest luminosity considered, $L=50$ fb${}^{-1}$, where the signal rates found at the parton level are too low as to survive the extra factors suppression due to the missing detector efficiencies, hadronization effects etc, the sensitivities found point towards the potential of VBS processes in order to obtain a precise measurement of $\lambda$. The values close to the SM value, are, again, very challenging to reach at the LHC, since the statistical significances of $\kappa\in[0.5,2]$ are always below 2$\sigma$ for this case as well. However, the HL-LHC should be able to probe deviations in $\lambda$ very efficiently in this channel.

\subsection{Discussion}
\label{discussion}
Finally, to close this section of results, we find pertinent to discuss on how the precision of our predictions could be improved by including additional considerations. We comment here just on those that we consider are the most relevant ones. 
\begin{itemize}
\item[1.-] Our computation of the $HHjj$ signal rates incorporates just those coming from the subprocess $qq \to HH jj$, which includes VBS, but this is not the only contributing channel. It is well known that also the subprocess $gg \to HH jj$, initiated by gluons, does contribute to these signal rates, and it is also sensitive to large BSM $\lambda$ values~\cite{Dolan:2013rja}. Although it is a one-loop subprocess, mediated mainly by top quark loops, it provides  a sizable contribution to the total $HHjj$ signal cross section. For instance, for the case of $\lambda=\lambda_{SM}$, the total cross section at the LHC with $\sqrt{s}$ = 14 TeV is,  according to \cite{Dolan:2015zja},  $5.5$ fb from $gg \to HH jj$ to be compared with $2$ fb from VBS. Therefore, when  considering both contributions to the signal, the sensitivity to $\lambda$ presumably increases. However, we have explicitly checked that once we apply our optimized VBS selection cuts summarized in Eq.~(\ref{VBSselectioncuts}) and in \tabref{table:VBS}, we get a notably reduced cross section for this $gg$ subprocess. In particular, our estimate of the signal rates at the LHC with $\sqrt{s}$ = 14 TeV from $gg \to HH jj$, after applying the stringent $M_{jj}>500$ GeV cut and using the results in \cite{Dolan:2015zja} for the $M_{jj}$ distribution, gives a strong reduction in the corresponding cross section, and leads to smaller rates for $gg$ than those from VBS  by about a factor of 20. Therefore its contribution to the signal rates studied here can be safely neglected, and no much better precision will be obtained by including this new contribution in the signal rates. We have also checked that this finding is true for other BSM values of $\lambda$.   
\item[2.-] When considering next to leading order (NLO) QCD corrections in our estimates of both the signal and background rates, we expect some modifications in our results. These can be very easily estimated, as usual, by using the corresponding $K$-factors. Thus, for instance, for the leading $b \bar b b \bar b j j$ final state,  we can include these NLO corrections by taking into account the $K$-factors for the VBS signal and for the main background from multijet QCD. For the signal we take the $K$-factor  from \cite{Frederix:2014hta}, given  by $K_{\rm VBS}=1.09$. For the QCD-multijet background the corresponding $K$-factor is, to our knowledge, not available in the literature, and different choices are usually assumed. We consider here two choices: $K_{\rm QCD}=1.5$, and another more conservative one $K_{\rm QCD}=3$. This implies that our predictions for the signal rates are practically unchanged, but those for the background rates are increased by a factor of 1.5 and 3 respectively. This modifies our predictions for the statistical significance  of the $b \bar b b \bar b j j$ signal, from the $\mathcal{S}_{\rm stat}$ results given \figref{fig:significances_4b2j} to $\mathcal{S}_{\rm stat}^{\rm NLO}\sim K_{\rm VBS}/\sqrt{K_{\rm QCD}}\,\,\mathcal{S}_{\rm stat} \sim 0.9\,\, \mathcal{S}_{\rm stat}$  and $0.6\,\, \mathcal{S}_{\rm stat}$ for $K_{\rm QCD}=1.5$ and $K_{\rm QCD}=3$,  respectively.  For instance, for the high luminosity considered of 
$1000 \,{\rm fb}^{-1}$ we get sensitivities of $\kappa >3.8 (4.3)$ for $K_{\rm QCD}=1.5$, and of $\kappa >4.5 (4.8)$ for $K_{\rm QCD}=3$, both at the $3\sigma$ ($5\sigma$) level, to be compared with our benchmark result in  \tabref{Tab:range_4b2j} of $\kappa >3.7 (4.2)$. Therefore ignoring these NLO corrections does not provide large uncertainties either.      
\item[3.-] When including $b$-tagging efficiencies in our estimates of the $b \bar b b \bar b j j$ signal and background rates, our predictions of the statistical significance do also change. However, an estimate of this change can be easily done by adding the corresponding modifying factors. For instance, by assuming well known $b$-tagging efficiencies of $70\%$, that apply to both the signal and background, the two rates are reduced by a factor of ${0.7}^ 4 \sim 0.24$ . Therefore we get a reduced  statistical significance of $\mathcal{S}_{\rm stat}^{b-{\rm tag}}\sim 0.24/\sqrt{0.24}\,\, \mathcal{S}_{\rm stat} \sim 0.5 \,\,\mathcal{S}_{\rm stat}$ with respect to the ones that we have reported previously. This factor of $0.5$ will change our predicted sensitivities to BSM $\lambda$ values. Again, as an example, for the considered luminosity of
$1000 \,{\rm fb}^{-1}$ we get sensitivities of $\kappa >4.3 (4.9)$ at the $3\sigma$ ($5\sigma$) level to be compared with our benchmark result in \tabref{Tab:range_4b2j} of $\kappa >3.7 (4.2)$. 

Similarly, considering also photon-identification efficiencies (also called in this work $\gamma$- tagging) of $95\%$, as presented in the literature, we get reduced signal and background rates for the $b \bar b \gamma \gamma j j$ final state by a factor of ${0.7}^ 2 \times {0.95}^ 2 \sim 0.44$. Accordingly, we obtain a reduction in the statistical significance of the $b \bar b \gamma \gamma j j$ events, given by  $\mathcal{S}_{\rm stat}^{b, \gamma-{\rm tag}}\sim 0.44/\sqrt{0.44} \,\,\mathcal{S}_{\rm stat} \sim 0.7 \,\,\mathcal{S}_{\rm stat}$ with respect to our results reported in the pages above. The changes in the sensitivities to $\kappa$ can be easily derived. using the same illustrative example, for $1000 \,{\rm fb}^{-1}$ of luminosity, we get sensitivities of $\kappa > 6.0 (8.0)$ at the $3\sigma$ ($5\sigma$) level to be compared with our benchmark result in \tabref{Tab:range_2b2a2j} of $\kappa >4.6 (6.0)$. 
\item[4.-] One of the largest uncertainties comes from the choice of the energy resolution needed for the reconstruction of the $HHjj$ signal events from the corresponding final state. This basically can be translated into the choice for the particular definition of the $\chi_{HH}$ variable which is very relevant for the selection of the $HH$ candidates. Thus, for the  
$b \bar b b \bar b j j$ final state, in our benchmark scenario we have taken $0.05 \times m_H$ around $m_H$ in the definition of $X_{HH}$ in Eq.~(\ref{cutsHH}), i.e.\footnote{Equivalently in the case of $pp\to b\bar{b}\gamma\gamma jj$ substituting $M_{bb^s}$ by $M_{\gamma\gamma}$.},
\begin{align}
\chi_{HH}\equiv\sqrt{\left(\dfrac{M_{bb^l}-m_H}{ \Delta_{E}\,m_H}\right)^2+\left(\dfrac{M_{bb^s}-m_H}{  \Delta_{E}\,m_H}\right)^2}<1\,,
\label{cutXHH}
\end{align}
with $ \Delta_{E}$ being the energy resolution, which in this case was set to 0.05, leading to a mass resolution of 5\% of the Higgs mass, since this value optimizes the selection efficiency and could be useful for future experiments with better energy resolution. However a more realistic choice, given the current energy resolution at the LHC experiments, could rather be $\Delta_E\cdot m_H = 0.1 \times m_H$ GeV $\sim 12.5$ GeV. We have redone the analysis with this alternative and more conservative choice and we have obtained, as expected,  a reduced statistical significance. The signal rates do not  change (we still get $4.1 \times 10^{-5}$ pb) , but the main QCD-background does (we get 
$1.8 \times 10^{-2}$ pb  instead of our benchmark value of $6.8 \times 10^{-3}$ pb). This translates in a reduction of the significance given by a factor $\mathcal{S}_{\rm stat}^{\chi_{HH}}\sim 1/\sqrt{18/6.8}\,\, \mathcal{S}_{\rm stat} \sim 0.7 \,\,\mathcal{S}_{\rm stat}$.  The implication of this reduction can directly be seen as a modification of the sensitivity to $\kappa$. Once again, for our benchmark case of $1000 \,{\rm fb}^{-1}$, we obtain sensitivities of $\kappa > 4.0 (4.5)$ at the $3\sigma$ ($5\sigma$) level to be compared with our benchmark result in \tabref{Tab:range_4b2j} of $\kappa >3.7 (4.2)$.

Apart from redoing the analysis for this 10\% resolution\footnote{When mentioning a percentage for the energy resolution we refer to that percentage of the Higgs mass.}, we have also studied other possible and realistic values such as $\Delta_E=20\%$ and $\Delta_E=30\%$, to have a better idea of the implications of the value of the mass determination uncertainty in our predictions. The results for both of our signals are shown in \figref{fig:XHHcut} by the green lines and green shaded areas, where we present the values for the statistical significance at $1000 \,{\rm fb}^{-1}$ as a function of the value of $\kappa$ for different energy resolutions of $\Delta_E=$ 5\% (original scenario throughout the work), 10\%, 20\% and 30\% (the purple lines and purple areas of this figure will be discussed in the next point of this discussion section).  One can see that, as expected, the statistical significance decreases as the energy resolution worsens, but in any case, from the most optimistic case ($\Delta_E=5\%$) to the less optimistic one ($\Delta_E=30\%$), we only obtain a reduction factor of at most  0.4 in the statistical significance.

 \begin{figure}[t!]
\begin{center}
\includegraphics[width=0.49\textwidth]{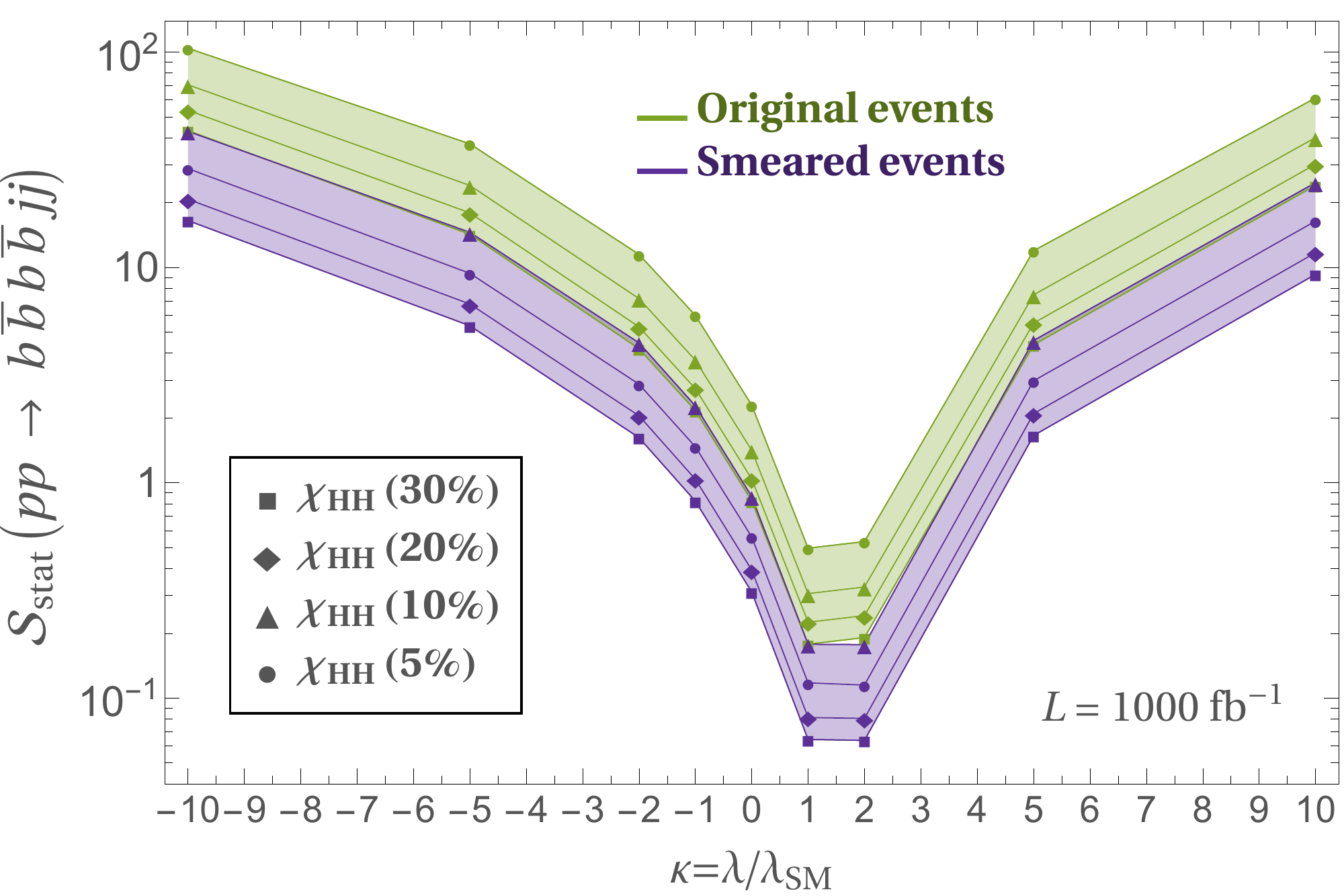}
\includegraphics[width=0.49\textwidth]{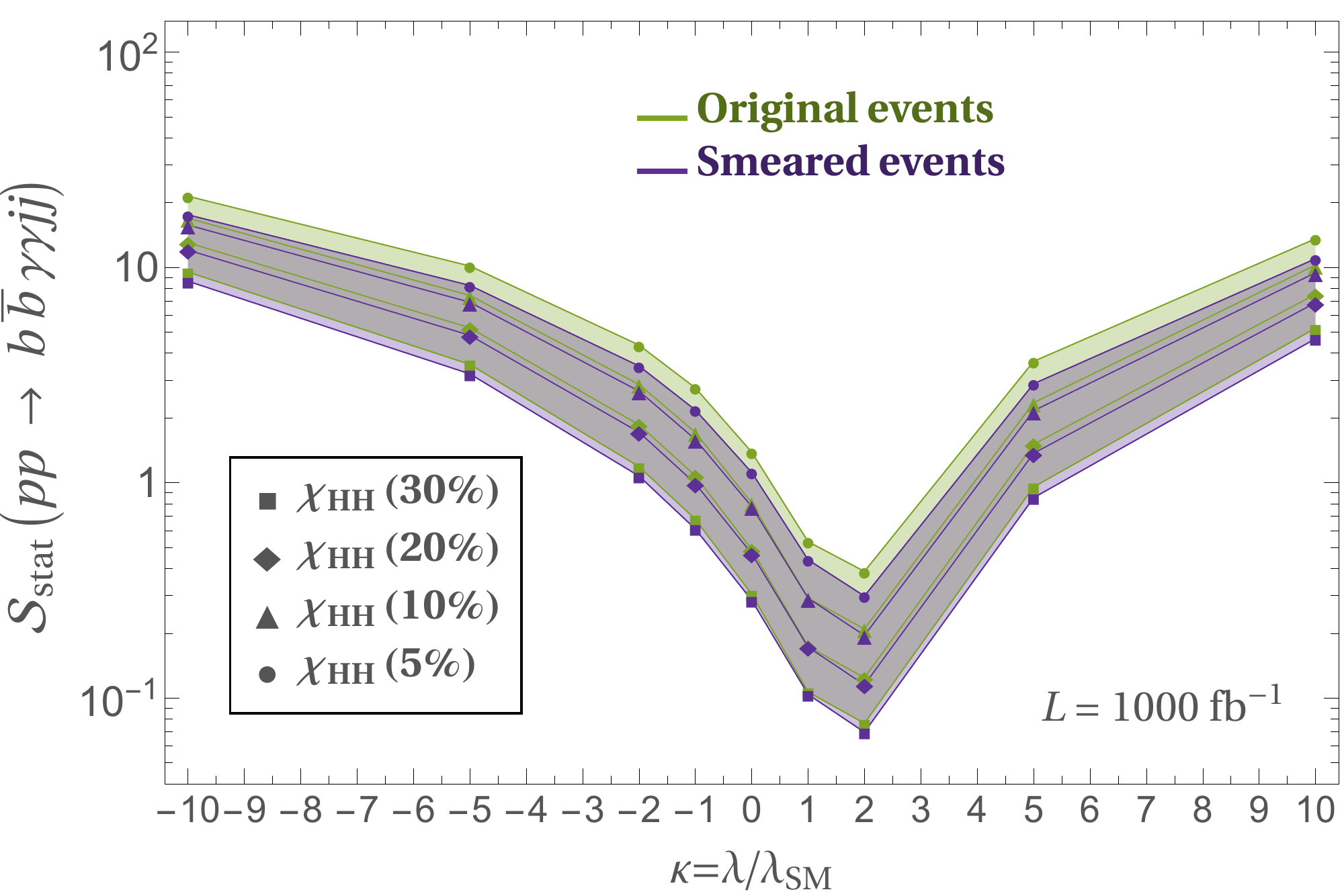}
\caption{Prediction of the statistical significance, $\mathcal{S}_{\rm stat}$, of the process $p p\to b\bar{b} b\bar{b} jj$ (left panel) and of the process $p p\to b\bar{b}\gamma\gamma jj$ (right panel) for $L=1000$ fb${}^{-1}$  as a function of the value of $\kappa$ for different values of the energy resolution, ($\Delta_E$\%), applied through the variable $\chi_{HH}$ defined in \eqref{cutXHH}. These different values are marked with different symbols. We show the predictions for the original events (green lines and green shaded areas; notice that the upper green line corresponds to the green line presented in \figref{fig:significances_4b2j} (left panel) and \figref{fig:significances_2b2a2j} (right panel)), and for the events with a Gaussian smearing applied in order to account for detector effects (purple lines and purple shaded area). The marked points represent our evaluations. The center of mass energy has been set to $\sqrt{s}=14$ TeV.}
\label{fig:XHHcut}
\end{center}
\end{figure}

\begin{figure}[t!]
\begin{center}
\includegraphics[width=0.49\textwidth]{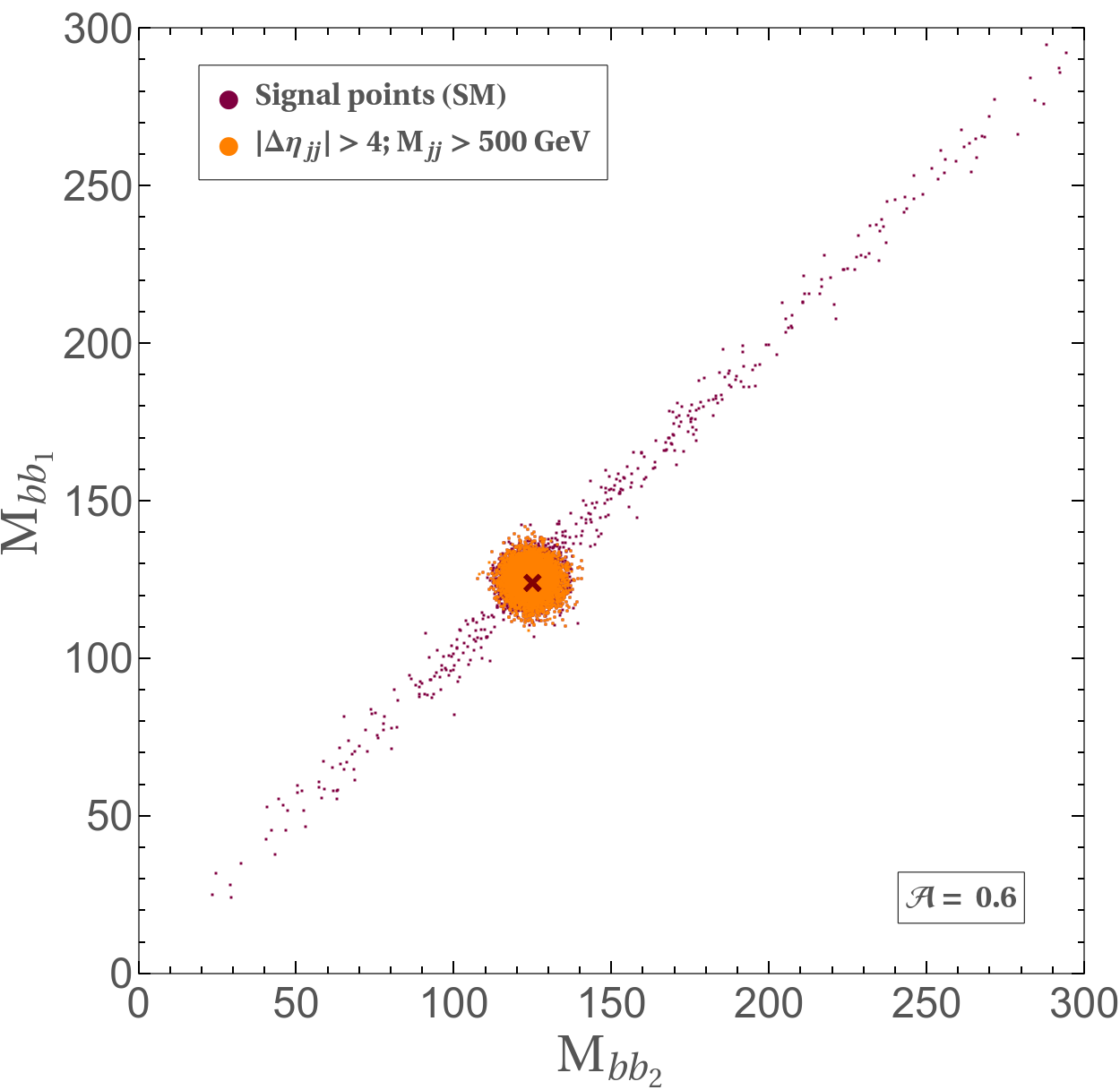}
\includegraphics[width=0.49\textwidth]{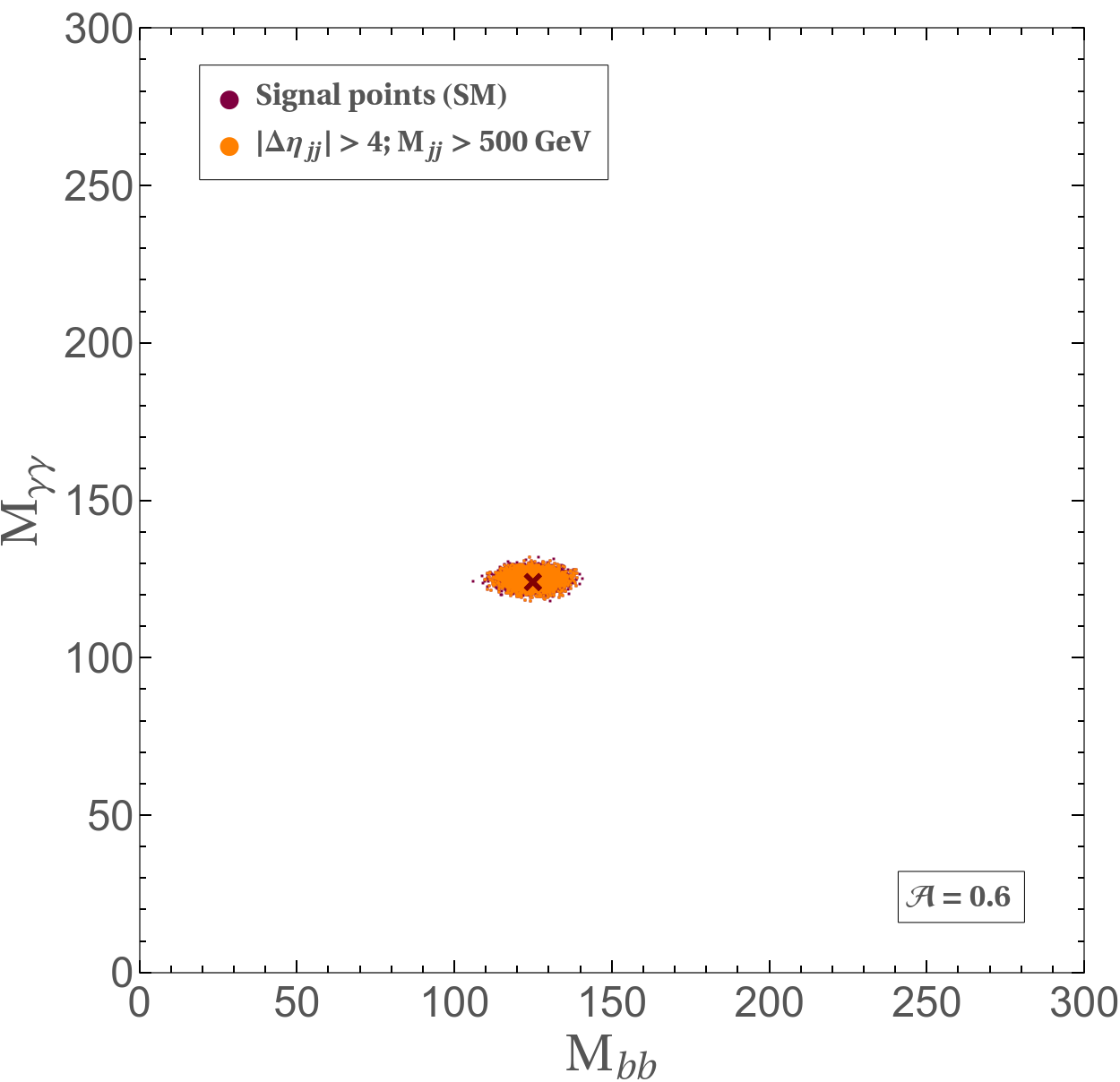}
\caption{Distribution of 10000 Monte Carlo signal events  of $pp\to HHjj\to b\bar{b}b\bar{b}jj$ (left panel) and of $pp\to HHjj\to b\bar{b}\gamma \gamma jj$ (right panel) in the plane of the invariant mass of one the Higgs candidates ($M_{bb_1}$ in the left panel and $M_{bb}$ in the right panel) versus the invariant mass of the other Higgs candidate ($M_{bb_2}$ in the left panel and $M_{\gamma\gamma}$ in the right panel) after applying a Gaussian smearing to the energy of all final state partons as explained in the text. See details of $HH$ candidate selection in the text.  Orange dots correspond to those events that pass the implemented VBS selection cuts given in Eq.(\ref{VBSselectioncuts}). Cuts in \eqref{basiccuts4b2j} (left panel) and in  \eqref{basiccuts2b2a2j} (right panel) have been implemented. The value of the acceptance $\mathcal{A}$ of the VBS cuts is also included. The red cross represents the value of the Higgs mass. The center of mass energy has been set to $\sqrt{s}=$14 TeV.}
\label{fig:MMplanesmearing}
\end{center}
\end{figure}

\item[5.-]  
Another important point that might change significantly our predictions is that introduced by the Higgs mass reconstruction uncertainty coming from detector effects. To estimate this uncertainty, we have applied a Gaussian smearing to the energy of all final state partons. Following~\cite{Pascoli:2018heg}, this gaussian dispersion has been introduced as $1/\sqrt{2\pi\sigma}\cdot e^{-x^2/(2\sigma^2)}$, with $\sigma=0.05\cdot E_{j,b}$ for the energy dispersion of the final light and $b$ jets and with $\sigma=0.02\cdot E_\gamma$ for the energy dispersion of the final photons. We have performed this for each studied signal and for their corresponding backgrounds in order to characterize the impact that these detector effects have regarding the distribution of our events on the relevant kinematical variables. In \figref{fig:MMplanesmearing} we show the distribution of  10000 Monte Carlo signal events  of $pp\to HHjj\to b\bar{b}b\bar{b}jj$ (left panel) and of $pp\to HHjj\to b\bar{b}\gamma \gamma jj$ (right panel) in the plane of the invariant mass of one the Higgs candidates ($M_{bb_2}$ in the left panel and $M_{bb}$ in the right panel) versus the invariant mass of the other Higgs candidate ($M_{bb_1}$ in the left panel and $M_{\gamma\gamma}$ in the right panel). No other cuts than those of the basic selection, given in \eqref{basiccuts4b2j} (left panel) and in  \eqref{basiccuts2b2a2j} (right panel) have been implemented. The orange points correspond to those events that fulfill the VBS selection criteria. The impact of these VBS cuts does not change appreciably after the smearing, not on the signal nor on the background events. The selection of the Higgs candidates in the case of the $b\bar{b}b\bar{b}jj$ signal is performed as explained in the text, following the minimization of $|M_{bb_1}-M_{bb_2}|$. This is the reason why we obtain several points distributed in the diagonal of the left panel. As expected, the detector effects translate into a dispersion of the signal points from the Higgs mass point outwards. In the $b\bar{b}b\bar{b} jj$ case, the dispersion is isotropic, since the smearing affects all four $b$-quarks in the same way, whereas in the $b\bar{b}b\gamma\gamma jj$ case, the dispersion in the $M_{bb}$ direction is bigger with respect to that in the $M_{\gamma\gamma}$ direction, accordingly to the difference in the energy resolution of $b$-quarks and photons in the detector. These results are compatible to those obtained in reference \cite{Kling:2016lay}. In any case, both signals seem to lie inside a circle of radius around 12 GeV, which corresponds to a 10\% of the Higgs mass value. This suggests that the effects of the smearing on our predictions of the statistical significance will severely depend on the $\Delta_E$ we use in the $\chi_{HH}$ selection cut, and, in principle, we expect that for $\Delta_E=10\%$ we will obtain the best sensitivities. This is so because, for this $\Delta_E=10\%$, we select the minimum possible number of background events compatible with selecting all of our signal events simultaneously. 

In order to better understand the impact of the $\Delta_E$ value once the detector effects have been taken into account, we present in the purple lines and purple shaded areas of \figref{fig:XHHcut}  the values for the statistical significance at $1000 \,{\rm fb}^{-1}$ as a function of the value of $\kappa$ for different energy resolutions of 5\% (original scenario throughout the work), 10\%, 20\% and 30\%  after the smearing on the energy of all final state partons has been applied. Is it clear from this figure that, indeed, taking $\Delta_E=10\%$ in the $b\bar{b}b\bar{b}jj$ case maximizes the statistical significance once the detector effects are included. In the $b\bar{b}\gamma\gamma jj $ case (notice that the purple area overlaps with the green one) the $\Delta_E=5\%$ is still the value that gives the best sensitivities, since the signal to background ratio is larger. In any case, from the upper green line to the upper purple line, there is at most a reduction factor of 0.4 in the statistical significance.

  \begin{figure}[t!]
\begin{center}
\includegraphics[width=0.49\textwidth]{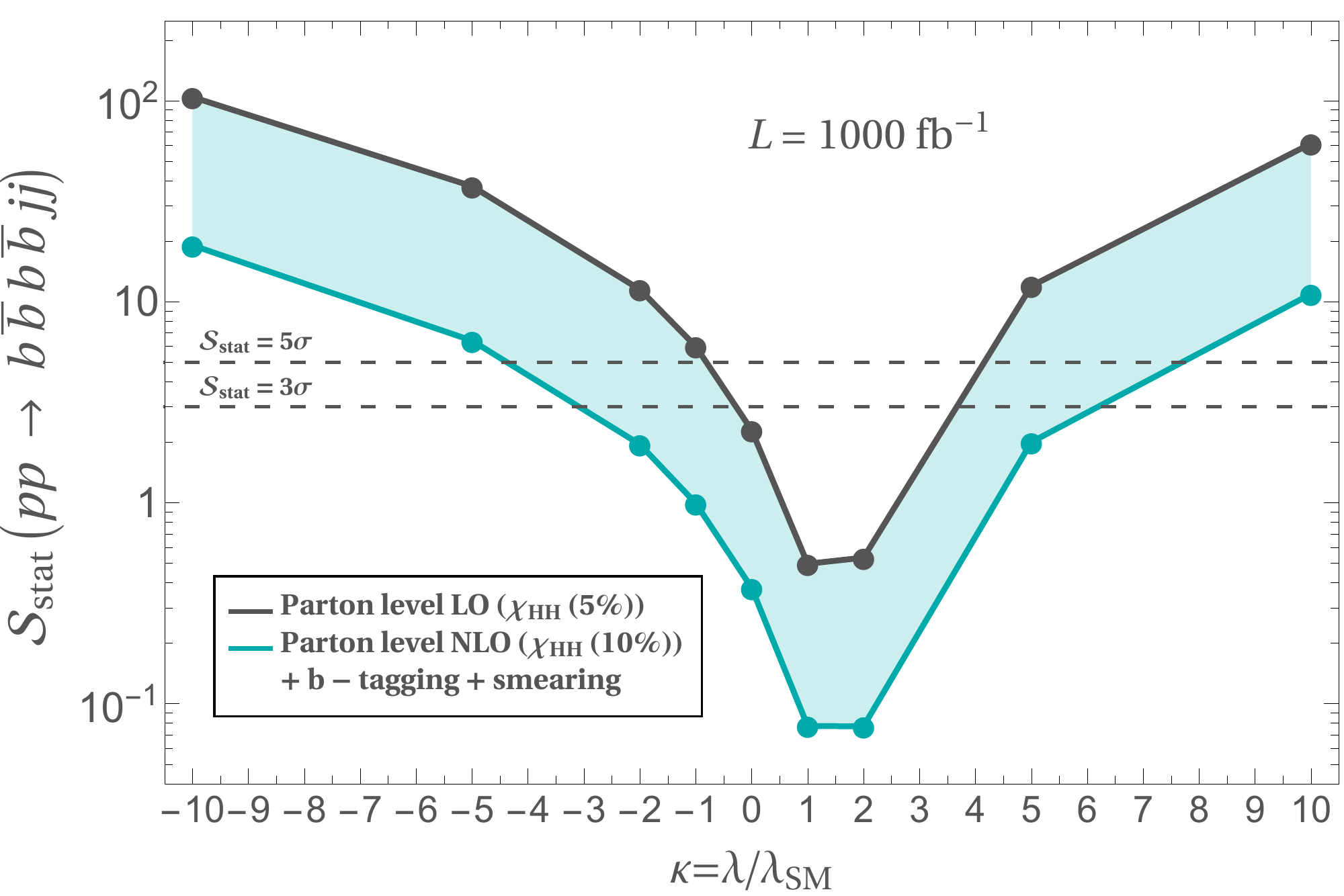}
\includegraphics[width=0.49\textwidth]{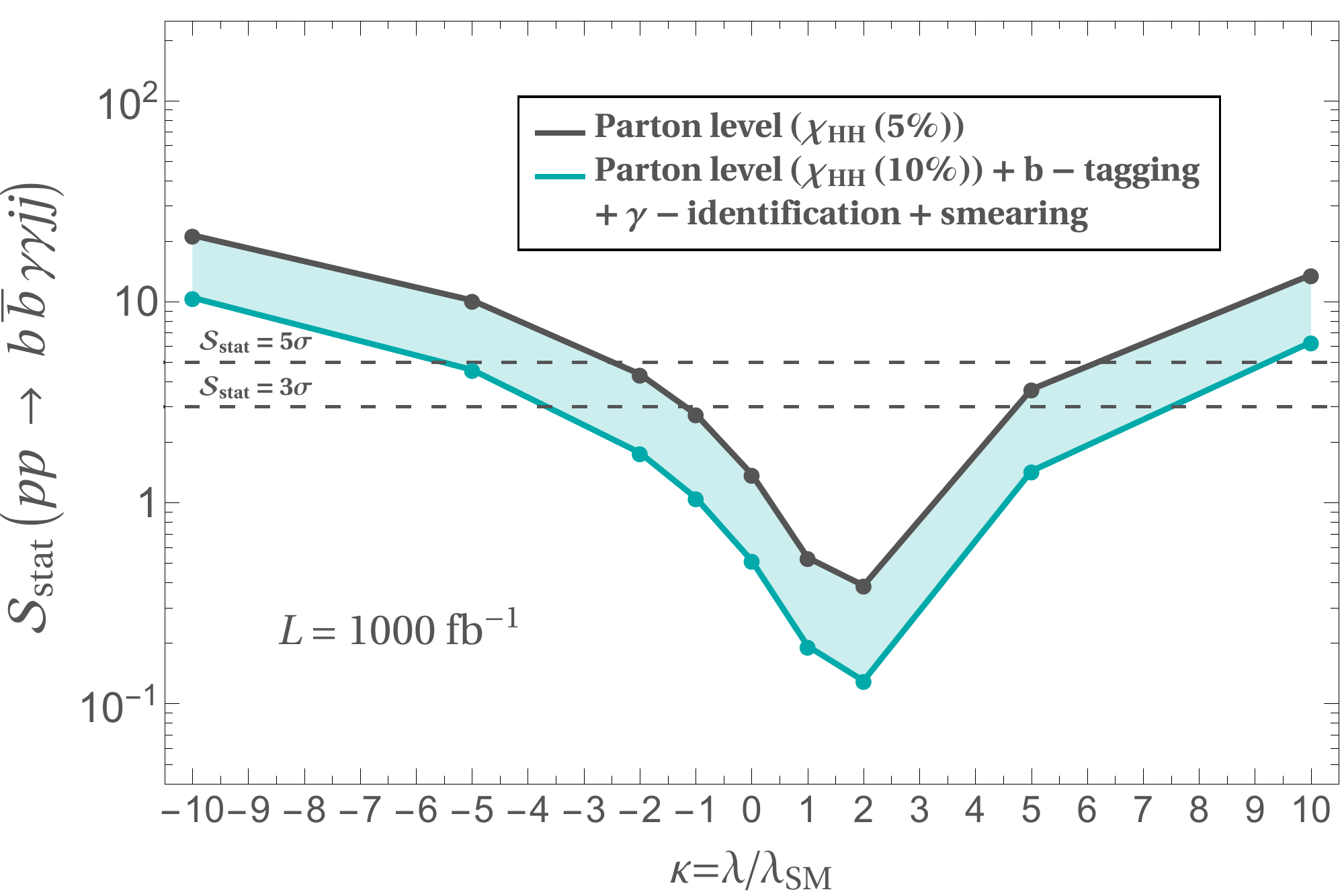}
\caption{Prediction of the statistical significance, $\mathcal{S}_{\rm stat}$, of the process $p p\to b\bar{b} b\bar{b} jj$ (left panel) and of the process $p p\to b\bar{b}\gamma\gamma jj$ (right panel) for $L=1000$ fb${}^{-1}$ as a function of the value of $\kappa$ for two different scenarios: the original parton level analysis (dark grey line, corresponding to the green lines in \figref{fig:significances_4b2j} (left panel) and \figref{fig:significances_2b2a2j} (right panel)) and the analysis performed taking into account the tagging efficiencies of the final state particles, the NLO corrections, the estimation of the detector effects via a Gaussian smearing on the energy of all final state partons and with a 10\% Higgs mass determination uncertainty (blue line, see details of these considerations in the text). The marked points represent our evaluations. The center of mass energy has been set to $\sqrt{s}=14$ TeV.}
\label{fig:errorband}
\end{center}
\end{figure}

\item[6.-]  Considering in addition the effects from showering and clustering of the final jets  will presumably change our naive parton level predictions. As we have said, it is not the purpose of this paper to provide a full complete analysis including these  important effects. It is clearly beyond the scope of this work and they will require a more sophisticated and devoted analysis with full computing power and the use of additional techniques like Boost Decision Trees (BDT) and others. This is particularly involved  if we wish to control efficiently the background form QCD-multijets and, consequently, we have postponed this full analysis for a future work in collaboration with our experimental colleagues\footnote{We wish to thank our experimental colleague  Aurelio Juste for his interesting discussions on this issue and for his involvement in this future project in collaboration with us.}. However, to get a first indication of the importance of these effects in the signal rates, we have performed a computation of the $b \bar b b \bar b j j$ signal MadGraph events after showering with PYTHIA8 \cite{Sjostrand:2014zea} and  clustering with MadAnalysis5 \cite{Conte:2012fm,Conte:2014zja,Dumont:2014tja,Conte:2018vmg} by using the anti-$k_t$ algorithm with $R=0.4$. We have performed this signal estimate for one BSM example with $\kappa=5$ and we have obtained that the cross-section after applying our basic and VBS cuts is $3.0. 10^{-3}$ pb if we include showering+clustering, which should be compared with our parton level estimate of $3.7.10^{-3}$ pb , i.e. without applying showering+clustering. Therefore, the effect from showering+clustering at this signal level is not very relevant. However, it is expected that it could be relevant in the $HH$ selection candidates and, as we have said, in the reduction efficiency of the QCD-multijet background. Nevertheless this is left for our future project. 

\end{itemize}

Finally, to conclude this discussion section and in order to give a more accurate and realistic prediction, all the above mentioned considerations must be taken into account simultaneously. To this aim, we present, in \figref{fig:errorband}, the predictions of the statistical significance as a function of the value of $\kappa$ at $1000~ {\rm fb}^{-1}$ for two comparative scenarios: the original analysis from LO parton level predictions (dark grey line) and the analysis performed after taking into account the main distorting effects which are the tagging efficiencies of the final state particles, described in point 3.- of this discussion, the NLO corrections described in point 2.- and the estimation of the detector effects, introduced in point 5.- with a 10\% Higgs mass determination uncertainty. We give these predictions for both of the studied signals: $p p\to b\bar{b} b\bar{b} jj$ (left panel) and $p p\to b\bar{b}\gamma\gamma jj$ (right panel). The main conclusion is that the biggest uncertainty in our predictions comes from the fact that we are not taking into account, a priori, detector effects.  We have already seen that this can reduce the statistical significance by a factor of 0.4. The second biggest source of uncertainty is the choice of the value of the Higgs mass resolution, $\Delta_E$. Taking a 10\% mass resolution instead of a 5\% can account for a reduction of 0.7 in the statistical significance. Similarly, the $b$-tagging efficiencies in the $b\bar{b}b\bar{b}jj$ case can lead to a similar reduction factor of 0.7. Finally, the NLO corrections play the less relevant role when estimating the uncertainties of the calculation. All the main effects together lead to a reduction factor of at most 0.2 in the statistical significance for $pp\to b\bar{b}b\bar{b}jj$ and of at most 0.5 for $pp\to b\bar{b}\gamma\gamma jj$.
The corresponding changes in the sensitivities to $\kappa$ can be easily derived from \figref{fig:errorband}. Using the same illustrative example, for $1000 \,{\rm fb}^{-1}$ of luminosity, we get sensitivities to $\kappa > 6.2 (7.7)$ at the $3\sigma$ ($5\sigma$) level to be compared with our benchmark result in \tabref{Tab:range_4b2j} of $\kappa >3.7 (4.2)$ for the $b\bar{b}b\bar{b} jj$ case and of $\kappa > 7.7 (9.4)$ at the $3\sigma$ ($5\sigma$) level to be compared with our benchmark result in \tabref{Tab:range_2b2a2j} of $\kappa >4.6 (6.0)$ for the $b\bar{b}\gamma\gamma jj$ one. 

Based on the discussion above we believe that a more dedicated analysis, including more accurately all the considerations above with showering, clustering, and detector effects, and optimizing the selection criteria accordingly\footnote{As previously said, we have postponed this full analysis for a future work in collaboration with our experimental colleagues.}, might lead to a sensitivity to the Higgs self-coupling of the same order of magnitude, although a bit smaller, than the one obtained with our naive original analysis. We believe that our findings indicate that double Higgs production via vector boson scattering is a viable and promising observable to measure the Higgs self-coupling in BSM scenarios.

\section{Conclusions}
\label{Conclusions}

Being able to determine with precision the value of the Higgs self-coupling $\lambda$ would allow us to understand the true nature of the Higgs mechanism and, therefore, of the scalar sector of the SM. In particular, an independent measurement of $\lambda$ and $m_H$ will be crucial in this understanding. 
At the LHC, the most sensitive channel to this coupling $\lambda$ is that of double Higgs production, that can take place through several initial configurations. Most of the theoretical and experimental studies of $HH$ production focus on gluon-gluon fusion since it benefits from the largest rates. Nevertheless, double Higgs production by vector boson scattering has important advantages with respect to gluon-gluon fusion that, despite its lower statistics, make of it a very promising and competitive channel to probe the Higgs self coupling at the LHC. These features have motivated our study here.

In the present work, we have analyzed the sensitivity to $\lambda$ in double Higgs production via vector boson scattering at the LHC, taking advantage of the fact that these processes have very characteristic kinematics that allow us to select them very efficiently against competing SM backgrounds. We have first explored and characterized the VBS subprocesses of our interest, $WW\to HH$ and $ZZ\to HH$, both for the SM with $\lambda=\lambda_{SM}$, and for BSM scenarios with $\lambda=\kappa\,\lambda_{SM}$, considering  values of $\kappa$ between -10 and 10, to move afterwards to the LHC scenario. We have then studied the process $pp\to HHjj$, in order to understand the properties of this scattering, and finally we have explored and provided quantitative results for the sensitivity to the Higgs self-coupling after the Higgs decays.

We have focused mainly on the $pp\to b\bar{b}b\bar{b}jj$ process since it benefits from the largest rates. After applying all our selection criteria, based on the VBS characteristic kinematical configuration and in the $HH$ candidates reconstruction, we give predictions for the sensitivity to $\lambda$ in $pp\to b\bar{b}b\bar{b}jj$ events at the parton level for $\sqrt{s}=14$ TeV and for different future expected luminosities: $L=50,300,1000,3000$ fb${}^{-1}$. Our main results for this channel are summarized in \tabref{Tab:range_4b2j} and in \figref{fig:significances_4b2j}, in which we present the values of $\kappa$ that the LHC would be sensitive to at the $3\sigma$ and at the $5\sigma$ level. The sensitivities we obtain here, at the parton level, even for the lowest luminosity, are very encouraging and clearly invite to explore this channel further with the new technology applied to control the QCD-multijet background, including hadronization and detector effects, which will allow us to get a fully realistic result. Furthermore, our predictions show that the HL-LHC should be able to probe small deviations in $\lambda$ respect to the SM value, reaching very good sensitivities with the highest luminosity to up to $\kappa > 3.2\,(3.7) $ at the $3\sigma\,(5\sigma)$ level in the best scenario for positive values. In the case of negative values, the HL-LHC would be sensitive to all the $\kappa<0$ values that have been considered in this work.

We give as well predictions for $pp\to b\bar{b}\gamma\gamma jj$ events, also at the parton level and for $\sqrt{s}=14$ TeV, due to the fact that it provides a cleaner, although with smaller rates, signature. The results of the sensitivities to the Higgs self-coupling in this channel, after applying the proper selection criteria, are collected in \tabref{Tab:range_2b2a2j} and in 
\figref{fig:significances_2b2a2j}. Again, we obtain very promising results for $b\bar{b}\gamma\gamma$ events, except for the lowest luminosity considered, where the signal rates found are too low. Interestingly, we show that  the statistical significance grows faster with luminosity in this channel. This would imply that for bigger luminosities, very small deviations in $\lambda$ could also be measured in $pp\to b\bar{b}\gamma\gamma jj$ events. In particular, for the highest luminosity considered, our predictions show that the HL-LHC could reach sensitivities to $\kappa > 3.8\,(4.7) $ at the $3\sigma\,(5\sigma)$ level in this channel.

Furthermore, we give predictions for the interesting case of $L=1000$ fb${}^{-1}$ of how the sensitivity to $\lambda$ will change from our naive parton level results when taking into account the main distorting effects. We have discussed the impact of $b$-tagging and of $\gamma$ identification efficiencies, of detector effects and of Higgs mass reconstruction resolution. All these main effects together lead to a reduction factor of at most 0.2 in the statistical significance for $pp\to b\bar{b}b\bar{b}jj$ and of at most 0.5 for $pp\to b\bar{b}\gamma\gamma jj$, as it can be seen in \figref{fig:errorband}. The corresponding changes in the sensitivities to $\kappa$ translate into the fact that, at this luminosity, the LHC will be sensitive to $\kappa > 6.2 \,(7.7)$ at the $3\sigma$ ($5\sigma$) level for the $b\bar{b}b\bar{b} jj$ case and of $\kappa > 7.7 \,(9.4)$ at the $3\sigma$ ($5\sigma$) level for the $b\bar{b}\gamma\gamma jj$ one. In the also interesting case of $L=3000$ fb${}^{-1}$, we get similar reduction factors. The reachable values of $\kappa$ at this last HL-LHC stage via VBS configurations are predicted to be $\kappa > 5.0 \,(6.3)$ at the $3\sigma$ ($5\sigma$) level for the $b\bar{b}b\bar{b} jj$ case and of $\kappa > 6.1 \,(8.0)$ at the $3\sigma$ ($5\sigma$) level for the $b\bar{b}\gamma\gamma jj$ one.

In both cases we have seen that the values of $\kappa$ that are closer to the SM value, say, $\kappa\in[0.5,2]$, are the most challenging ones to reach at the LHC. Even for the largest luminosity considered in this work, their corresponding statistical significances are always below 2$\sigma$. Hopefully, in this case gluon gluon fusion will be undoubtedly the only way to reach enough sensitivity, see refs~\cite{Baur:2003gp, Baglio:2012np, Yao:2013ika, Barger:2013jfa, Huang:2015tdv, Kling:2016lay}. It is predicted, that, at  $L=3000$ fb${}^{-1}$, statistical significances above 2$\sigma$ will be always achieved in the gluon gluon fusion channel for $\lambda\sim\lambda_{\rm SM}$.

The present study shows that double Higgs production via vector boson scattering is a viable and promising window to measure BSM deviations to the Higgs self-coupling and to deeply understand the scalar sector of the SM. Although all simulations are performed at the parton level, without hadronization or detector response simulation, and should be understood as a naive first approximation, we obtain very competitive results for the sensitivity to $\lambda$ at the LHC. Because of this, we believe that the vector boson scattering $HH$ production channel will lead to very interesting (and complementary to those of gluon-gluon fusion) findings about the true nature of the Higgs boson.


\section*{Acknowledgments}
We would like to warmly thank Michelangelo L. Mangano for his invaluable help and generosity in guiding us with the use of AlpGen, taking active part in the running of this Monte Carlo, which has been a very important part of the QCD background evaluation for this work.  We would also like to thank Juan Antonio Aguilar Saavedra for fruitful discussions and Richard Ruiz for his help and suggestions in the use of MadGraph. C.G.G. wishes to thank V\'ictor Mart\'in Lozano, Javier Quilis and especially Xabier Marcano for supportive and helpful conversations. E.A. warmly thanks IFT of Madrid for its hospitality hosting him during the completion of this work. This work is supported by the European Union through the ITN ELUSIVES H2020-MSCA-ITN-2015//674896 and the RISE INVISIBLESPLUS H2020-MSCA-RISE-2015//690575, by the CICYT through the projects FPA2016-78645-P, by the Spanish Consolider-Ingenio 2010 Programme CPAN (CSD2007-00042) and by the Spanish MINECO's ``Centro de Excelencia Severo Ochoa''  Programme under grant SEV-2016-0597. This work has also been partially supported by CONICET and ANPCyT projects no. PICT 2016-0164 and no. PICT-2017-2765 (E. A.).


\bibliographystyle{JHEP}

\begin{thebibliography}{10}



\bibitem{Aad:2012tfa}
{\scshape ATLAS} collaboration, G.~Aad et~al., \emph{{Observation of a new
  particle in the search for the Standard Model Higgs boson with the ATLAS
  detector at the LHC}},
  \href{https://doi.org/10.1016/j.physletb.2012.08.020}{\emph{Phys. Lett.}
  {\bfseries B716} (2012) 1} [\href{https://arxiv.org/abs/1207.7214}{{\ttfamily
  1207.7214}}].

\bibitem{Chatrchyan:2012xdj}
{\scshape CMS} collaboration, S.~Chatrchyan et~al., \emph{{Observation of a new
  boson at a mass of 125 GeV with the CMS experiment at the LHC}},
  \href{https://doi.org/10.1016/j.physletb.2012.08.021}{\emph{Phys. Lett.}
  {\bfseries B716} (2012) 30}
  [\href{https://arxiv.org/abs/1207.7235}{{\ttfamily 1207.7235}}].

\bibitem{Higgs:1964ia}
P.~W. Higgs, \emph{{Broken symmetries, massless particles and gauge fields}},
  \href{https://doi.org/10.1016/0031-9163(64)91136-9}{\emph{Phys. Lett.}
  {\bfseries 12} (1964) 132}.

\bibitem{Englert:1964et}
F.~Englert and R.~Brout, \emph{{Broken Symmetry and the Mass of Gauge Vector
  Mesons}}, \href{https://doi.org/10.1103/PhysRevLett.13.321}{\emph{Phys. Rev.
  Lett.} {\bfseries 13} (1964) 321}.

\bibitem{Higgs:1964pj}
P.~W. Higgs, \emph{{Broken Symmetries and the Masses of Gauge Bosons}},
  \href{https://doi.org/10.1103/PhysRevLett.13.508}{\emph{Phys. Rev. Lett.}
  {\bfseries 13} (1964) 508}.

\bibitem{Higgs:1966ev}
P.~W. Higgs, \emph{{Spontaneous Symmetry Breakdown without Massless Bosons}},
  \href{https://doi.org/10.1103/PhysRev.145.1156}{\emph{Phys. Rev.} {\bfseries
  145} (1966) 1156}.

\bibitem{Simon:2012ik}
F.~Simon, \emph{{Prospects for Precision Higgs Physics at Linear Colliders}},
  {\emph{PoS} {\bfseries ICHEP2012} (2013) 066}
  [\href{https://arxiv.org/abs/1211.7242}{{\ttfamily 1211.7242}}].

\bibitem{Dawson:2013bba}
S.~Dawson et~al., \emph{{Working Group Report: Higgs Boson}},  in
  \emph{{Proceedings, 2013 Community Summer Study on the Future of U.S.
  Particle Physics: Snowmass on the Mississippi (CSS2013): Minneapolis, MN,
  USA, July 29-August 6, 2013}}, 2013,
  \href{https://arxiv.org/abs/1310.8361}{{\ttfamily 1310.8361}},
  \href{http://inspirehep.net/record/1262795/files/arXiv:1310.8361.pdf}{http://inspirehep.net/record/1262795/files/arXiv:1310.8361.pdf}.

\bibitem{Baer:2013cma}
H.~Baer, T.~Barklow, K.~Fujii, Y.~Gao, A.~Hoang, S.~Kanemura et~al., \emph{{The
  International Linear Collider Technical Design Report - Volume 2: Physics}},
  \href{https://arxiv.org/abs/1306.6352}{{\ttfamily 1306.6352}}.

\bibitem{Abramowicz:2016zbo}
H.~Abramowicz et~al., \emph{{Higgs physics at the CLIC electron--positron
  linear collider}},
  \href{https://doi.org/10.1140/epjc/s10052-017-4968-5}{\emph{Eur. Phys. J.}
  {\bfseries C77} (2017) 475}
  [\href{https://arxiv.org/abs/1608.07538}{{\ttfamily 1608.07538}}].

\bibitem{deFlorian:2016spz}
{\scshape LHC Higgs Cross Section Working Group} collaboration, D.~de~Florian
  et~al., \emph{{Handbook of LHC Higgs Cross Sections: 4. Deciphering the
  Nature of the Higgs Sector}},
  \href{https://arxiv.org/abs/1610.07922}{{\ttfamily 1610.07922}}.

\bibitem{Glover:1987nx}
E.~W.~N. Glover and J.~J. van~der Bij, \emph{{Higgs boson pair production via gluon fusion}},
  \href{https://doi.org/10.1016/0550-3213(88)90083-1}{\emph{Nucl. Phys.}
  {\bfseries B309} (1988) 282}.

\bibitem{Dicus:1987ic}
D.~A. Dicus, C.~Kao and S.~S.~D. Willenbrock, \emph{{Higgs Boson Pair
  Production From Gluon Fusion}},
  \href{https://doi.org/10.1016/0370-2693(88)90202-X}{\emph{Phys. Lett.}
  {\bfseries B203} (1988) 457}.

\bibitem{Plehn:1996wb}
T.~Plehn, M.~Spira and P.~M. Zerwas, \emph{{Pair production of neutral Higgs
  particles in gluon-gluon collisions}},
  \href{https://doi.org/10.1016/0550-3213(96)00418-X,
  10.1016/S0550-3213(98)00406-4}{\emph{Nucl. Phys.} {\bfseries B479} (1996) 46}
  [\href{https://arxiv.org/abs/hep-ph/9603205}{{\ttfamily hep-ph/9603205}}].

\bibitem{Dawson:1998py}
S.~Dawson, S.~Dittmaier and M.~Spira, \emph{{Neutral Higgs boson pair
  production at hadron colliders: QCD corrections}},
  \href{https://doi.org/10.1103/PhysRevD.58.115012}{\emph{Phys. Rev.}
  {\bfseries D58} (1998) 115012}
  [\href{https://arxiv.org/abs/hep-ph/9805244}{{\ttfamily hep-ph/9805244}}].

\bibitem{Djouadi:1999rca}
A.~Djouadi, W.~Kilian, M.~Muhlleitner and P.~M. Zerwas, \emph{{Production of
  neutral Higgs boson pairs at LHC}},
  \href{https://doi.org/10.1007/s100529900083}{\emph{Eur. Phys. J.} {\bfseries
  C10} (1999) 45} [\href{https://arxiv.org/abs/hep-ph/9904287}{{\ttfamily
  hep-ph/9904287}}].

\bibitem{Baur:2003gp}
U.~Baur, T.~Plehn and D.~L. Rainwater, \emph{{Probing the Higgs selfcoupling at
  hadron colliders using rare decays}},
  \href{https://doi.org/10.1103/PhysRevD.69.053004}{\emph{Phys. Rev.}
  {\bfseries D69} (2004) 053004}
  [\href{https://arxiv.org/abs/hep-ph/0310056}{{\ttfamily hep-ph/0310056}}].

\bibitem{Grober:2010yv}
R.~Grober and M.~Muhlleitner, \emph{{Composite Higgs Boson Pair Production at
  the LHC}}, \href{https://doi.org/10.1007/JHEP06(2011)020}{\emph{JHEP}
  {\bfseries 06} (2011) 020} [\href{https://arxiv.org/abs/1012.1562}{{\ttfamily
  1012.1562}}].

\bibitem{Dolan:2012rv}
M.~J. Dolan, C.~Englert and M.~Spannowsky, \emph{{Higgs self-coupling
  measurements at the LHC}},
  \href{https://doi.org/10.1007/JHEP10(2012)112}{\emph{JHEP} {\bfseries 10}
  (2012) 112} [\href{https://arxiv.org/abs/1206.5001}{{\ttfamily 1206.5001}}].

\bibitem{Papaefstathiou:2012qe}
A.~Papaefstathiou, L.~L. Yang and J.~Zurita, \emph{{Higgs boson pair production
  at the LHC in the $b \bar{b} W^+ W^-$ channel}},
  \href{https://doi.org/10.1103/PhysRevD.87.011301}{\emph{Phys. Rev.}
  {\bfseries D87} (2013) 011301}
  [\href{https://arxiv.org/abs/1209.1489}{{\ttfamily 1209.1489}}].

\bibitem{Baglio:2012np}
J.~Baglio, A.~Djouadi, R.~Gr{\"o}ber, M.~M. M{\"u}hlleitner, J.~Quevillon and
  M.~Spira, \emph{{The measurement of the Higgs self-coupling at the LHC:
  theoretical status}},
  \href{https://doi.org/10.1007/JHEP04(2013)151}{\emph{JHEP} {\bfseries 04}
  (2013) 151} [\href{https://arxiv.org/abs/1212.5581}{{\ttfamily 1212.5581}}].

\bibitem{Yao:2013ika}
W.~Yao,
 \emph{{Studies of measuring Higgs self-coupling with $HH\rightarrow b\bar b \gamma\gamma$ at the future hadron colliders}},
[\href{https://arxiv.org/abs/1308.6302}{{\ttfamily 1308.6302}}].

\bibitem{deFlorian:2013jea}
D.~de~Florian and J.~Mazzitelli, \emph{{Higgs Boson Pair Production at
  Next-to-Next-to-Leading Order in QCD}},
  \href{https://doi.org/10.1103/PhysRevLett.111.201801}{\emph{Phys. Rev. Lett.}
  {\bfseries 111} (2013) 201801}
  [\href{https://arxiv.org/abs/1309.6594}{{\ttfamily 1309.6594}}].
  

\bibitem{Dolan:2013rja}
M.~J. Dolan, C.~Englert, N.~Greiner and M.~Spannowsky, \emph{{Further on up the
  road: $hhjj$ production at the LHC}},
  \href{https://doi.org/10.1103/PhysRevLett.112.101802}{\emph{Phys. Rev. Lett.}
  {\bfseries 112} (2014) 101802}
  [\href{https://arxiv.org/abs/1310.1084}{{\ttfamily 1310.1084}}].
  
  \bibitem{Barger:2013jfa}
V.~Barger, L.~L.~Everett, C.~B.~Jackson and G.~Shaughnessy,
 \emph{{Higgs-Pair Production and Measurement of the Triscalar Coupling at LHC(8,14)}}, \href{https://doi.org/10.1016/j.physletb.2013.12.013}{\emph{Phys.
  Lett.} {\bfseries B728} (2014) 433}
  [\href{https://arxiv.org/abs/1311.29310}{{\ttfamily 1311.2931}}].
  
  \bibitem{Frederix:2014hta}
R.~Frederix, S.~Frixione, V.~Hirschi, F.~Maltoni, O.~Mattelaer, P.~Torrielli
  et~al., \emph{{Higgs pair production at the LHC with NLO and parton-shower
  effects}}, \href{https://doi.org/10.1016/j.physletb.2014.03.026}{\emph{Phys.
  Lett.} {\bfseries B732} (2014) 142}
  [\href{https://arxiv.org/abs/1401.7340}{{\ttfamily 1401.7340}}].

\bibitem{Liu-Sheng:2014gxa}
L.-S. Ling, R.-Y. Zhang, W.-G. Ma, L.~Guo, W.-H. Li and X.-Z. Li, \emph{{NNLO
  QCD corrections to Higgs pair production via vector boson fusion at hadron
  colliders}}, \href{https://doi.org/10.1103/PhysRevD.89.073001}{\emph{Phys.
  Rev.} {\bfseries D89} (2014) 073001}
  [\href{https://arxiv.org/abs/1401.7754}{{\ttfamily 1401.7754}}].

\bibitem{Goertz:2014qta}
F.~Goertz, A.~Papaefstathiou, L.~L. Yang and J.~Zurita, \emph{{Higgs boson pair
  production in the D=6 extension of the SM}},
  \href{https://doi.org/10.1007/JHEP04(2015)167}{\emph{JHEP} {\bfseries 04}
  (2015) 167} [\href{https://arxiv.org/abs/1410.3471}{{\ttfamily 1410.3471}}].

\bibitem{Azatov:2015oxa}
A.~Azatov, R.~Contino, G.~Panico and M.~Son, \emph{{Effective field theory
  analysis of double Higgs boson production via gluon fusion}},
  \href{https://doi.org/10.1103/PhysRevD.92.035001}{\emph{Phys. Rev.}
  {\bfseries D92} (2015) 035001}
  [\href{https://arxiv.org/abs/1502.00539}{{\ttfamily 1502.00539}}].

\bibitem{Dicus:2015yva}
D.~A. Dicus, C.~Kao and W.~W. Repko, \emph{{Interference effects and the use of
  Higgs boson pair production to study the Higgs trilinear self coupling}},
  \href{https://doi.org/10.1103/PhysRevD.92.093003}{\emph{Phys. Rev.}
  {\bfseries D92} (2015) 093003}
  [\href{https://arxiv.org/abs/1504.02334}{{\ttfamily 1504.02334}}].

\bibitem{Dawson:2015oha}
S.~Dawson, A.~Ismail and I.~Low, \emph{{What's in the loop? The anatomy of
  double Higgs production}},
  \href{https://doi.org/10.1103/PhysRevD.91.115008}{\emph{Phys. Rev.}
  {\bfseries D91} (2015) 115008}
  [\href{https://arxiv.org/abs/1504.05596}{{\ttfamily 1504.05596}}].

\bibitem{He:2015spf}
H.-J. He, J.~Ren and W.~Yao, \emph{{Probing new physics of cubic Higgs boson
  interaction via Higgs pair production at hadron colliders}},
  \href{https://doi.org/10.1103/PhysRevD.93.015003}{\emph{Phys. Rev.}
  {\bfseries D93} (2016) 015003}
  [\href{https://arxiv.org/abs/1506.03302}{{\ttfamily 1506.03302}}].

\bibitem{Dolan:2015zja}
M.~J. Dolan, C.~Englert, N.~Greiner, K.~Nordstrom and M.~Spannowsky,
  \emph{{$hhjj$ production at the LHC}},
  \href{https://doi.org/10.1140/epjc/s10052-015-3622-3}{\emph{Eur. Phys. J.}
  {\bfseries C75} (2015) 387}
  [\href{https://arxiv.org/abs/1506.08008}{{\ttfamily 1506.08008}}].

\bibitem{Cao:2015oaa}
Q.-H. Cao, B.~Yan, D.-M. Zhang and H.~Zhang, \emph{{Resolving the Degeneracy in
  Single Higgs Production with Higgs Pair Production}},
  \href{https://doi.org/10.1016/j.physletb.2015.11.045}{\emph{Phys. Lett.}
  {\bfseries B752} (2016) 285}
  [\href{https://arxiv.org/abs/1508.06512}{{\ttfamily 1508.06512}}].

\bibitem{Cao:2015oxx}
Q.-H. Cao, Y.~Liu and B.~Yan, \emph{{Measuring trilinear Higgs coupling in WHH
  and ZHH productions at the high-luminosity LHC}},
  \href{https://doi.org/10.1103/PhysRevD.95.073006}{\emph{Phys. Rev.}
  {\bfseries D95} (2017) 073006}
  [\href{https://arxiv.org/abs/1511.03311}{{\ttfamily 1511.03311}}].

\bibitem{Huang:2015tdv}
  P.~Huang, A.~Joglekar, B.~Li and C.~E.~M.~Wagner,
  \emph{{Probing the Electroweak Phase Transition at the LHC}},
  \href{https://doi.org/10.1103/PhysRevD.93.055049}{\emph{Phys. Rev.}
  {\bfseries D93} (2016) no.5,  055049}
 [\href{https://arxiv.org/abs/1512.00068}{{\ttfamily 1512.00068}}].
 

  
\bibitem{Behr:2015oqq}
J.~K. Behr, D.~Bortoletto, J.~A. Frost, N.~P. Hartland, C.~Issever and J.~Rojo,
  \emph{{Boosting Higgs pair production in the $b\bar{b}b\bar{b}$ final state
  with multivariate techniques}},
  \href{https://doi.org/10.1140/epjc/s10052-016-4215-5}{\emph{Eur. Phys. J.}
  {\bfseries C76} (2016) 386}
  [\href{https://arxiv.org/abs/1512.08928}{{\ttfamily 1512.08928}}].


 \bibitem{Kling:2016lay}
  F.~Kling, T.~Plehn and P.~Schichtel,
  \emph{{`Maximizing the significance in Higgs boson pair analyses}},
  \href{https://doi.org/10.1103/PhysRevD.95.035026}{\emph{Phys. Rev.}
  {\bfseries D95} (2017) no.3,  035026}
 [\href{https://arxiv.org/abs/1607.07441}{{\ttfamily 1607.07441}}].


\bibitem{Borowka:2016ypz}
  S.~Borowka, N.~Greiner, G.~Heinrich, S.~P.~Jones, M.~Kerner, J.~Schlenk and T.~Zirke,
  \emph{{Full top quark mass dependence in Higgs boson pair production at NLO}},
  \href{https://doi.org/10.1007/JHEP10(2016)107}{\emph{JHEP.}
  {\bfseries 1610} (2016) 107 }
 [\href{https://arxiv.org/abs/1608.04798}{{\ttfamily 1608.04798}}].
 
\bibitem{Bishara:2016kjn}
F.~Bishara, R.~Contino and J.~Rojo, \emph{{Higgs pair production in
  vector-boson fusion at the LHC and beyond}},
  \href{https://doi.org/10.1140/epjc/s10052-017-5037-9}{\emph{Eur. Phys. J.}
  {\bfseries C77} (2017) 481}
  [\href{https://arxiv.org/abs/1611.03860}{{\ttfamily 1611.03860}}].

\bibitem{Cao:2016zob}
Q.-H. Cao, G.~Li, B.~Yan, D.-M. Zhang and H.~Zhang, \emph{{Double Higgs
  production at the 14 TeV LHC and a 100 TeV $pp$ collider}},
  \href{https://doi.org/10.1103/PhysRevD.96.095031}{\emph{Phys. Rev.}
  {\bfseries D96} (2017) 095031}
  [\href{https://arxiv.org/abs/1611.09336}{{\ttfamily 1611.09336}}].

\bibitem{Adhikary:2017jtu}
A.~Adhikary, S.~Banerjee, R.~K. Barman, B.~Bhattacherjee and S.~Niyogi,
  \emph{{Revisiting the non-resonant Higgs pair production at the HL-LHC}},
  \href{https://doi.org/10.1007/JHEP07(2018)116}{\emph{Physics} {\bfseries
  2018} (2018) 116} [\href{https://arxiv.org/abs/1712.05346}{{\ttfamily
  1712.05346}}].
  
  \bibitem{Kim:2018uty}
J.~H.~Kim, Y.~Sakaki and M.~Son,
 \emph{{Combined analysis of double Higgs production via gluon fusion at the HL-LHC in the effective field theory approach}},
  \href{https://doi.org/10.1103/PhysRevD.98.015016}{\emph{Phys. Rev.}
  {\bfseries D98} (2018) no.1, 015016}
  [\href{https://arxiv.org/abs/1801.06093}{{\ttfamily 1801.06093}}].

\bibitem{Banerjee:2018yxy}
S.~Banerjee, C.~Englert, M.~L. Mangano, M.~Selvaggi and M.~Spannowsky,
  \emph{{$hh+\text{jet}$ production at 100 TeV}},
  \href{https://doi.org/10.1140/epjc/s10052-018-5788-y}{\emph{Eur. Phys. J.}
  {\bfseries C78} (2018) 322}
  [\href{https://arxiv.org/abs/1802.01607}{{\ttfamily 1802.01607}}].

\bibitem{Goncalves:2018qas}
D.~Gon{\c c}alves, T.~Han, F.~Kling, T.~Plehn and M.~Takeuchi, \emph{{Higgs
  boson pair production at future hadron colliders: From kinematics to
  dynamics}}, \href{https://doi.org/10.1103/PhysRevD.97.113004}{\emph{Phys.
  Rev.} {\bfseries D97} (2018) 113004}
  [\href{https://arxiv.org/abs/1802.04319}{{\ttfamily 1802.04319}}].
  
  
  
  \bibitem{Bizon:2018syu}
  W.~Bizon, U.~Haisch and L.~Rottoli,
  \emph{{Constraints on the quartic Higgs self-coupling from double-Higgs production at future hadron colliders}},
 [\href{https://arxiv.org/abs/1810.04665}{{\ttfamily 1810.04665}}].
 
 
 \bibitem{Borowka:2018pxx}
  S.~Borowka, C.~Duhr, F.~Maltoni, D.~Pagani, A.~Shivaji and X.~Zhao,
  \emph{{Probing the scalar potential via double Higgs boson production at hadron colliders}},
 [\href{https://arxiv.org/abs/1811.12366}{{\ttfamily 1811.12366}}].
 
 
 \bibitem{Gorbahn:2019lwq}
 M.~Gorbahn and U.~Haisch,
   \emph{{Two-loop amplitudes for Higgs plus jet production involving a modified trilinear Higgs coupling}},
 [\href{https://arxiv.org/abs/1902.05480}{{\ttfamily 1902.05480}}].
 
  

\bibitem{Aaboud:2016xco}
{\scshape ATLAS} collaboration, M.~Aaboud et~al., \emph{{Search for pair
  production of Higgs bosons in the $b\bar{b}b\bar{b}$ final state using
  proton--proton collisions at $\sqrt{s} = 13$ TeV with the ATLAS detector}},
  \href{https://doi.org/10.1103/PhysRevD.94.052002}{\emph{Phys. Rev.}
  {\bfseries D94} (2016) 052002}
  [\href{https://arxiv.org/abs/1606.04782}{{\ttfamily 1606.04782}}].

\bibitem{CMS:2016foy}
{\scshape CMS} collaboration, C.~Collaboration, \emph{{Search for non-resonant
  pair production of Higgs bosons in the $\rm{b} \bar{\rm{b}} \rm{b}
  \bar{\rm{b}}$ final state with 13 TeV CMS data}}, {\emph{CMS-PAS-HIG-16-026}
  (2016) }.

\bibitem{CMS:2017ihs}
{\scshape CMS} collaboration, C.~Collaboration, \emph{{Search for Higgs boson
  pair production in the final state containing two photons and two bottom
  quarks in proton-proton collisions at $\sqrt{s}=13~\mathrm{TeV}$}},
  {\emph{CMS-PAS-HIG-17-008} (2017) }.

\bibitem{Aaboud:2018knk}
{\scshape ATLAS} collaboration, M.~Aaboud et~al., \emph{{Search for pair
  production of Higgs bosons in the $b\bar{b}b\bar{b}$ final state using
  proton-proton collisions at $\sqrt{s} = 13$ TeV with the ATLAS detector}},
  \href{https://arxiv.org/abs/1804.06174}{{\ttfamily 1804.06174}}.

\bibitem{Sirunyan:2018iwt}
{\scshape CMS} collaboration, A.~M. Sirunyan et~al., \emph{{Search for Higgs
  boson pair production in the $\gamma\gamma\mathrm{b\overline{b}}$ final state
  in pp collisions at $\sqrt{s}=$ 13 TeV}},
  \href{https://arxiv.org/abs/1806.00408}{{\ttfamily 1806.00408}}.

\bibitem{Aaboud:2018ftw}
{\scshape ATLAS} collaboration, M.~Aaboud et~al., \emph{{Search for Higgs boson
  pair production in the $\gamma\gamma b\bar{b}$ final state with 13 TeV $pp$
  collision data collected by the ATLAS experiment}},
  \href{https://arxiv.org/abs/1807.04873}{{\ttfamily 1807.04873}}.

\bibitem{Doroba:2012pd}
K.~Doroba, J.~Kalinowski, J.~Kuczmarski, S.~Pokorski, J.~Rosiek, M.~Szleper
  et~al., \emph{{The $W_L W_L$ Scattering at the LHC: Improving the Selection
  Criteria}}, \href{https://doi.org/10.1103/PhysRevD.86.036011}{\emph{Phys.
  Rev.} {\bfseries D86} (2012) 036011}
  [\href{https://arxiv.org/abs/1201.2768}{{\ttfamily 1201.2768}}].

\bibitem{Szleper:2014xxa}
M.~Szleper, \emph{{The Higgs boson and the physics of $WW$ scattering before
  and after Higgs discovery}},
  \href{https://arxiv.org/abs/1412.8367}{{\ttfamily 1412.8367}}.

\bibitem{Fabbrichesi:2015hsa}
M.~Fabbrichesi, M.~Pinamonti, A.~Tonero and A.~Urbano, \emph{{Vector boson
  scattering at the LHC: A study of the WW $\to$ WW channels with the Warsaw
  cut}}, \href{https://doi.org/10.1103/PhysRevD.93.015004}{\emph{Phys. Rev.}
  {\bfseries D93} (2016) 015004}
  [\href{https://arxiv.org/abs/1509.06378}{{\ttfamily 1509.06378}}].

\bibitem{Delgado:2017cls}
R.~L. Delgado, A.~Dobado, D.~Espriu, C.~Garcia-Garcia, M.~J. Herrero,
  X.~Marcano et~al., \emph{{Production of vector resonances at the LHC via
  WZ-scattering: a unitarized EChL analysis}},
  \href{https://doi.org/10.1007/JHEP11(2017)098}{\emph{JHEP} {\bfseries 11}
  (2017) 098} [\href{https://arxiv.org/abs/1707.04580}{{\ttfamily
  1707.04580}}].

\bibitem{PDG2018}
{\scshape Particle Data Group} collaboration, M.~Tanabashi et~al.,
  \emph{{Review of Particle Physics}}, {\emph{Phys. Rev.} {\bfseries D98}
  (2018) 030001}.

\bibitem{Aad:2014zda}
{\scshape ATLAS} collaboration, G.~Aad et~al., \emph{{Evidence for Electroweak
  Production of $W^{\pm}W^{\pm}jj$ in $pp$ Collisions at $\sqrt{s}=8$ TeV with
  the ATLAS Detector}},
  \href{https://doi.org/10.1103/PhysRevLett.113.141803}{\emph{Phys. Rev. Lett.}
  {\bfseries 113} (2014) 141803}
  [\href{https://arxiv.org/abs/1405.6241}{{\ttfamily 1405.6241}}].

\bibitem{Aad:2016ett}
{\scshape ATLAS} collaboration, G.~Aad et~al., \emph{{Measurements of $W^\pm Z$
  production cross sections in $pp$ collisions at $\sqrt{s} = 8$ TeV with the
  ATLAS detector and limits on anomalous gauge boson self-couplings}},
  \href{https://doi.org/10.1103/PhysRevD.93.092004}{\emph{Phys. Rev.}
  {\bfseries D93} (2016) 092004}
  [\href{https://arxiv.org/abs/1603.02151}{{\ttfamily 1603.02151}}].

\bibitem{Aaboud:2016uuk}
{\scshape ATLAS} collaboration, M.~Aaboud et~al., \emph{{Search for anomalous
  electroweak production of $WW/WZ$ in association with a high-mass dijet
  system in $pp$ collisions at $\sqrt{s}=8$ TeV with the ATLAS detector}},
  \href{https://doi.org/10.1103/PhysRevD.95.032001}{\emph{Phys. Rev.}
  {\bfseries D95} (2017) 032001}
  [\href{https://arxiv.org/abs/1609.05122}{{\ttfamily 1609.05122}}].

\bibitem{Aaboud:2016ffv}
{\scshape ATLAS} collaboration, M.~Aaboud et~al., \emph{{Measurement of
  $W^{\pm}W^{\pm}$ vector-boson scattering and limits on anomalous quartic
  gauge couplings with the ATLAS detector}},
  \href{https://doi.org/10.1103/PhysRevD.96.012007}{\emph{Phys. Rev.}
  {\bfseries D96} (2017) 012007}
  [\href{https://arxiv.org/abs/1611.02428}{{\ttfamily 1611.02428}}].

\bibitem{Aaboud:2017pds}
{\scshape ATLAS} collaboration, M.~Aaboud et~al., \emph{{Studies of $Z\gamma$
  production in association with a high-mass dijet system in $pp$ collisions at
  $\sqrt{s}=$ 8 TeV with the ATLAS detector}},
  \href{https://doi.org/10.1007/JHEP07(2017)107}{\emph{JHEP} {\bfseries 07}
  (2017) 107} [\href{https://arxiv.org/abs/1705.01966}{{\ttfamily
  1705.01966}}].

\bibitem{ATLAS:2018ogo}
{\scshape ATLAS} collaboration, T.~A. collaboration, \emph{{Observation of
  electroweak production of a same-sign $W$ boson pair in association with two
  jets in $pp$ collisions at $\sqrt{s}=13$ TeV with the ATLAS detector}},
  {\emph{ATLAS-CONF-2018-030} (2018) }.

\bibitem{ATLAS:2018ucv}
{\scshape ATLAS} collaboration, T.~A. collaboration, \emph{{Observation of
  electroweak $W^{\pm}Z$ boson pair production in association with two jets in
  pp collisions at $\sqrt{s}$ = 13TeV with the ATLAS Detector}},
  {\emph{ATLAS-CONF-2018-033} (2018) }.

\bibitem{Khachatryan:2014sta}
{\scshape CMS} collaboration, V.~Khachatryan et~al., \emph{{Study of vector
  boson scattering and search for new physics in events with two same-sign
  leptons and two jets}},
  \href{https://doi.org/10.1103/PhysRevLett.114.051801}{\emph{Phys. Rev. Lett.}
  {\bfseries 114} (2015) 051801}
  [\href{https://arxiv.org/abs/1410.6315}{{\ttfamily 1410.6315}}].

\bibitem{Khachatryan:2016vif}
{\scshape CMS} collaboration, V.~Khachatryan et~al., \emph{{Measurement of
  electroweak-induced production of W$\gamma$ with two jets in pp collisions at
  $ \sqrt{s}=8 $ TeV and constraints on anomalous quartic gauge couplings}},
  \href{https://doi.org/10.1007/JHEP06(2017)106}{\emph{JHEP} {\bfseries 06}
  (2017) 106} [\href{https://arxiv.org/abs/1612.09256}{{\ttfamily
  1612.09256}}].

\bibitem{Khachatryan:2017jub}
{\scshape CMS} collaboration, V.~Khachatryan et~al., \emph{{Measurement of the
  cross section for electroweak production of Z$\gamma$ in association with two
  jets and constraints on anomalous quartic gauge couplings in proton--proton
  collisions at $\sqrt{s} = 8$ TeV}},
  \href{https://doi.org/10.1016/j.physletb.2017.04.071}{\emph{Phys. Lett.}
  {\bfseries B770} (2017) 380}
  [\href{https://arxiv.org/abs/1702.03025}{{\ttfamily 1702.03025}}].

\bibitem{Sirunyan:2017fvv}
{\scshape CMS} collaboration, A.~M. Sirunyan et~al., \emph{{Measurement of
  vector boson scattering and constraints on anomalous quartic couplings from
  events with four leptons and two jets in proton--proton collisions at
  $\sqrt{s}=$ 13 TeV}},
  \href{https://doi.org/10.1016/j.physletb.2017.10.020}{\emph{Phys. Lett.}
  {\bfseries B774} (2017) 682}
  [\href{https://arxiv.org/abs/1708.02812}{{\ttfamily 1708.02812}}].

\bibitem{Sirunyan:2017ret}
{\scshape CMS} collaboration, A.~M. Sirunyan et~al., \emph{{Observation of
  electroweak production of same-sign W boson pairs in the two jet and two
  same-sign lepton final state in proton-proton collisions at $\sqrt{s} = $ 13
  TeV}}, \href{https://doi.org/10.1103/PhysRevLett.120.081801}{\emph{Phys. Rev.
  Lett.} {\bfseries 120} (2018) 081801}
  [\href{https://arxiv.org/abs/1709.05822}{{\ttfamily 1709.05822}}].

\bibitem{Sirunyan:2017jej}
{\scshape CMS} collaboration, A.~M. Sirunyan et~al., \emph{{Electroweak
  production of two jets in association with a Z boson in proton-proton
  collisions at $\sqrt{s}= $ 13 TeV}}, {\emph{Submitted to: Eur. Phys. J. C}
  (2017) } [\href{https://arxiv.org/abs/1712.09814}{{\ttfamily 1712.09814}}].

\bibitem{CMS:2018hlo}
{\scshape CMS} collaboration, C.~Collaboration, \emph{{Measurements of the
  $\mathrm{pp}\to\mathrm{WZ}$ inclusive and differential production cross
  section and constraints on charged anomalous triple gauge couplings at
  $\sqrt{s} = 13~\mathrm{TeV}$.}}, {\emph{CMS-PAS-SMP-18-002} (2018) }.

\bibitem{CMS:2018ysc}
{\scshape CMS} collaboration, C.~Collaboration, \emph{{Measurement of
  electroweak WZ production and search for new physics in pp collisions at
  sqrt(s) = 13 TeV}}, {\emph{CMS-PAS-SMP-18-001} (2018) }.

\bibitem{Alwall:2014hca}
J.~Alwall, R.~Frederix, S.~Frixione, V.~Hirschi, F.~Maltoni, O.~Mattelaer
  et~al., \emph{{The automated computation of tree-level and next-to-leading
  order differential cross sections, and their matching to parton shower
  simulations}}, \href{https://doi.org/10.1007/JHEP07(2014)079}{\emph{JHEP}
  {\bfseries 07} (2014) 079} [\href{https://arxiv.org/abs/1405.0301}{{\ttfamily
  1405.0301}}].

\bibitem{Ball:2013hta}
{\scshape NNPDF} collaboration, R.~D. Ball, V.~Bertone, S.~Carrazza,
  L.~Del~Debbio, S.~Forte, A.~Guffanti et~al., \emph{{Parton distributions with
  QED corrections}},
  \href{https://doi.org/10.1016/j.nuclphysb.2013.10.010}{\emph{Nucl. Phys.}
  {\bfseries B877} (2013) 290}
  [\href{https://arxiv.org/abs/1308.0598}{{\ttfamily 1308.0598}}].

\bibitem{Mangano:2002ea}
M.~L. Mangano, M.~Moretti, F.~Piccinini, R.~Pittau and A.~D. Polosa,
  \emph{{ALPGEN, a generator for hard multiparton processes in hadronic
  collisions}},
  \href{https://doi.org/10.1088/1126-6708/2003/07/001}{\emph{JHEP} {\bfseries
  07} (2003) 001} [\href{https://arxiv.org/abs/hep-ph/0206293}{{\ttfamily
  hep-ph/0206293}}].

\bibitem{Lai:1999wy}
{\scshape CTEQ} collaboration, H.~L. Lai, J.~Huston, S.~Kuhlmann, J.~Morfin,
  F.~I. Olness, J.~F. Owens et~al., \emph{{Global QCD analysis of parton
  structure of the nucleon: CTEQ5 parton distributions}},
  \href{https://doi.org/10.1007/s100529900196}{\emph{Eur. Phys. J.} {\bfseries
  C12} (2000) 375} [\href{https://arxiv.org/abs/hep-ph/9903282}{{\ttfamily
  hep-ph/9903282}}].

\bibitem{Cowan:2010js}
G.~Cowan, K.~Cranmer, E.~Gross and O.~Vitells, \emph{{Asymptotic formulae for
  likelihood-based tests of new physics}},
  \href{https://doi.org/10.1140/epjc/s10052-011-1554-0,
  10.1140/epjc/s10052-013-2501-z}{\emph{Eur. Phys. J.} {\bfseries C71} (2011)
  1554} [\href{https://arxiv.org/abs/1007.1727}{{\ttfamily 1007.1727}}].

\bibitem{Barachetti:2016chu}
A.~Barachetti, L.~Rossi and A.~Szeberenyi, \emph{{Final Project Report:
  Deliverable D1.14}}, {\emph{CERN-ACC-2016-0007} (2016) }.
  
  \bibitem{Pascoli:2018heg}
  S.~Pascoli, R.~Ruiz and C.~Weiland,
  \emph{{Heavy Neutrinos with Dynamic Jet Vetoes: Multilepton Searches at $\sqrt{s} = 14,~27,$ and $100$ TeV}},
   [\href{https://arxiv.org/abs/1812.08750}{{\ttfamily 1812.08750}}].
  
  
\bibitem{Sjostrand:2014zea}
  T.~Sjöstrand {\it et al.},
  \emph{An Introduction to PYTHIA 8.2,}
  Comput.\ Phys.\ Commun.\  {\bf 191}, 159 (2015)
  [arXiv:1410.3012 [hep-ph]].


\bibitem{Conte:2012fm}
  E.~Conte, B.~Fuks and G.~Serret,
   \emph{MadAnalysis 5, A User-Friendly Framework for Collider Phenomenology,}
  Comput.\ Phys.\ Commun.\  {\bf 184} (2013) 222
  [arXiv:1206.1599 [hep-ph]].


\bibitem{Conte:2014zja}
  E.~Conte, B.~Dumont, B.~Fuks and C.~Wymant,
   \emph{Designing and recasting LHC analyses with MadAnalysis 5,}
  Eur.\ Phys.\ J.\ C {\bf 74} (2014)  3103
  [arXiv:1405.3982 [hep-ph]].
  
\bibitem{Dumont:2014tja}
  B.~Dumont {\it et al.},
   \emph{Toward a public analysis database for LHC new physics searches using MADANALYSIS 5,}
  Eur.\ Phys.\ J.\ C {\bf 75} (2015) 56
  [arXiv:1407.3278 [hep-ph]].
 

\bibitem{Conte:2018vmg}
  E.~Conte and B.~Fuks,
   \emph{Confronting new physics theories to LHC data with MadAnalysis 5,}
  arXiv:1808.00480 [hep-ph].
  

\end{thebibliography}

\providecommand{\href}[2]{#2}\begingroup\raggedright\endgroup


\end{document}